\documentclass[12pt]{article}

\renewcommand{\baselinestretch}{1.75}

\usepackage{amsmath, amssymb, amsthm, bm, bbm,graphicx, mathtools, enumerate,multirow,booktabs}
\usepackage[shortlabels]{enumitem}
\usepackage{psfrag,epsf}
\usepackage[affil-it]{authblk}
\usepackage{natbib}
\usepackage{hyperref}
\usepackage[usenames]{color}
\usepackage[dvipsnames]{xcolor}
\usepackage[linesnumbered, ruled, lined, commentsnumbered]{algorithm2e}
\usepackage{prettyref,soul}
\usepackage{setspace}
\usepackage{float}
\usepackage[section]{placeins}
\usepackage{caption,subcaption}
\usepackage{rotating}
\usepackage{xr}

\DeclareMathOperator*{\argmin}{arg\,min}

\newcommand{\wkconv}{\rightsquigarrow}

\newcommand{\E}{\mbox{{\rm E}}}

\newcommand{\tr}{\mbox{{\rm tr}}}

\newcommand{\RPD}{\mbox{{\rm RPD}}}

\newcommand{\ve}{\mbox{{\rm vec}}}
\newcommand{\pen}{\mbox{{\rm pen}}}
\newcommand{\dd}{\mathrm{d}}

\newcommand{\bmat}[1]{\begin{bmatrix} #1 \end{bmatrix}}
\newcommand{\bzero}{{\bf 0}}
\newcommand{\bone}{{\bf 1}}

\newcommand{\bA}{{\bf A}}
\newcommand{\bB}{{\bf B}}

\newcommand{\bH}{{\bf H}}

\newcommand{\bba}{{\bf a}}
\newcommand{\bI}{{\bf I}}

\newcommand{\bz}{{\bf z}}

\newcommand{\bP}{{\bf P}}

\newcommand{\barrho}{\bar{\rho}}
\newcommand{\eps}{{\epsilon}}

\newcommand{\thetam}{\mbox{\boldmath $\theta$}}

\newcommand{\Sigmam}{\mbox{\boldmath $\Sigma$}}

\newcommand{\R}{{\mathbb{R}}}
\newcommand{\N}{{\mathbb{N}}}

\newcommand{\oj}{\overline{j}}

\newcommand{\GamX}{\Gammam^{X}}
\newcommand{\GamhX}{\Gammam^{\hX}}
\newcommand{\cGamX}{\check{\Gammam}^{\mX}}
\newcommand{\tGamX}{\tilde{\Gammam}^{\mX}}
\newcommand{\GamY}{\Gammam^{Y}}
\newcommand{\GamhY}{\Gammam^{\hY}}
\newcommand{\cGamY}{\check{\Gammam}^{\mY}}
\newcommand{\tGamY}{\tilde{\Gammam}^{\mY}}
\newcommand{\cGamZ}{\check{\bm \Gamma}^{Z}}

\newcommand{\hX}{{\hat{X}}}
\newcommand{\hY}{{\hat{Y}}}
\newcommand{\hZ}{{\hat{Z}}}
\newcommand{\tX}{{\tilde{X}}}

\newcommand{\tY}{{\tilde{Y}}}

\newcommand{\tZ}{{\tilde{Z}}}

\newcommand{\hthX}[1]{\thetam^{\hX_{#1}}}
\newcommand{\hthY}[1]{\thetam^{\hY_{#1}}}

\newcommand{\hthetam}{{\hat{\thetam}}}

\newcommand{\mH}{{\mathcal{H}}}

\newcommand{\mD}{{\mathcal{D}}}
\newcommand{\hmD}{{\hat{\mathcal{D}}}}

\newcommand{\mS}{{\mathcal{S}}}

\newcommand{\mR}{{\mathcal{R}}}

\newcommand{\mX}{{\mathcal{X}}}
\newcommand{\mY}{{\mathcal{Y}}}
\newcommand{\mZ}{{\mathcal{Z}}}

\newcommand{\kx}{\kappa_{\mathcal{X}}}

\newcommand{\ckx}{{\check{\kappa}_{\mathcal{X}}}}
\newcommand{\bkx}{{\bm \kappa_{\mathcal{X}}}}

\newcommand{\hbkx}{{\hat{\bm \kappa}_{\mathcal{X}}}}
\newcommand{\ky}{\kappa_{\mathcal{Y}}}

\newcommand{\cky}{{\check{\kappa}_{\mathcal{Y}}}}
\newcommand{\bky}{{\bm \kappa_{\mathcal{Y}}}}

\newcommand{\hbky}{{\hat{\bm \kappa}_{\mathcal{Y}}}}
\newcommand{\kz}{\kappa_{\mathcal{Z}}}

\newcommand{\ckz}{{\check{\kappa}_{\mathcal{Z}}}}
\newcommand{\bkz}{{\bm \kappa_{\mathcal{Z}}}}

\newcommand{\hbkz}{{\hat{\bm \kappa}_{\mathcal{Z}}}}
\newcommand{\Hkx}{{\mH(\kappa_{\mathcal{X}})}}
\newcommand{\Hky}{{\mH(\kappa_{\mathcal{Y}})}}
\newcommand{\Hkz}{{\mH(\kappa_{\mathcal{Z}})}}
\newcommand{\Hkxy}{{\mH(\kappa_{\mathcal{X}}\otimes \kappa_{\mathcal{Y}})}}

\newcommand{\Xp}{{X^{\prime}}}

\newcommand{\Bpqa}{B_{p,q}^{\alpha}}

\newcommand{\bpqa}{b_{p,q}^{\alpha}}

\newcommand{\ip}[1]{\langle #1 \rangle}

\newcommand{\norm}[1]{\| #1 \|}

\newcommand{\nmB}[1]{\| #1 \|_{\Bpqa}}

\newcommand{\nmb}[1]{\| #1 \|_{\bpqa}}

\newcommand{\nmbaZ}[1]{\| #1 \|_{b^{\alpha_Z}}}

\newcommand{\nmba}[1]{\| #1 \|_{b^{\alpha}}}
\newcommand{\nmbb}[1]{\| #1 \|_{b^{\beta}}}
\newcommand{\nmbbZ}[1]{\| #1 \|_{b^{\beta_Z}}}

\newcommand{\ray}[1]{{\color{cyan}\bf [Raymond: #1]}}

\newcommand{\blue}[1]{{\color{blue}#1}}
\definecolor{Area4}{RGB}{163,163,255}
\definecolor{Area3a}{RGB}{159,159,0}
\definecolor{Area3b}{RGB}{201,255,255}
\definecolor{Area1}{RGB}{255,0,0}
\definecolor{Area2}{RGB}{255,195,135}

\newtheorem{thm}{Theorem}
\newtheorem{defi}{Definition}
\newtheorem{lem}{Lemma}

\newtheorem{prop}{Proposition}
\newtheorem{ex}{Example}
\newtheorem{rmk}{Remark}

\begin{document}
\def\spacingset#1{\renewcommand{\baselinestretch}%
{#1}\small\normalsize} \spacingset{1}

{ 
  \title{\vspace{-2cm}\bf A Wavelet-Based Independence Test for Functional Data with an Application to MEG Functional Connectivity}
  \author{Rui Miao,
    Xiaoke Zhang 
\thanks{The research of Xiaoke Zhang is partially supported by National Science Foundation grant DMS-1832046.}\\
    Department of Statistics, The George Washington University\\
    and \\
    Raymond K. W. Wong
\thanks{The research of Raymond K. W. Wong is partially supported by National Science Foundation grants DMS-1806063, DMS-1711952 and CCF-1934904.}\\
    Department of Statistics, Texas A\&M University
    \vspace{-1cm}
    }
\date{}
  \maketitle
} 

\begin{abstract}
Measuring and testing the dependency between multiple random functions is often
an important task in functional data analysis.
In the literature, a model-based method relies on a model which is subject to the risk of model misspecification, while 
a model-free method only provides a correlation measure which is inadequate to test independence. In this paper, we adopt the Hilbert-Schmidt Independence Criterion (HSIC) to measure the dependency between two random functions.
 We develop a two-step procedure by first pre-smoothing each function based on its %
 discrete and noisy measurements 
 and then applying the HSIC to recovered functions. 
To ensure the compatibility between the two steps such that the effect of the
pre-smoothing error on the subsequent HSIC is asymptotically negligible when the data are densely measured, we
propose {a new wavelet thresholding method for pre-smoothing and to use Besov-norm-induced kernels for HSIC.} 
We also provide the corresponding asymptotic analysis.
  The superior numerical performance of the proposed method over existing ones is demonstrated in a simulation study. Moreover, in an magnetoencephalography (MEG) data application, the functional connectivity patterns identified by the proposed method are more anatomically interpretable than those by existing methods.
\end{abstract}

\noindent%

\vfill

\newpage
\spacingset{1.45} %

\section{Introduction} \label{sec:intro}

In recent decades, functional data analysis (FDA)
 has developed rapidly due to
 a huge and increasing number of
   datasets collected in the form of curves, surfaces and volumes. General introductions to
   the subject
     may be found in a few monographs \citep[e.g.,][]{RamsS05, FerrV06}.
In many scientific fields, %
measurements are taken from multiple random functions per subject and the dependency between these functions is of interest.  
For instance, neuroscientists are interested in functional connectivity patterns between signals at multiple brain regions, which are measured over time in functional magnetic resonance imaging data. 
It is thus an important task in FDA to measure their dependency and to further test the significance of the dependency.
Among extensive relevant research endeavors,
most dependency test methods can be categorized as either model-based or model-free.

A model-based method typically infers the dependency between multiple functions by first assuming a functional regression model \citep[see, e.g.,][for a survey]{Morr15} which characterizes their structural relationship, and then testing the significance of the assumed model. See examples of model-based methods by \citet{Guo02, HuanWZ02, ShenF04, AntoS07} for concurrent/varying-coefficient models and by \citet{KokoMS08, ChenJF20} for function-on-function regression models. 
The main disadvantage of a model-based method is its reliance on correct model specification. If the model is misspecified, the inference is not well grounded and might be inaccurate.

A model-free method can avoid the misspecification issue associated with model-based methods since it typically quantifies the dependency between random functions by a correlation measure, without assuming any particular model. As a natural extension of the canonical correlation for multivariate data, the functional canonical correlation is a popular correlation measure for functional data \citep[e.g.,][]{LeurMS93, %
HeMW03, %
EubaH08, %
ShinL15}. However, %
it is plagued by 
the involvement of inverting a covariance operator, which is an ill-posed problem and often requires proper regularizations. The dynamical correlation \citep{DubiM05, SangWC19} 
and temporal correlation \citep{ZhouLW18} are two functional correlation measures without the aforementioned inverse problem. The former measures the angle between two random functions in the $L^2$ space. The latter essentially computes the Pearson correlation between two random functions at each time point %
and then averages all pointwise Pearson correlations over the time domain. However, since uncorrelatedness does not imply independence, these functional correlations are insufficient to test independence. Recently a few model-free approaches have been developed to test 
mean independence for functional data \citep[e.g.,][]{%
PatiSS16, LeeZS20}, %
but they can only test a weaker notion of independence.

In this paper we develop a model-free independence test for functional data. Under the reproducing kernel Hilbert space (RKHS) framework, we propose to 
use the Hilbert-Schmidt Independence Criterion \citep[HSIC, e.g.,][]{GretBS05, GretFTCSS2008} 
to measure the dependency between two random functions.
An appealing property is that HSIC endowed with characteristic kernels is zero if and only if the two random functions are independent.
However, the application of HSIC requires fully observed and noiseless functional data,
while in practice functional data are always discretely measured and contaminated by noise.
To tackle this problem, one may perform a two-step procedure: first pre-smooth the data, and then apply HSIC to the resulting functions.
Clearly, %
pre-smoothing will affect the performance of HSIC.
Indeed, the functional distance with respect to which the asymptotic convergence of the pre-smoothing procedure is measured 
is crucial, as HSIC is fundamentally based on a functional distance.
Some common pre-smoothing procedures do not have existing convergence results on the required functional distance, and hence may not be compatible;
namely, the pre-smoothing error may have a profound effect on the subsequent HSIC.
See Section \ref{sec:method} for more discussion.
In this work, we carefully design our procedure to ensure that 
 the two steps are compatible.
{For the first step, we propose a new 
wavelet thresholding method}
while we use 
Besov-norm-induced kernels for HSIC in the second step.
  We can show that these choices in the two steps 
  are theoretically compatible if the functional data are sufficiently densely measured.
  See Section \ref{sec:theory} for details.
Our work is motivated by the Human Connectome Project (HCP, \url{https://www.humanconnectome.org}) from which various brain imaging datasets are publicly accessible. 
In Section \ref{sec:data}, the application of our method to a magnetoencephalography (MEG) dataset from HCP is capable of identifying anatomically interpretable functional connectivity patterns,
{suggesting a great potential of the proposed method in the study of functional connectivity
between brain regions.}

The main contribution of this paper is three-fold.
First, we design some suitable kernels such that the corresponding HSIC can identify
the independence of a pair of random functions of which sample paths belong to Besov spaces,
a larger class of functions than Sobolev spaces which are popular in RKHS modeling.
We propose to use the Besov sequence norm for the wavelet coefficients of these random functions %
induce such kernel, which is shown to be characteristic.
Second, for dense
functional data, we develop the asymptotic distribution of the empirical HSIC
based on pre-smoothed functions by wavelet thresholding. {To theoretically guarantee the compatibility between the pre-smoothing and empirical HSIC, we propose a new wavelet thresholding method that can efficiently reduce the pre-smoothing error measured by the Besov sequence norm used in the empirical HSIC when the noise is nearly independent.}
Since the asymptotic distribution involves many unknown quantities, we suggest a permutation test in practice and prove that not only can the test control the Type I error probability but also it is consistent. The theoretical results show that the two steps in our proposed procedure are compatible. Finally, we propose a data-adaptive approach to tuning the smoothness parameter for the Besov norm needed to induce the kernel for HSIC. It is numerically shown that this approach is able to enhance the sensitivity of HSIC to detecting dependencies at high frequencies. 

The rest of the paper proceeds as follows. Section \ref{sec:background} provides a brief introduction to 
HSIC. The two-step procedure for the proposed wavelet-based HSIC test is given in Section \ref{sec:method}. Its asymptotic properties are presented in Section \ref{sec:theory}.
Section \ref{sec:select} discusses 
tuning parameter selection. The numerical performance of the proposed method is illustrated in a simulation study in Section \ref{sec:simu} and an MEG functional connectivity study in Section \ref{sec:data} where it is also compared with representative existing methods. Section \ref{sec:disc} concludes the paper. %
The code to implement the proposed method is publicly available on GitHub (\url{https://github.com/rui-miao/wavHSIC}).

\section{Hilbert-Schmidt Independence Criterion} \label{sec:background}

In this section we give a brief introduction to %
HSIC. Let $X$ and $Y$ be two random functions of which sample paths belong to %
function spaces $\mX$ and $\mY$ respectively, 
and 
$\Hkx$ and $\Hky$ be the RKHS equipped with %
kernels $\kx$ and $\ky$ defined on %
$\mX\times \mX$ and $\mY\times\mY$ respectively. 

HSIC requires that both $\kx$ and $\ky$ are characteristic, in the sense that two probability measures $P = P'$ if and only if $
\bP^{\kz}(P) =  \bP^{\kz}(P')$ where $\bP^{\kz}(P) = E_{P} \{\kz(Z, \cdot) \}$ for a random function $Z \in \mZ$ which follows $P$ and $(Z, \mZ) = (X, \mX)$ or $(Y, \mY)$. A characteristic kernel may be induced by a strong negative type semi-metric (see Definition \blue{S1} and Proposition \blue{S1} in the supplementary material). 
Denote the joint probability measure of $X$ and $Y$ by $P_{XY}$ and their marginal probability measures by $P_X$ and $P_Y$ respectively. 
Since $\kx$ and $\ky$ are characteristic, $P_X$ and $P_Y$ are fully characterized by 
$
\bP^{\kx}(P_X) = E_{P_X} \{\kx(X, \cdot) \}$ and 
$\bP^{\ky}(P_Y) = E_{P_Y} \{\ky(Y, \cdot)\}
$
respectively.
Let $\bP^{\kx \otimes \ky}(P_{XY}) = E_{P_{XY}} \{(\kx\otimes\ky)((X, Y), (*, \cdot))\}$,
where the tensor product kernel $\kx\otimes\ky$ is defined by 
$(\kx\otimes\ky)((x, y), (x', y')) =
\kx(x,x')\ky(y, y'), x, x' \in \mX, y, y' \in \mY$. 

\citet{SejdSG13} showed that $X$ and $Y$ are independent, i.e., $P_{XY}=P_X P_Y$, if and only if
$\bP^{\kx\otimes\ky}(P_{XY})=\bP^{\kx}(P_X)\bP^{\ky}(P_Y)$, although $\kx\otimes\ky$ is not 
characteristic 
for all probability
measures on $\Hky\times\Hky$. Therefore, to test the independence between $X$ and $Y$, it suffices to study the difference between $\bP^{\kx\otimes\ky}(P_{XY})$ and $\bP^{\kx}(P_X)\bP^{\ky}(P_Y)$. Since $\bP^{\kx}(P_X) \in \Hkx$, $\bP^{\ky}(P_Y) \in
\Hky$ and $\bP^{\kx \otimes \ky}(P_{XY}) \in \Hkxy$ where $\Hkxy$ is the RKHS equipped with 
$\kx\otimes\ky$, HSIC may be used to measure this difference under the norm of $\Hkxy$.
\begin{defi}[HSIC]
  \label{def:HSIC}
  Suppose that $\int_{\mX}
  \kx(x,x)\dd P_X(x) < \infty$ and $\int_{\mY} \ky(y,y)\dd P_Y(y)<\infty$. %
  The HSIC of $P_{XY}$ %
  is defined by 
\vspace{-0.4cm}
\begin{eqnarray*}
  \gamma(P_{XY},\kx,\ky) &=& \norm{\bP^{\kx \otimes \ky}(P_{XY})- \bP^{\kx}(P_X) \bP^{\ky}(P_Y)}_{\Hkxy}^2\\
  &=& 4\int_{\mX\times\mY}\int_{\mX\times\mY} \kx(x,x^{\prime})\ky(y,y^{\prime})\dd (P_{XY}-P_XP_Y)(x,y)\dd (P_{XY}-P_XP_Y)(x^{\prime},y^{\prime}).
\vspace{-0.4cm}
  \end{eqnarray*}
\end{defi}

\vspace{-0.4cm}
In practice with $\{(X_i, Y_i): i=1, \ldots, n\}$ which are independently and identically distributed (i.i.d.) copies of $(X, Y)$, the sample versions of $\bP^{\kx}(P_X)$, $\bP^{\ky}(P_Y)$ and $\bP^{\kx \otimes \ky}(P_{XY})$ are defined by 
$
\bP^{\kx}(P_{n, X}) = n^{-1}\sum_{i=1}^n \kx(X_i, \cdot)
$, 
$
\bP^{\ky}(P_{n, Y}) = n^{-1}\sum_{i=1}^n \ky(Y_i, \cdot)$,  and 
$
\bP^{\kx \otimes \ky}(P_{n, XY}) = n^{-1}\sum_{i=1}^n\{(\kx\otimes\ky)((X_i, Y_i), (*, \cdot))\}.
$
Obviously $\bP^{\kx}(P_{n, X}) \in \Hkx$, $\bP^{\ky}(P_{n, Y}) \in \Hky$ and $\bP^{\kx \otimes \ky}(P_{n, XY}) \in \Hkxy$, so we can obtain a sample version of HSIC as follows. 

\begin{defi}[Empirical HSIC]\label{def:EmpHSIC}
Under the same setting in Definition \ref{def:HSIC}, the empirical HSIC, which is an estimator of HSIC, 
is defined by
\vspace{-0.4cm}
\begin{align*}
  &\gamma(P_{n,XY},\kx,\ky) = \norm{\bP^{\kx \otimes \ky}(P_{n, XY})- \bP^{\kx}(P_{n,X}) \bP^{\ky}(P_{n,Y})}_{\Hkxy}^2\\
  &= 4\int_{\mX\times\mY}\int_{\mX\times\mY} \kx(x,x^{\prime})\ky(y,y^{\prime})\dd (P_{n,XY}-P_{n,X}P_{n,Y})(x,y)\dd (P_{XY}-P_{n,X}P_{n,Y})(x^{\prime},y^{\prime}).
\end{align*}
\end{defi}
\vspace{-0.4cm}
By \citet{SejdSG13}, the empirical HSIC can be rewritten as
\vspace{-0.4cm}
\begin{equation*}%
\gamma(P_{n,XY},\kx,\ky)= n^{-2} \tr(\GamX \bH\GamY \bH),
\vspace{-0.4cm}
\end{equation*}
where $\GamX = (\kx(X_i,X_j))_{1 \leq i, j \leq n}$ and $\GamY=(\ky(Y_i,Y_j))_{1 \leq i, j \leq n}$ 
are Gram matrices, and $\bH = \bI_n - n^{-1}\bone_n \bone_n^{\top}$ is the centering matrix with the $n \times n$ identity matrix $\bI_n$ and $\bone_n=(1, \ldots, 1)^\top$ of dimension $n$.

\section{Methodology} \label{sec:method}

Suppose that bivariate functional data $\{(X_i, Y_i): i=1, \ldots, n\}$ 
collected from $n$ subjects are i.i.d.~copies of a pair of random functions $(X,Y)$, which, without loss of generality, is defined on the domain $[0, 1] \times [0, 1]$.
Let the sample paths of $X$ and $Y$ belong to function spaces $\mX$ and $\mY$ respectively.
In many applications such as brain imaging analysis, the measurements of each function 
are sampled at a discrete and regular grid and subject to noise contamination. Hence we assume that the observations are
$
\{(\tX_{il}, \tY_{il}):=(X_i(T_l) + e_{il}^X, Y_i(T_l) + e_{il}^Y):
i=1, \ldots, n; l=1, \ldots, m\}, 
$
where $\{T_l= (l-1)/m: l=1, \ldots,
m\}$ is a regular grid with $m=2^{J+1}$ for some integer $J > 0$ and the two sets of %
mean-zero 
random noise, $\{e_{il}^X: i=1, \ldots, n; l=1, \ldots, m\}$
and $\{e_{il}^Y: i=1, \ldots, n; l=1, \ldots, m\}$, 
are independent of each
other and of $\{(X_i, Y_i): i=1, \ldots, n\}$. {The error terms in each set are further assumed to be identically distributed, independent across subjects, but %
\textit{possibly} dependent within each subject.}
{We defer the discussion on the error dependence structures to Section \ref{sec:theory}. %
}
Our goal is to formulate an HSIC-based test for the independence
between $X$ and $Y$ via %
$\{(\tX_{il}, \tY_{il}): i=1, \ldots, n; l=1, %
\ldots, m\}$.
For simplicity we assume that all functions share the same measurement grid and $m=2^{J+1}$, but the proposed method is applicable with minor modifications if the grid is irregular, the functions are measured at different grids, or $m \neq 2^{J+1}$ (see Remark \ref{rmk:method}).

Due to the success of existing HSIC-based independence tests for multivariate data,
it is tempted to treat the discretized observations as multivariate data and directly apply
existing methods.
However, there are two issues with this approach.
First, in order to capture enough information, $m$ should be large enough, which naturally leads to high-dimensional data. Without reasonable structure across these $m$ dimensions,
HSIC does not perform well.
In the FDA literature, %
modeling the sample paths with certain form of smoothness has been shown an empirically successful
strategy in many applications.
It is beneficial to incorporate smoothness structure during the design of a tailor-made HSIC method.
Second, the discretized observations are contaminated by noise. Hence these raw observations
are indeed not ``smooth" but the noiseless ones are.

The proposed method is directly based on the definition of HSIC (Definition \ref{def:HSIC})
when applied to random functions.
Clearly, the application of such HSIC requires the trajectories of all random functions to be fully observed and noiseless.
Thus, with discrete and noisy measurements in practice, a natural idea is to perform pre-smoothing to recover these trajectories followed by
applying HSIC to random functions.
However, the compatibility of these two steps is generally unclear. Namely, it is non-trivial to know whether the pre-smoothing error (measured in a certain norm) would have a profound effect on the subsequent HSIC-based test.
For instance, if the sample paths of all random functions are assumed to belong to a Sobolev space, it is seemingly reasonable to pre-smooth each trajectory by a smoothing spline %
followed by the HSIC based on Sobolev-norm-induced kernels. However, the compatibility of the two steps is unknown since there is no theoretical result to guarantee that the pre-smoothing error under a Sobolev norm converges to zero, although the corresponding results with respect to the $L^2$ or empirical norm exist. 

To address this compatibility issue, we propose to use HSIC based on Besov-norm-induced kernels for testing independence
under the assumption that the sample paths of all random functions belong to Besov spaces, a larger class of functions than Sobolev spaces. %
To recover each trajectory,
{we develop a new wavelet thresholding method for pre-smoothing.} 
Its theoretical compatibility with the proposed HSIC is given in Section \ref{sec:theory}. 
In the rest of this section, we first introduce wavelets \citep[e.g.,][]{Ogde97, Vida09, MorePV17} together with  
other related results
and then %
the details of the proposed two-step procedure. %
\vspace{-0.3cm}

\subsection{Wavelets and Besov Sequence Norms} \label{sec:wavelets}
Following the Cohen-Daubechies-Jawerth-Vial (CDJV) construction
\citep{CoheDJV93}, let father and mother wavelets be $\phi, \psi \in C^R[0,1]$ respectively 
with $D$ vanishing moments \citep[e.g.,][]{Daub1992} where $C^R[0,1]$ is the space of all functions on $[0, 1]$ with $R$-th
order continuous derivatives. 
We consider a Besov space $\Bpqa[0,1]$ with norm $\norm{\cdot}_{\Bpqa[0,1]}$ of which smoothness parameter $\alpha$ satisfies $1/p < \alpha < \min\{R, D\}$ such that $\Bpqa[0,1]$ can be embedded continuously in
$C[0,1]$. Formal definitions of $\Bpqa[0,1]$ and its norm
$\norm{\cdot}_{\Bpqa[0,1]}$ are given in Section \blue{S1.2} in the supplementary material. 
Then %
for any function $f\in \Bpqa[0,1]\cap L^2[0,1]$ and a fixed coarse scale $L$, 
we have the following decomposition
\vspace{-0.4cm}
\begin{equation}
\label{eq:waveExpan}
\vspace{-0.4cm}
f(t) = \sum_{k=0}^{2^L-1} \xi_k  %
\{2^{L/2}\phi(2^L t  - k)\} + \sum_{j\geq L}\sum_{k=0}^{2^j-1} \theta_{j,k} %
\{2^{j/2}\psi(2^j t-k)\}, \quad t \in [0, 1].
\end{equation}
Denote $\theta_{j,k}=\xi_{2^j+k}, 0\leq j<L,0\leq k <2^j$ and
$\theta_{-1,0}=\xi_0$. Based on the wavelet coefficients of $f$, 
$
\thetam^{f}=((\thetam_{-1}^{f})^{\top},(\thetam_{0}^{f})^{\top}, \dots,(\thetam_{L}^{f})^{\top}, (\thetam_{L+1}^{f})^{\top},\dots)^\top
$
where $\thetam_{j}^f =
(\theta_{j,0},\theta_{j,1},\dots,\theta_{j,2^j-1})^{\top}$ and $\thetam_{-1}^f =
\theta_{-1,0}$,
the Besov sequence norm $\norm{\cdot}_{b_{p, q}^{\alpha}}$ \citep[e.g.,][]{DJKP1995,JohnS05} is defined by 
\vspace{-0.4cm}
\begin{equation} \label{eq:Besovseqnorm}
\vspace{-0.4cm}
\nmb{\thetam^f} = \left( \sum_{j\geq -1} 2^{jsq} \norm{\thetam_{j}^f}_p^q \right)^{1/q}, \quad s = \alpha+1/2 - 1/p,
\end{equation}
where 
$\norm{\cdot}_p$ refers to the $\ell_p$-norm for vectors. Denote the corresponding space by $b_{p, q}^{\alpha} = \{\bba: \nmb{\bba} < \infty\}$. 
Note that the two norms $\nmB{\cdot}$ and $\nmb{\cdot}$ are equivalent \citep[e.g.,][]{DevoL93,DJKP1995} and 
obviously $b_{p, q}^{\alpha} \subset b_{p, q}^{\beta}$ if $\beta \leq \alpha$.
{
  In practice, if $f$ is observed at $m=2^{J+1}$ dyadic time points
  $\{0/m,1/m,\dots, (m-1)/m\}$, the discrete wavelet transformation can be used to
  calculate the wavelet coefficients $\thetam^f$ with $\thetam_j^f=\bzero$ when
  $j>J$. Then 
  we can %
  denote
  $\thetam^f = ((\thetam_{-1}^{f})^{\top},(\thetam_{0}^{f})^{\top}, \dots,(\thetam_{J}^{f})^{\top})^\top$.
}

We can show that some Besov sequence norm can induce a characteristic kernel, which is required by HSIC.
\begin{thm}\label{thm:BesovNegType}
  For $0<q'<q\leq p\leq 2$, $0\leq \alpha\leq \alpha'$ and $\alpha'>1/p$, let the semi-metric $\rho_{b_{p, q}^{\alpha}}(f,g) = \nmb{\thetam^f-\thetam^g}^{q'}$ 
for $f,g\in B_{p,q}^{\alpha'}[0,1]$, where $\thetam^f$ and $\thetam^g$ are the wavelet
coefficients of $f$ and $g$ respectively. The function induced by $\rho_{b_{p,
q}^{\alpha}}$, which is $\kappa\left(z,z^{\prime}\right) = \rho_{b_{p,
q}^{\alpha}}(z, 0) +
  \rho_{b_{p, q}^{\alpha}} \left( z^{\prime}, 0 \right) - \rho_{b_{p, q}^{\alpha}} \left( z,z^{\prime} \right)$,
   $z, z^{\prime} \in B_{p,q}^{\alpha'}[0,1]$, is a characteristic kernel. 
\end{thm}
\vspace{-0.2cm}
The proof of Theorem \ref{thm:BesovNegType} is given in Section \blue{S2.1} in the supplementary material. 
By Theorem \ref{thm:BesovNegType}, we can define HSIC properly based on kernels induced by Besov sequence norms. 
For simplicity, hereafter we focus on popular choices of $p=q=2$ and
$q'=1$. Accordingly we
abbreviate $B_{2, 2}^\alpha[0, 1]$ and $b_{2, 2}^{\alpha}$ to $B^\alpha$ and 
$b^\alpha$ respectively, and 
the kernel functions are
\vspace{-0.4cm}
$$\kz(z_1,z_2) = \norm{\thetam^{z_1}}_{b^{\beta_Z}} +
\norm{\thetam^{z_2}}_{b^{\beta_Z}} - \norm{\thetam^{z_1} -
\thetam^{z_2}}_{b^{\beta_Z}},\quad z_1,z_2\in \mZ, 0\leq \beta_Z\leq \alpha_Z, 
\vspace{-0.4cm}
$$
for $(Z,\mZ)=(X,\mX)$ and $(Y,\mY)$.

\subsection{Two-Step Procedure} \label{sec:twostep}
Let $Z=X$ or $Y$. Under the setting in Section \ref{sec:wavelets}, %
we assume $Z \in B^{\alpha_Z}$ where $1/2 < \alpha_Z < \min\{R, D\}$. Note that
$B^{\alpha_Z} \subset B^{\beta_Z}$ for $0 < \beta_Z < \alpha_Z$ so $Z \in
B^{\beta_Z}$ as well. 
To test the
independence between $X$ and $Y$ based on their discretely measured and noisy
observations, we propose to first denoise each function and then apply HSIC
to the recovered functions. The two-step procedure is explicitly stated as follows:

\vspace{-0.4cm}

\paragraph{Step 1} By the decomposition \eqref{eq:waveExpan} and the resolution limitation due
to a finite number of measurements $m=2^{J+1}$ taken for each subject,  %
{we obtain the initial wavelet coefficient estimates for each $Z_i$, denoted by $\thetam^{\tZ_i}=((\thetam_{-1}^{\tZ_i})^{\top}, (\thetam_{0}^{\tZ_i})^{\top}, \dots, (\thetam_{J}^{\tZ_i})^{\top})^{\top}$, via the discrete wavelet transformation %
with the coarse scale $L_Z$. The coarse scale $L_Z$ may be selected by cross-validation or domain knowledge.}
We propose to denoise $\thetam^{\tZ_i}$ and accordingly obtain 
$\thetam^{\hZ_i}=((\thetam_{-1}^{\hZ_i})^{\top}, (\thetam_{0}^{\hZ_i})^{\top},
\dots, (\thetam_{J}^{\hZ_i})^{\top})^{\top}$ as follows. First, we let $\thetam_j^{\hZ_i}=\thetam_j^{\tZ_i}$ for $j=-1,\dots,L_Z-1$. Moreover, we apply the following penalized least squares to obtain $\thetam_{j}^{\hZ_i}, j=L_Z,\dots,J$: 
\vspace{-0.4cm}
\begin{equation}\label{eq:plse2}
\thetam_{j}^{\hZ_i} = \argmin_{\thetam_j} \left\{\norm{\thetam_j^{\tZ_i} - \thetam_j}_2^2 + \delta_{Z,j}^2 \pen_j(\norm{\thetam_j}_0) \right\}, \quad j=L_Z,\dots,J, %
\vspace{-0.4cm}
\end{equation}
where $\norm{\cdot}_2$ denotes the Euclidean norm, $\norm{\cdot}_0$ denotes the
number of non-zero elements, %
$\delta_{Z,j}=2^{\varsigma_Z
  j}\delta_Z$ is the noise standard deviation at the resolution level $j$ with $\delta_Z > 0$, 
 and the penalty 
$\pen_j(k) = k\zeta_Z \{1+\sqrt{2(1+2\varsigma_Z)\log (\tau_j m_j/k)}\}^2$ that depends on %
  $\zeta_Z>1$, $\varsigma_Z > -1/2$, %
$ m_j = 2^j$, and 
  $\tau_j = \tau_Z 2^{2\alpha_Z(j-j_{\#}^Z)_+}$ with $\tau_Z>e$ and 
  $j_{\#}^Z=(1+(\varsigma_Z+1/2)/\alpha_Z)/(\alpha_Z+\varsigma_Z+1/2) \cdot
  \log_2 \delta_Z^{-1}$. %

The proposed procedure in (\ref{eq:plse2}) is capable of denoising %
a certain type of correlated noise (see technical assumptions in Theorem \ref{thm:denoise} in Section \ref{sec:theory}). Compared to the penalty (12.34) in \citet{John19}, we employ a different 
$\tau_j$ in the penalty in (\ref{eq:plse2})
such that the pre-smoothing error measured by the Besov sequence norm used in the empirical HSIC in Step 2 below converges to zero if $m$ diverges to infinity (see Theorem \ref{thm:denoise} in Section \ref{sec:theory}). This can guarantee the compatibility between this and the next steps. %

Similar to %
\citet{JohnP14}, to %
obtain the estimate $\thetam_{j}^{\hZ_i}$, $L_Z \leq j \leq J$, defined in \eqref{eq:plse2}, 
one may apply the level-wise hard thresholding 
as follows: For each level $L_Z \leq j \leq J$, 
let $|\theta_{j,(k)}^{\tZ_i}|$ be the $k$-th term after the elements of $\thetam^{\tZ_i}_j$ are sorted 
in a decreasing
order of their absolute values, namely
$|\theta_{j,(0)}^{\tZ_i}|\geq|\theta_{j,(1)}^{\tZ_i}|\geq\dots\geq|\theta_{j,(2^j-1)}^{\tZ_i}|$. 
Then the hard threshold at level $j$ is %
$
\delta_{Z,j}\sqrt{\pen_j(\hat k_j) - \pen_j(\hat k_j-1)}, 
$
where $
\hat k_j = \argmin_{k\geq0} \left\{\sum_{k'\geq k}|\theta^{\tZ_i}_{j,(k')}|^2 + \delta_{Z,j}^2\pen_j(k) \right\}. 
$
Detailed steps of solving \eqref{eq:plse2} are summarized in Algorithm
\ref{alg:step1}. The discussion of tuning parameter selection
is deferred to
Section \ref{sec:select}.

\vspace{-0.4cm}
\paragraph{Step 2} Since the %
wavelet coefficient estimates $\hthX{i} \in b^{\alpha_X} \subset b^{\beta_X}$ and $\hthY{i} \in b^{\alpha_Y} \subset b^{\beta_Y}$, $i=1,\ldots, n$, for $\beta_X < \alpha_X$ and $\beta_Y < \alpha_Y$, 
we may apply HSIC to the denoised functions where the kernels $\kx$ and $\ky$ are induced by $\rho_{b^{\beta_X}}$ and $\rho_{b^{\beta_Y}}$
respectively as defined in Theorem \ref{thm:BesovNegType}. Explicitly, we have 
$
\gamma(P_{n,\hX\hY},\kx,\ky) = n^{-2} \tr(\GamhX \bH\GamhY \bH),
$
where
\vspace{-0.4cm}
\begin{align*}
\GamhX &= \left( \norm{\hthX{i}}_{b^{\beta_X}} + \norm{\hthX{j}}_{b^{\beta_X}}
- \norm{\hthX{i} - \hthX{j}}_{b^{\beta_X}} \right)_{1 \leq i, j \leq n},
\quad\mbox{and}\\
\GamhY &= \left( \norm{\hthY{i}}_{b^{\beta_Y}}  + \norm{\hthY{j}}_{b^{\beta_Y}}
- \norm{\hthY{i} - \hthY{j}}_{b^{\beta_Y}} \right)_{1 \leq i, j \leq n}.
\end{align*}
By adopting 
$\rho_{b^{\beta_X}}$ and $\rho_{b^{\beta_Y}}$ where $\beta_X < \alpha_X$ and $\beta_Y < \alpha_Y$ to construct kernels, we are able to make the pre-smoothing step theoretically compatible with the HSIC. As revealed in Theorems \ref{thm:denoise} and \ref{thm:wkconvD} in Section \ref{sec:theory} below, if the observations of all functions are sufficiently dense, the denoising error %
is asymptotically negligible in the asymptotic distribution of the HSIC. This is a key benefit of using wavelets and Besov norms for pre-smoothing.

In Section \ref{sec:theory}, the asymptotic distribution of $\gamma(P_{n,\hX\hY},\kx,\ky)$ is developed in Theorem \ref{thm:wkconvD} under the independence hypothesis.
Despite its theoretical appeal,
the asymptotic distribution unfortunately involves many unknown quantities. Therefore, we suggest using permutations to perform the independence test %
which, as shown in Theorem \ref{thm:perm}, can control the Type I error probability and is also consistent.

\begin{rmk}\label{rmk:method}
Since denoising is performed separately for each function and subject, the proposed method is applicable when the functions of different subjects
are not measured at the same grid.
For $m \neq 2^{J+1}$ at %
possibly irregular and uncommon 
designs, %
linear interpolation may be applied if the original measurement
  resolution is sufficiently high \citep[e.g.,][]{KovaS00}. 
  We demonstrate the satisfactory performance of this strategy via a simulation study, and the corresponding results are given in Section \blue{S3} of the supplementary material.
\end{rmk}

\begin{algorithm}[H]
  \SetKwInOut{Input}{Input}\SetKwInOut{Output}{Output}
  \SetKwFunction{DescendingSort}{DescendingSort}
  \Input{$\{\thetam_j^{\tZ_i}: j=L_Z,\dots,J\}$; \\
  fixed tuning parameters $\zeta_Z>1,\tau_Z>e,\varsigma_Z>-1/2,\delta_Z>0$.}
  \Output{$\{\thetam_j^{\hZ_i}: j=L_Z,\dots,J\}$.}

  \For{$j\leftarrow L_Z$ \KwTo $J$}{
  $|\theta_{j,(0)}^{\tZ_i}|\geq |\theta_{j,(1)}^{\tZ_i}|\geq\dots\geq
  |\theta_{j,(2^j-1)}^{\tZ_i}|\leftarrow$ sort $\{
    |\theta_{j,k}^{\tZ_i}| \}_{k=0}^{2^j-1}$ in descending order\;
$\theta_j^{\text{Thresh}}\leftarrow\delta_{Z,j}\sqrt{\pen_j(\hat k_j) - \pen_j(\hat k_j-1)},$
where $ \hat k_j = \argmin_{k\geq0} \left\{\sum_{k'\geq k}|\theta^{\tZ_i}_{j,(k')}|^2 + \delta_{Z,j}^2\pen_j(k) \right\}$\;
\For{$k\leftarrow 0$ \KwTo $2^j-1$}{
$\theta_{j,k}^{\hZ_i}\leftarrow\theta_{j,k}^{\tZ_i}\mathbb{I}(|\theta_{j,k}^{\tZ_i}|\geq\theta_j^{\text{Thresh}})$
}
  }
  \caption{
  Solving \eqref{eq:plse2} by wavelet thresholding.}\label{alg:step1}
\end{algorithm}
\begin{rmk}\label{rmk:complexity}
  In Step 1, the time complexity for the
   discrete wavelet transformation
  is $O(m)$ for each subject \citep{CoheDJV93} and so is that for denoising. In Step 2, the
  time complexity for calculating Gram matrices is $O(mn^2)$ and so is that for calculating the empirical HSIC. 
  Therefore, the permutation test based on $B$ 
  permutations requires %
  $O(Bmn^2)$ of time. 
  In addition to the time complexity analysis, we report
  the computing time for the proposed method when applied to the
  MEG data in Section \ref{sec:data}.
\end{rmk}

\section{Asymptotic Theory}\label{sec:theory}
In this section we
show that the proposed two-step procedure
can lead to an asymptotically valid test,
which addresses the compatibility issue raised in Section \ref{sec:method}.
Explicitly, we first provide
the rate of convergence for the denoising error involved in Step 1 in Theorem \ref{thm:denoise}, then
the asymptotic distribution of HSIC $\gamma(P_{n,\hX\hY},\kx,\ky)$ in Step 2 in
Theorem \ref{thm:wkconvD},
and finally the asymptotic properties of the permutation test in Theorem \ref{thm:perm}.
Hereafter, the kernels $\kx$ and $\ky$ are induced by $\rho_{b^{\beta_X}}$ and
$\rho_{b^{\beta_Y}}$ respectively. %
{For the noise terms $\{e_{il}^Z: i=1, \ldots, n; l=1, \ldots, m\}$ where $Z=X$ or $Y$, we assume that $e_{il}^Z = e_i^Z(T_l), l=1,\dots,m$ where $\{e_i^Z: i=1,\dots,n\}$ are i.i.d. copies of a stationary stochastic process $e^Z$. }

\begin{thm} \label{thm:denoise}
Assume that %
$\beta_Z<\alpha_Z$,  $\nmbaZ{\thetam^Z}\leq C_Z$ for a constant $C_Z>0$, 
 and $\thetam^{e^Z}=(\theta_{j,k}^{e^Z} )_{-1\leq j\leq J_Z; k=0,\dots,2^j-1}$, the discrete wavelet coefficients of $e^Z$,  %
  satisfy $\theta_{j,k}^{e^Z} = 2^{\varsigma_Z
  j}\delta_Z z_{j,k}$ %
where   
  $\varsigma_Z>-1/2$ 
  and 
  $\bz=(z_{j,k})_{-1\leq j\leq J_Z; k=0,\dots,2^j-1}$ 
  is a zero mean Gaussian random vector %
  that is weakly correlated, 
  i.e., its covariance matrix $\Sigmam$ satisfies $\xi_0^Z\bI\preceq\Sigmam\preceq\xi_1^Z\bI$ where $\bI$ is the identity matrix, $0<\xi_0^Z\leq 1\leq \xi_1^Z < \infty$ are constants, and $\bA \preceq \bB$ means that $\bB-\bA$ is positive semidefinite. 
Then for $\thetam^{Z_i}$ obtained by (\ref{eq:plse2}), we have 
\vspace{-0.4cm}
\begin{equation*} \label{eq:convrate}
\sup_{\nmbaZ{\thetam^{Z_i}}\leq C_Z} \E \left(\nmbbZ{\thetam^{\hZ_i} - \thetam^{Z_i}}^2\mid \thetam^{Z_i}\right) = O(\delta_Z^{2r}) = O(m^{-r}), \quad \text{uniformly for $i=1, \ldots, n$, }
\vspace{-0.4cm}
\end{equation*}
as $m\rightarrow \infty$, where $r = (\alpha_Z -
\beta_Z)/(\alpha_Z+\varsigma_Z+1/2)$. 
This implies that
$\nmbbZ{\thetam^{\hZ_i} - \thetam^{Z_i}}^2 = O_p(m^{-r})$ uniformly for $i=1, \ldots, n$ as $m\rightarrow \infty$.
\end{thm}

The proof of Theorem \ref{thm:denoise} is given in Section \blue{S2.2} in the supplementary material. 
{Theorem \ref{thm:denoise} shows that the pre-smoothing error under the Besov sequence norm $b^{\beta_Z}$ converges to zero uniformly for all subjects if $m$ diverges to infinity. 
This 
theoretical guarantee is achieved due to the new penalty in the proposed wavelet thresholding method (\ref{eq:plse2}). }
{The assumption on the noise $\theta_{j,k}^{e^Z} = 2^{\varsigma_Z
  j}\delta_Z z_{j,k}$ where $\bz=(z_{j,k})_{-1\leq j\leq J_Z; k=0,\dots,2^j-1}$ 
  is weakly correlated Gaussian is a generalization of the Gaussian white noise model by
allowing correlation among noise terms to some extent. 
First, the assumption encompasses both short- and long-range dependences of the noise process when it is a stationary and Gaussian \citep{JohnS97}. 
For the 
short-range dependence case where $\varsigma_Z=0$, there is no variance inflation with
the increase of level $j$.
For the long-range dependence case where $-1/2<\varsigma_Z<0$, the process $e_i^m(t) =
m^{-1}\sum_{l=1}^{\lfloor mt \rfloor} e_{il}^Z$ 
can be approximated by a
fractional Brownian motion $\delta_Z^{2-2H}B_H(t)$, $H = 1/2 - \varsigma_Z$ \citep{Taqq75}, which is
widely used for modeling long-range dependence. Then the convergence rate (with
$\delta_Z$ replaced by $\delta_Z^{2-2H}=\delta_Z^{1+2\varsigma_Z}$ in \eqref{eq:convrate}) is
asymptotically minimax up to a constant. When $\beta_Z=0$ in particular, this rate
coincides with those of \citet{Wang96} and \citet{JohnS97}.
Second, when $\varsigma_Z>0$, %
$\varsigma_Z$ captures noise amplification as reflected in the noise level $\delta_{Z,j} = 2^{\varsigma_Z j}\delta_Z$, which is common in the linear
inverse problem \citep{AbraS98,JohnP14}, e.g., $\varsigma_Z=1/2$ for %
the two-dimensional Radon
transformation \citep{Dono95}.
}

Since the HSIC is constructed based on the kernels induced by $\rho_{b^{\beta_X}}$ and $\rho_{b^{\beta_Y}}$, the same norms used to evaluate the denoising error as in Theorem \ref{thm:denoise}, the compatibility between the pre-smoothing by wavelet soft-thresholding and HSIC is 
theoretically guaranteed. As shown in Theorem \ref{thm:wkconvD}, the effect of the denoising error on the distribution of the HSIC is asymptotically negligible for dense functional data. 

To develop the asymptotic distribution of $\gamma(P_{n,\hX\hY},\kx,\ky)$, we further 
define the centered kernel for $\kx$ by 
$
\ckx(X,\Xp)=\ip{\kx(X,\cdot) - \bP^{\kx}(P_X), \kx(\Xp,\cdot) - \bP^{\kx}(P_X)}_{\Hkx}.
$
Furthermore define an integral kernel operator $S_{\ckx}: \Hkx \rightarrow \Hkx$
by $S_{\ckx}(g) = \int_{\mX}\ckx(x,\cdot)g(x)\dd P_X(x)$ for any $g \in \Hkx$.
An integral kernel operator $S_{\cky}$ for $Y$ can be similarly defined.

\begin{thm}\label{thm:wkconvD}
Under the same assumptions of Theorem \ref{thm:denoise}, 
  if $m$ satisfies %
\vspace{-0.4cm}
\begin{equation}\label{eq:smoothrate}
\vspace{-0.4cm}
m^{-(\alpha_Z - \beta_Z)/(2\alpha_Z+2\varsigma_Z+1)} = 
o(n^{-1}), 
\end{equation}
\vspace{-0.4cm}
for both $Z=X$ and $Z=Y$, 
then
\begin{equation*}
  n \gamma(P_{n,\hX\hY},\kx,\ky) \wkconv 
  \begin{cases}
  \sum_{u=1}^{\infty}\sum_{v=1}^{\infty}\mu_u\nu_v N_{uv}^2, &\text{if $X$ and $Y$ are independent},\\
 \infty, &\text{otherwise},
  \end{cases}
\end{equation*}
where ``$\wkconv$'' represents weak convergence, $N_{uv}\sim N(0,1),u, v \geq 1$ %
are i.i.d. and $\left\{ \mu_u: u \geq 1
\right\}$ %
and $\left\{\nu_v: v \geq 1\right\}$
are eigenvalues of 
$S_{\ckx}$ and $S_{\cky}$ respectively.
\end{thm}
The proof of Theorem
\ref{thm:wkconvD} is given in Section \blue{S2.3} in the supplementary material. 
The asymptotic distribution of $\gamma(P_{n,\hX\hY},\kx,\ky)$ in Theorem
\ref{thm:wkconvD} is the same as that for fully observed $\{(X_i, Y_i): X_i \in
B^{\beta_X}, Y_i \in B^{\beta_Y}, i=1, \ldots, n\}$ \citep{SejdSG13}. 
{
The requirement (\ref{eq:smoothrate}) ensures that the error due to the denoising procedure
is asymptotically negligible under $b^{\beta_Z}$ norm 
if the measurements are sufficiently dense.
In general, for fixed $\alpha_Z$, $\beta_Z$ and $\varsigma_Z$, 
the order of
$m$ should be higher than $n^{1/r}$ where 
$r = (\alpha_Z -
\beta_Z)/(2\alpha_Z+2\varsigma_Z+1)$
which, for example, is 
$n^{10/3}$ if $(\alpha_Z, \beta_Z, \varsigma_Z)=(2, 1/2, 0)$
and %
$n^4$ if $(\alpha_Z, \beta_Z, \varsigma_Z)=(3, 1, 1/2)$. 
}

Since the asymptotic reference distribution of $\gamma(P_{n,\hX\hY},\kx,\ky)$ when $X$ and $Y$ are assumed independent involves many unknown quantities, in practice we perform the test by permutation. As shown in Theorem \ref{thm:perm}, 
 the permutation test can control the Type I error probability and is also consistent.

\begin{thm}[Permutation Test]
  \label{thm:perm}
  Let the level of significance be $\alpha \in (0, 1)$. If the null hypothesis
  that $X$ and $Y$ are independent is true, 
  the permutation test of $\gamma(P_{n,\hX\hY},\kx,\ky)$ based on a finite number of permutations rejects the null hypothesis with probability at most $\alpha$. If the alternative hypothesis that $X$ and $Y$ are dependent is true and the assumptions of Theorem \ref{thm:denoise} and 
\eqref{eq:smoothrate} hold, the permutation test of $\gamma(P_{n,\hX\hY},\kx,\ky)$ based on $B \geq 1/\alpha -1$ %
permutations is consistent, i.e., $P(\hat p_{\hX\hY}\leq \alpha) \rightarrow 1$ as $n\rightarrow\infty$, where $\hat p_{\hX\hY}$ is the p-value.
\end{thm}
The proof of Theorem \ref{thm:perm} is given in Section \blue{S2.4} in the supplementary material. Theorem \ref{thm:perm} shows that the proposed permutation test is
also theoretically compatible with {the proposed wavelet thresholding
method in Step 1.}

\section{Tuning Parameter Selection} \label{sec:select}

In this section, we discuss the selection of tuning parameters 
involved in the two-step procedure proposed in Section \ref{sec:method}. They include $\zeta_Z$, $\tau_Z$, $\varsigma_Z$ and $\delta_Z$ in Step 1 and $\beta_Z$ in Step 2, where $Z=X$ or $Y$. 

First, to guarantee $\zeta_Z>1$ and $\tau_Z>e$, we suggest $\zeta_Z=1.0001$ and $\tau_Z = 1.0001e$ which are slightly larger than their respective lower bounds, unless domain knowledge is available. 

Second, for $\varsigma_Z$ which captures noise amplification and $\delta_Z$ which reflects the noise level, we adopt %
crude estimates for them based on the top two levels of the wavelet coefficients \citep{JohnS97}. Explicitly, we obtain $\hat\varsigma_Z = \log_2 (\hat\delta_{Z,J}/\hat\delta_{Z,J-1})$ and $\hat\delta_Z = \hat\delta_{Z,J}/2^{\hat\varsigma_Z J}$, where
  $\hat \delta_{Z,j} = \text{median} \left\{ \sqrt{m} \theta_{j,k}^{\tZ_i}:
  k=0,\dots,2^j-1 \right\}/\text{median}(\left| W \right|)$ for $j=J-1,J$, and $W$ is a 
standard normal random variable.

Finally, for the smoothness parameter $\beta_Z$, we will first discuss its role in dependency detection and then propose a data-adaptive selection method for it. 

In Section \ref{sec:theory}, 
Theorem \ref{thm:denoise} seems to imply that given
$\alpha_X$ and $\alpha_Y$, the best choice is $\beta_X=\beta_Y=0$ because the corresponding denoising error attains the best rate of convergence. However, this choice of $\beta_X$ and $\beta_Y$ may result in a poor dependency detection 
especially when the dependency of $X$ and $Y$ originates from their high frequency bands. 

For illustration, by Definition \ref{def:HSIC} and (\ref{eq:Besovseqnorm}), we
consider the first-order approximation \citep[][Theorem 5.1]{ChakZ19}
\vspace{-0.4cm}
\begin{equation}
  \label{eq:HSIC_decomp}
\vspace{-0.4cm}
  \gamma(P_{XY},\kx,\ky) 
  \approx c_{XY}^{-1}\sum_{j_X\geq -1}\sum_{j_Y\geq -1}\gamma\Big(P_{XY},2^{2\beta_X j_X}\kx^{(j_X)},2^{2\beta_Y j_Y}\ky^{(j_Y)}\Big),
\end{equation}
where
$\kz^{(j_Z)}(z,z')=\norm{\thetam_{j}^{z}}_2^2+\norm{\thetam_{j}^{z'}}_2^2-\norm{\thetam_{j}^{z}-\thetam_{j}^{z'}}_2^2$
for $j_Z \geq -1$,
with $(z, Z, \mZ)=(x, X, \mX)$ or $(y, Y, \mY)$ and Euclidean norm
$\norm{\cdot}_2$, and 
$c_{XY} = 4\sqrt{\E\norm{X-X'}_{b^{\alpha_X}}^2\E\norm{Y-Y'}_{b^{\alpha_Y}}^2}$
with $X'$ and $Y'$ being the independent copies of $X$ and $Y$, respectively.
Apparently 
$%
\gamma\Big(P_{XY},2^{2\beta_X j_X}\kx^{(j_X)},2^{2\beta_Y j_Y}\ky^{(j_Y)}\Big)$ measures the dependency contribution to the HSIC 
at $j_X$ and $j_Y$ of $X$ and $Y$ respectively,
which is zero if and only if $X$ and $Y$ are independent at $j_X$ and $j_Y$. 
If $\beta_X = \beta_Y =0$,
the scaling factors $2^{\beta_X j_X} = 2^{\beta_Y j_Y}=1$ for all $j_X \geq -1, j_Y \geq -1$ and 
it will be very difficult to detect the dependency between $X$ and $Y$ at high frequencies since 
the dependency contributions contained at high frequencies are very likely to be overwhelmed by the 
independent signals at low frequencies. 
Therefore, we aim to select $\beta_X$
and $\beta_Y$ such that the dependency contributions at high frequencies, if any, are detectable.

The idea of the proposed tuning method is to balance the dependency
contributions to HSIC at all frequency scales
such that they are approximately the same.
To lessen the computational burden, a marginal selection algorithm is proposed in the sense that the optimal $\beta_X$ is selected only based on $X$ without reliance on $Y$. 
Note that, by Appendix A in \citet{SejdSG13} and the properties of distance covariance, %
the dependency contribution at each $j_X,  j_Y \geq -1$ satisfies
\vspace{-0.4cm}
\begin{align*}
\gamma\Big(P_{XY},2^{2\beta_X j_X}\kx^{(j_X)},2^{2\beta_Y j_Y}\ky^{(j_Y)}\Big)
&\leq %
2^{2(\beta_X j_X+\beta_Y j_Y)}\sqrt{\gamma(P_X,\kx^{(j_X)})\gamma(P_Y,\ky^{(j_Y)})},
\vspace{-0.4cm}
\end{align*}
where
$\gamma(P_Z,\kz^{(j_Z)}) = \norm{\bP^{\kz^{(j_Z)}\otimes\kz^{(j_Z)}}(P_{ZZ})(*,\cdot) - \bP^{\kz^{(j_Z)}}(P_Z)(*)\bP^{\kz^{(j_Z)}}(P_Z)(\cdot)}_{\mH(\kz^{(j_Z)}\otimes\kz^{(j_Z)})}^2,$
$ j_Z \geq -1,$
is essentially a distance variance \citep{SzekRB07} 
with $(z, Z, \mZ)=(x, X, \mX)$ or
$(y, Y, \mY)$ \citep{SejdSG13}.
Thus we propose to select $\beta_X$ by balancing $2^{2\beta_X j_X}\sqrt{\gamma(P_X,\kx^{(j_X)})}$ at all $j_X \geq -1$. If 
 $
 2^{2\beta_X j_X}\sqrt{\gamma(P_X,\kx^{(j_X)})} \approx C
 $ 
 where $C > 0$ is a constant, then 
 $
 2 \beta_X j_X + \frac{1}{2}\log_2 \gamma(P_X,\kx^{(j_X)}) \approx \log_2 C,  
 $
so $\beta_X$ may be selected as the estimated slope of the linear regression on $(-2 j_X, \log_2 \gamma(P_X,\kx^{(j_X)})/2)$.
 
In practice, we could estimate $\gamma(P_X,\kx^{(j_X)})$ by
$\gamma(P_{n,\hX},\kx^{(j_X)})$ for each $j_X \geq -1$, but its
accuracy is poor for very high frequencies due to noise contamination. 
Thus we only consider $j_X$ up to $\oj_X=\max
\{j_X \geq L_X: \gamma(P_{n,\hX},\kx^{(j_X)})\geq \gamma(P_{n,\hat{e}^X},\kx^{(j_X)})\}$ 
where $\hat{e}^X = \tX - \hX$ is the residual, such that the distance variances of all $j_X \leq \oj_X$ are not smaller than that of the residual. %
If a known frequency band is of interest in the context of a study, e.g., the alpha 
band of brain signals, one may alternatively select $\beta_X$ by balancing $2^{2\beta_X j_X}\sqrt{\gamma(P_X,\kx^{(j_X)})}$ over that frequency band.
Last, we remark that
the computational benefit of the proposed marginal approach for tuning parameter selection
is substantial when many tests have to be performed, such as in the functional connectivity analysis (Section \ref{sec:data}).

\section{Simulation}\label{sec:simu}

In this section we evaluate the numerical performance of our proposed
wavelet-based HSIC method \textsf{wavHSIC} 
in both controlling the Type I error probability and statistical power. 
We also compare it with a few representative existing methods, including 
\begin{enumerate}[label=(\alph*)]
\itemsep0em
\item Pearson Correlation (\textsf{Pearson}). %
It is a one-sample t-test based on Fisher-Z transformed correlation coefficients of all subjects. 
The correlation coefficient for each subject is obtained by applying the Pearson correlation formula to the bivariate time series of the subject,
without adjusting for any possible dependence within the time series.
It is a popular functional connectivity measure in neuroscience %
\citep[e.g.,][]{HeCFSI2012}.
\item Dynamical Correlation \citep[\textsf{dnm},][]{DubiM05}. It is defined as the expectation of the cosine of the $L^2$ angle between the standardized versions of two random functions. %
\item Global Temporal Correlation \citep[\textsf{gtemp},][]{ZhouLW18}. It is the integral of the Pearson correlation obtained at each time point.
\item Bias-Corrected Distance Covariance \citep[\textsf{dCov-c},][]{SzekR131}. 
It is a t-test designed to correct the bias of distance covariance for high-dimensional multivariate data. We apply it by treating the discrete %
measurements of two random functions as multivariate data.
If the bias is not corrected, it is equivalent to \textsf{wavHSIC} with $\beta_X=\beta_Y=0$.

\item Functional Principle Component Analysis (FPCA) Based Distance Covariance
  \citep[\textsf{FPCA},][]{Koso09}. The distance covariance \citep{SzekRB07} is
  applied to top Functional Principle Component (FPC) scores which cumulatively account for 
95\% of the variation of each random function. When all FPC scores are used, it is
equivalent to \textsf{wavHSIC} when $\beta_X=\beta_Y=0$. %

\item Functional Linearity Test \citep[\textsf{KMSZ},][]{KokoMS08}. It is an approximate chi-squared test for the nullity of the coefficient function by assuming a functional linear model between the two random functions. The model fitting requires a satisfactory approximation of each random function by its top FPC scores and we select those which cumulatively account for 95\% of variation of each random function.

\item Permutation-Based Functional Linearity Test (\textsf{KMSZ-p}). It is the same as \textsf{KMSZ} except that the p-value is obtained by permutation.
Such a modification can be regarded as a
finite-sample correction of \textsf{KMSZ}.

{
\item Projection-based Mean Independence Test \citep[\textsf{PSS},][]{PatiSS16}. For a functional 
  response $Y$ and a functional predictor $X$, \textsf{PSS} aims to test the conditional mean independence 
  of $Y$ given $X$, i.e., $E(Y \mid X) = E(Y), a.s.$ \textsf{PSS} is a model-free test that does not specify
  a model for $E(Y \mid X)$. It requires a finite-dimensional projection of $X$ and uses wild bootstrap to 
  find critical values. To implement \textsf{PSS}, we used the R
  package \texttt{fdapss}  \footnote{The package is only for Windows platform. For the user-chosen
  parameters required by this package, we followed the
  recommendation in Section 4.1 of \citet{PatiSS16} and set the bandwidth $h=n^{-2/9}$, penalty coefficient
  $\alpha=2$, grid size $n_q=50$ and number of FPCs which cumulatively account
  for 95\% of the variation of the functional predictor.}, which is publicly available at \url{http://webspersoais.usc.es/persoais/cesar.sanchez/}.

\item Functional Martingale Difference Divergence Based Mean Independence Test \citep[\textsf{FMDD},][]{LeeZS20}. \textsf{FMDD} is also a model-free mean independence test. It measures the conditional mean independence using the metric of functional martingale difference divergence and uses wild bootstrap to find critical values. To implement \textsf{FMDD}, we used the 
  R code publicly available at \url{https://publish.illinois.edu/xshao/files/2019/06/CodeCMDexample1.txt}.
}
\end{enumerate}

The first five (a-e) in comparison are model-free methods. 
\textsf{KMSZ} 
is one of the most popular model-based methods in the FDA literature, 
{but it can only test for linearity}. 
{\textsf{PSS} and \textsf{FMDD} can handle nonlinear effects of the functional predictor, 
but only on the mean of the functional response, so they can only test a weaker notion of independence. 
Hereafter, for bivariate random functions $(X, Y)$, \textsf{PSS}$(Y\sim X)$ denotes testing 
$E(Y \mid X) = E(Y), a.s.$ using \textsf{PSS}. Moreover, \textsf{PSS}(Omnibus) denotes the omnibus test 
which takes the smaller p-value between those obtained by \textsf{PSS}$(Y\sim X)$
and \textsf{PSS}$(X\sim Y)$ respectively. \textsf{FMDD}$(Y\sim X)$, \textsf{FMDD}$(X\sim Y)$ 
and \textsf{FMDD}(Omnibus) are similarly defined.} 
To obtain p-values, 1,999 permutations were
used 
for \textsf{wavHSIC},
\textsf{dnm}, \textsf{gtemp}, \textsf{FPCA} and \textsf{KMSZ-p} {while 1,999 bootstrap samples were used for \textsf{PSS} and \textsf{FMDD}}.
{We declare statistical significance in each simulated data based on the level of significance $0.05$.}

We generated $199$ simulated datasets,
where the number $199$ is chosen to prevent empirical Type I and Type II error probabilities from coinciding with the level of significance $0.05$. 
In each simulated dataset $n=50$ or $200$ independent subjects with bivariate functions 
$\{(X_i(t),Y_i(t)): t\in [0, 1], i=1, \ldots, n\}$ were generated where for the $i$-th subject, $X_i(t) =
\sum_{k=1}^{16}\eta_{ik}\phi_k(t)$ and $Y_i(t) =
\sum_{k=1}^{16}\zeta_{ik}\phi_{k}(t+0.2)$ with $\phi_{2k-1}(t) = \sqrt{2} \cos (2\pi
kt), \phi_{2k}(t) = \sqrt{2} \sin (2\pi kt)$ for $k=1,\dots,8$. 
We considered three settings with different dependency structures of the bivariate functional data which are controlled by the FPC scores $\{(\eta_{ik},\zeta_{ik}):k=1,\dots,16;i=1,\dots,n\}$.
\vspace{-0.2cm}
\begin{itemize}
  \itemsep0em
\item Setting 1. We generated $\eta_{ik} \sim N(0, k^{-1.05}), k=1, \ldots, 16$ and $\zeta_{ik} \sim N(0, k^{-1.2}), k=1, \ldots, 16$ independently. %
\item Setting 2. With $\rho = 0$ for $k=1,\dots,8$ and $\rho = 0.6$ for $k=9,\dots,16$, we generated 
\vspace{-0.4cm}
\begin{equation*}
\bmat{\eta_{ik}\\ \zeta_{ik}} \sim N\left(\bmat{0\\0}, \bmat{k^{-1.05} & \rho k^{-1.125}\\ \rho  k^{-1.125} & k^{-1.2}}\right).
\vspace{-0.4cm}
\end{equation*}
\item Setting 3. For $k=1,\dots,8$, $\eta_{ik} \sim N(0, k^{-1.05})$ was generated independently of $\zeta_{ik} \sim N(0, k^{-1.2})$. For $k=9,\dots,16$, $\eta_{ik} \sim N(0, k^{-1.05})$ and $\zeta_{ik} =\eta_{ik}^2 - \E \eta_{ik}^2$. %
\end{itemize}

Apparently $X$ and $Y$ are independent in Setting 1 and dependent in Settings 2 and 3. In Setting 2, the FPC scores of $X$ and $Y$ are linearly correlated but only at high spectral 
frequencies, 
while in Setting 3 they are
 linearly uncorrelated but dependent only at high spectral frequencies, so it is more difficult to detect dependency for all methods in Setting 3 than Setting 2.

Both functions are measured at $m=64$ or $256$ equidistant points on the time
domain $[0, 1]$. We added Gaussian noise to all measurements with
signal-to-noise ratio $\text{SNR}=4$ or $8$, which 
is the variance of all measurements over the noise variance.
The noise terms were generated independently across subjects. Within each subject, we experimented with both independent (white noise) and dependent (correlated noise) settings. For the dependent setting, the Gaussian noise was generated by differencing the fractional Brownian
motion with Hurst exponent
$0.7$. %

Since all methods in comparison require noiseless functions, we used the
same denoising procedure {in Step 1} for all of them for fairness.
We chose the CDJV wavelet basis functions %
with vanishing moment $D=10$ %
for both $X$ and $Y$, which leads to $\alpha_X = \alpha_Y\approx 2.902$ \citep{Daub1992}. 
The tuning parameters $\beta_X$ and $\beta_Y$ were selected by the method in Section \ref{sec:select}. 
The results are given in Tables \ref{tab:case1}--\ref{tab:case3c}.

Tables \ref{tab:case1} and \ref{tab:case1c} show that all methods are almost
always able to control type I error probabilities 
{except for \textsf{PSS}($X\sim Y$) and the two omnibus tests}
when the two random functions are truly independent.
Relatively, \textsf{KMSZ} is very conservative in many cases and \textsf{KMSZ-p} corrects its p-values to some extent. 
However,
\textsf{KMSZ-p} seems more likely to detect spurious dependency when
$(n,m)=(50,64)$, so does \textsf{dCov-c} when $(n,m)=(200,256)$.

Tables \ref{tab:case2}, \ref{tab:case3},  \ref{tab:case2c} and \ref{tab:case3c}
show that the statistical powers of all methods typically improve when one of
$n$, $m$ and SNR increases under Setting 2, but unnecessarily under Setting 3
except for \textsf{KMSZ}, \textsf{KMSZ-p} and \textsf{wavHSIC}. This
demonstrates the difficulty of Setting 3 in detecting dependency to some extent.
Except \textsf{wavHSIC}, all model-free methods have very low powers in all
scenarios under either Setting 2 or 3, which indicates their poor performances
in detecting linear dependency in high frequencies or nonlinear dependency.
The performance of \textsf{KMSZ} is satisfactory for $n=200$ under Setting 2
when the relationship between $X$ and $Y$ is truly linear.
\textsf{KMSZ-p} improves the statistical power of \textsf{KMSZ} further for $n=50$ under Setting 2
by permutation. However, both \textsf{KMSZ} and \textsf{KMSZ-p} are poor at
testing nonlinear dependency in Setting 3.
{ The performances of \textsf{PSS} and \textsf{FMDD}, which can detect nonlinear mean dependency, are comparable with those of \textsf{dCov-c} and \textsf{FPCA} in Settings 2 and 3, but worse than those of \textsf{KMSZ}
  and \textsf{KMSZ-p} in Setting 2 where $X$ and $Y$ are linearly dependent.}

Tables \ref{tab:case2}, \ref{tab:case3},  \ref{tab:case2c} and \ref{tab:case3c}
also demonstrate the appealing performance of \textsf{wavHSIC}. It is always the
most powerful method, and substantially better than the other methods.
Only the powers of \textsf{KMSZ} and \textsf{KMSZ-p} are comparable with those of \textsf{wavHSIC} when the
sample size $n=200$ is large and the linearity assumption is valid under Setting 2.
For fixed $(n, m, \text{SNR})$, the medians of the selected parameters
$\beta_X$ and $\beta_Y$ for \textsf{wavHSIC} are always similar between Settings
2 and 3 since they were tuned marginally regardless of the dependency structure.
On average, both $\beta_X$ and $\beta_Y$ were considerably away from zero, which
confirms the need and benefit of choosing them properly to enhance the detection
sensitivity of \textsf{wavHSIC}.

{We also performed an additional simulation study described in Section \blue{S3.2} of the supplementary material, which follows the same settings in Section 1.2 of the supplementary material of \citet{LeeZS20}. The results also demonstrate the superiority of \textsf{wavHSIC}.}

{
\begin{rmk}\label{rmk:m-vs-n}
It is worth noting that the development of the asymptotic distribution of \textsf{wavHSIC} as in Theorem \ref{thm:wkconvD} requires the number of measurements per curve $m$ to be large compared to the sample size $n$ (see \eqref{eq:smoothrate}), but the simulation results here show that the finite sample performance of \textsf{wavHSIC} is still satisfactory, even when $m$ is small relatively to $n$. However, this is not entirely surprising. First, under the null hypothesis that $X$ and $Y$ are independent, a poor pre-smoothing due to a relatively small $m$ does not inflate the empirical Type I error probability since the remaining noise does not enhance the dependency between $X$ and $Y$ and the critical value is obtained by permutation. Second, under the alternative hypothesis that $X$ and $Y$ are dependent, as long as $m$ is sufficiently large such that the dependency signals can captured by the wavelet coefficients, \textsf{wavHSIC} can still detect dependency, but its power may be worse if \eqref{eq:smoothrate} is not satisfied. 
\end{rmk}
}

\begin{table}[h!] 
  \caption{
  Empirical Type I error probabilities for %
  all methods under Setting 1 {with white noise}. 
  The last two rows provide the medians of the selected $\beta_X$ and $\beta_Y$ for \textsf{wavHSIC}. 
} 
\label{tab:case1}
\centering
\scriptsize
\begin{tabular}{l|rrrrrrrr}
  \toprule
Setting 1 with             & \multicolumn{4}{c}{$n=50$} & \multicolumn{4}{c}{$n=200$}\\
                           \cmidrule(lr){2-5}           \cmidrule(lr){6-9}
white noise                & \multicolumn{2}{c}{$m=64$} & \multicolumn{2}{c}{$m=256$}& \multicolumn{2}{c}{$m=64$} & \multicolumn{2}{c}{$m=256$}\\
\cmidrule(lr){1-1}         \cmidrule(lr){2-3}\cmidrule(lr){4-5}\cmidrule(lr){6-7}\cmidrule(lr){8-9}
Type I error rate          & SNR=4 & SNR=8 & SNR=4 & SNR=8 & SNR=4 & SNR=8 & SNR=4 & SNR=8\\
\cmidrule(lr){1-1}         \cmidrule(lr){2-3}\cmidrule(lr){4-5}\cmidrule(lr){6-7}\cmidrule(lr){8-9}
\textsf{Pearson}           & 0.0452 & 0.0352 & 0.0503 & 0.0452 & 0.0704 & 0.0704 & 0.0553 & 0.0553\\
\textsf{dnm}               & 0.0452 & 0.0452 & 0.0653 & 0.0503 & 0.0553 & 0.0553 & 0.0653 & 0.0603\\
\textsf{gtemp}             & 0.0503 & 0.0603 & 0.0653 & 0.0603 & 0.0402 & 0.0352 & 0.0352 & 0.0352\\
\textsf{dCov-c}            & 0.0653 & 0.0603 & 0.0603 & 0.0603 & 0.0503 & 0.0603 & 0.0704 & 0.0754\\
\textsf{FPCA}              & 0.0503 & 0.0452 & 0.0503 & 0.0452 & 0.0452 & 0.0503 & 0.0452 & 0.0452\\
\cmidrule(lr){1-1}         \cmidrule(lr){2-3}\cmidrule(lr){4-5}\cmidrule(lr){6-7}\cmidrule(lr){8-9}
\textsf{KMSZ}              & 0.0201 & 0.0101 & 0.0101 & 0.0151 & 0.0201 & 0.0151 & 0.0402 & 0.0251\\
\textsf{KMSZ-p}            & 0.0804 & 0.0905 & 0.0553 & 0.0402 & 0.0302 & 0.0352 & 0.0402 & 0.0352\\
\cmidrule(lr){1-1}         \cmidrule(lr){2-3}\cmidrule(lr){4-5}\cmidrule(lr){6-7}\cmidrule(lr){8-9}
\textsf{PSS}$(X\sim Y)$    & 0.0804 & 0.0754 & 0.1005 & 0.0804 & 0.0653 & 0.0905 & 0.0402 & 0.0955\\
\textsf{PSS}$(Y\sim X)$    & 0.0402 & 0.0754 & 0.0704 & 0.0452 & 0.0302 & 0.0754 & 0.0352 & 0.0503\\
\textsf{PSS}(Omnibus)      & 0.1156 & 0.1307 & 0.1558 & 0.1106 & 0.0955 & 0.1407 & 0.0754 & 0.1407\\
\cmidrule(lr){1-1}         \cmidrule(lr){2-3}\cmidrule(lr){4-5}\cmidrule(lr){6-7}\cmidrule(lr){8-9}
\textsf{FMDD}$(X\sim Y)$   & 0.0553 & 0.0552 & 0.0704 & 0.0653 & 0.0503 & 0.0603 & 0.0553 & 0.0603\\
\textsf{FMDD}$(Y\sim X)$   & 0.0553 & 0.0552 & 0.0553 & 0.0553 & 0.0503 & 0.0603 & 0.0503 & 0.0503\\
\textsf{FMDD}(Omnibus)     & 0.0653 & 0.0603 & 0.0704 & 0.0704 & 0.0603 & 0.0704 & 0.0704 & 0.0704\\
\cmidrule(lr){1-1}         \cmidrule(lr){2-3}\cmidrule(lr){4-5}\cmidrule(lr){6-7}\cmidrule(lr){8-9}
\textsf{wavHSIC}           & 0.0452 & 0.0352 & 0.0503 & 0.0653 & 0.0302 & 0.0402 & 0.0251 & 0.0251\\
median$\{\beta_X\}$        & 0.948 & 0.989 & 0.983 & 0.991 & 0.959 & 1.000 & 0.990 & 1.001        \\
median$\{\beta_Y\}$        & 0.671 & 0.724 & 0.733 & 0.741 & 0.696 & 0.745 & 0.747 & 0.761        \\
  \bottomrule
\end{tabular}
\end{table}
\begin{table}[h!] 
  \caption{
  Empirical powers for %
  all methods under Setting 2 {with white noise}.
  The last rows provide the medians of the selected $\beta_X$ and $\beta_Y$ for \textsf{wavHSIC}. 
} 
\label{tab:case2}
\centering
\scriptsize
\begin{tabular}{l|rrrrrrrr}
  \toprule
Setting 2 with             & \multicolumn{4}{c}{$n=50$} & \multicolumn{4}{c}{$n=200$}\\
                           \cmidrule(lr){2-5}           \cmidrule(lr){6-9}
white noise                & \multicolumn{2}{c}{$m=64$} & \multicolumn{2}{c}{$m=256$}& \multicolumn{2}{c}{$m=64$} & \multicolumn{2}{c}{$m=256$}\\
\cmidrule(lr){1-1}         \cmidrule(lr){2-3}\cmidrule(lr){4-5}\cmidrule(lr){6-7}\cmidrule(lr){8-9}
Power                      & SNR=4 & SNR=8 & SNR=4 & SNR=8 & SNR=4 & SNR=8 & SNR=4 & SNR=8\\
\cmidrule(lr){1-1}         \cmidrule(lr){2-3}\cmidrule(lr){4-5}\cmidrule(lr){6-7}\cmidrule(lr){8-9}
\textsf{Pearson}           & 0.0854 & 0.0804 & 0.0804 & 0.0804 & 0.1357 & 0.1357 & 0.1357 & 0.1206\\
\textsf{dnm}               & 0.0704 & 0.0653 & 0.0704 & 0.0704 & 0.1608 & 0.1558 & 0.1508 & 0.1457\\
\textsf{gtemp}             & 0.0653 & 0.0653 & 0.0804 & 0.0754 & 0.0905 & 0.0854 & 0.0754 & 0.0754\\
\textsf{dCov-c}            & 0.1106 & 0.1055 & 0.0905 & 0.0804 & 0.2362 & 0.2462 & 0.2714 & 0.2663\\
\textsf{FPCA}              & 0.0854 & 0.0804 & 0.0804 & 0.0804 & 0.1709 & 0.1709 & 0.1859 & 0.1809\\
\cmidrule(lr){1-1}         \cmidrule(lr){2-3}\cmidrule(lr){4-5}\cmidrule(lr){6-7}\cmidrule(lr){8-9}
\textsf{KMSZ}              & 0.4221 & 0.4925 & 0.5025 & 0.5075 & 1.0000 & 1.0000 & 1.0000 & 1.0000\\
\textsf{KMSZ-p}            & 0.7035 & 0.7889 & 0.7688 & 0.7990 & 1.0000 & 1.0000 & 1.0000 & 1.0000\\
\cmidrule(lr){1-1}         \cmidrule(lr){2-3}\cmidrule(lr){4-5}\cmidrule(lr){6-7}\cmidrule(lr){8-9}
\textsf{PSS}$(X\sim Y)$    & 0.0955 & 0.1055 & 0.0804 & 0.0905 & 0.1256 & 0.0955 & 0.1106 & 0.0804\\
\textsf{PSS}$(Y\sim X)$    & 0.0653 & 0.0653 & 0.0553 & 0.0653 & 0.0804 & 0.0653 & 0.0503 & 0.0503\\
\textsf{PSS}(Omnibus)      & 0.1558 & 0.1658 & 0.1307 & 0.1508 & 0.2060 & 0.1608 & 0.1558 & 0.1206\\
\cmidrule(lr){1-1}         \cmidrule(lr){2-3}\cmidrule(lr){4-5}\cmidrule(lr){6-7}\cmidrule(lr){8-9}
\textsf{FMDD}$(X\sim Y)$   & 0.0854 & 0.0905 & 0.0905 & 0.0854 & 0.1859 & 0.1960 & 0.2915 & 0.2814\\
\textsf{FMDD}$(Y\sim X)$   & 0.0754 & 0.0804 & 0.0704 & 0.0653 & 0.1407 & 0.1759 & 0.2161 & 0.2211\\
\textsf{FMDD}(Omnibus)     & 0.0955 & 0.1005 & 0.0905 & 0.0854 & 0.1960 & 0.2211 & 0.3116 & 0.3116\\
\cmidrule(lr){1-1}         \cmidrule(lr){2-3}\cmidrule(lr){4-5}\cmidrule(lr){6-7}\cmidrule(lr){8-9}
\textsf{wavHSIC}           & 0.9548 & 0.9849 & 0.9849 & 0.9899 & 1.0000 & 1.0000 & 1.0000 & 1.0000\\
median$\{\beta_X\}$        & 0.942 & 0.987 & 0.975 & 0.983 & 0.955 & 0.996 & 0.994 & 1.001        \\
median$\{\beta_Y\}$        & 0.674 & 0.720 & 0.741 & 0.752 & 0.693 & 0.739 & 0.742 & 0.762        \\
  \bottomrule
\end{tabular}
\end{table}
\begin{table}[h!] 
\caption{%
  Empirical powers for %
  all methods under Setting 3 {with white noise}.
  The last two rows provide the medians of the selected $\beta_X$ and $\beta_Y$ for \textsf{wavHSIC}. 
} \label{tab:case3}
\centering
\scriptsize
\begin{tabular}{l|rrrrrrrr}
\toprule
Setting 3 with             & \multicolumn{4}{c}{$n=50$} & \multicolumn{4}{c}{$n=200$}\\
                           \cmidrule(lr){2-5}           \cmidrule(lr){6-9}
white noise                & \multicolumn{2}{c}{$m=64$} & \multicolumn{2}{c}{$m=256$}& \multicolumn{2}{c}{$m=64$} & \multicolumn{2}{c}{$m=256$}\\
\cmidrule(lr){1-1}         \cmidrule(lr){2-3}\cmidrule(lr){4-5}\cmidrule(lr){6-7}\cmidrule(lr){8-9}
Power                      & SNR=4 & SNR=8 & SNR=4 & SNR=8 & SNR=4 & SNR=8 & SNR=4 & SNR=8\\
\cmidrule(lr){1-1}         \cmidrule(lr){2-3}\cmidrule(lr){4-5}\cmidrule(lr){6-7}\cmidrule(lr){8-9}
\textsf{Pearson}           & 0.0452 & 0.0402 & 0.0603 & 0.0603 & 0.0503 & 0.0503 & 0.0402 & 0.0503\\
\textsf{dnm}               & 0.0804 & 0.0804 & 0.0754 & 0.0704 & 0.0704 & 0.0603 & 0.0754 & 0.0754\\
\textsf{gtemp}             & 0.0754 & 0.0804 & 0.0754 & 0.0704 & 0.0704 & 0.0653 & 0.0754 & 0.0704\\
\textsf{dCov-c}            & 0.0955 & 0.1055 & 0.1005 & 0.1005 & 0.0854 & 0.0905 & 0.0854 & 0.0854\\
\textsf{FPCA}              & 0.0704 & 0.0854 & 0.0955 & 0.1005 & 0.0704 & 0.0704 & 0.0653 & 0.0704\\
\cmidrule(lr){1-1}         \cmidrule(lr){2-3}\cmidrule(lr){4-5}\cmidrule(lr){6-7}\cmidrule(lr){8-9}
\textsf{KMSZ}              & 0.0101 & 0.0101 & 0.0201 & 0.0251 & 0.1206 & 0.1307 & 0.1206 & 0.1357\\
\textsf{KMSZ-p}            & 0.1106 & 0.0854 & 0.1307 & 0.1357 & 0.1558 & 0.1608 & 0.1407 & 0.1709\\
\cmidrule(lr){1-1}         \cmidrule(lr){2-3}\cmidrule(lr){4-5}\cmidrule(lr){6-7}\cmidrule(lr){8-9}
\textsf{PSS}$(X\sim Y)$    & 0.0754 & 0.0854 & 0.0905 & 0.1055 & 0.1005 & 0.0452 & 0.0553 & 0.0653\\
\textsf{PSS}$(Y\sim X)$    & 0.0653 & 0.0553 & 0.0754 & 0.0804 & 0.0603 & 0.0503 & 0.0603 & 0.0704\\
\textsf{PSS}(Omnibus)      & 0.1357 & 0.1307 & 0.1508 & 0.1658 & 0.1558 & 0.0905 & 0.1106 & 0.1256\\
\cmidrule(lr){1-1}         \cmidrule(lr){2-3}\cmidrule(lr){4-5}\cmidrule(lr){6-7}\cmidrule(lr){8-9}
\textsf{FMDD}$(X\sim Y)$   & 0.0804 & 0.0804 & 0.0804 & 0.0804 & 0.0704 & 0.0704 & 0.0754 & 0.0704\\
\textsf{FMDD}$(Y\sim X)$   & 0.0955 & 0.1005 & 0.0955 & 0.1005 & 0.0804 & 0.0754 & 0.0754 & 0.0754\\
\textsf{FMDD}(Omnibus)     & 0.1005 & 0.1005 & 0.0955 & 0.1005 & 0.0854 & 0.0804 & 0.0804 & 0.0804\\
\cmidrule(lr){1-1}         \cmidrule(lr){2-3}\cmidrule(lr){4-5}\cmidrule(lr){6-7}\cmidrule(lr){8-9}
\textsf{wavHSIC}           & 0.2613 & 0.3618 & 0.3367 & 0.407 & 0.804 & 0.9347 & 0.9347 & 0.9749  \\
median$\{\beta_X\}$        & 0.948 & 0.993 & 0.968 & 0.979 & 0.949 & 0.993 & 0.981 & 0.989        \\
median$\{\beta_Y\}$        & 0.724 & 0.771 & 0.773 & 0.790 & 0.723 & 0.769 & 0.775 & 0.785        \\
\bottomrule
\end{tabular}
\end{table}
\begin{table}[h!] 
\caption{%
  Empirical Type I error probabilities for %
  all methods under Setting 1 {with correlated noise}. 
  The last two rows provide the medians of the selected $\beta_X$ and $\beta_Y$ for \textsf{wavHSIC}. 
}
\label{tab:case1c}
\centering
\scriptsize
\begin{tabular}{l|rrrrrrrr}
  \toprule
Setting 1 with             & \multicolumn{4}{c}{$n=50$} & \multicolumn{4}{c}{$n=200$}\\
                           \cmidrule(lr){2-5}           \cmidrule(lr){6-9}
correlated noise           & \multicolumn{2}{c}{$m=64$} & \multicolumn{2}{c}{$m=256$}& \multicolumn{2}{c}{$m=64$} & \multicolumn{2}{c}{$m=256$}\\
\cmidrule(lr){1-1}         \cmidrule(lr){2-3}\cmidrule(lr){4-5}\cmidrule(lr){6-7}\cmidrule(lr){8-9}
Type I error rate          & SNR=4 & SNR=8 & SNR=4 & SNR=8 & SNR=4 & SNR=8 & SNR=4 & SNR=8\\
\cmidrule(lr){1-1}         \cmidrule(lr){2-3}\cmidrule(lr){4-5}\cmidrule(lr){6-7}\cmidrule(lr){8-9}
\textsf{Pearson}           & 0.0352 & 0.0352 & 0.0452 & 0.0452 & 0.0704 & 0.0553 & 0.0553 & 0.0553\\
\textsf{dnm}               & 0.0402 & 0.0452 & 0.0603 & 0.0603 & 0.0553 & 0.0503 & 0.0503 & 0.0503\\
\textsf{gtemp}             & 0.0603 & 0.0553 & 0.0553 & 0.0653 & 0.0553 & 0.0503 & 0.0452 & 0.0553\\
\textsf{dCov-c}            & 0.0553 & 0.0653 & 0.0603 & 0.0603 & 0.0553 & 0.0603 & 0.0754 & 0.0754\\
\textsf{FPCA}              & 0.0452 & 0.0452 & 0.0452 & 0.0402 & 0.0503 & 0.0452 & 0.0452 & 0.0503\\
\cmidrule(lr){1-1}         \cmidrule(lr){2-3}\cmidrule(lr){4-5}\cmidrule(lr){6-7}\cmidrule(lr){8-9}
\textsf{KMSZ}              & 0.0151 & 0.0000 & 0.0151 & 0.0151 & 0.0201 & 0.0251 & 0.0251 & 0.0251\\
\textsf{KMSZ-p}            & 0.0804 & 0.0754 & 0.0553 & 0.0452 & 0.0302 & 0.0352 & 0.0352 & 0.0402\\
\cmidrule(lr){1-1}         \cmidrule(lr){2-3}\cmidrule(lr){4-5}\cmidrule(lr){6-7}\cmidrule(lr){8-9}
\textsf{PSS}$(X\sim Y)$    & 0.0452 & 0.0553 & 0.0603 & 0.0955 & 0.0653 & 0.0452 & 0.0452 & 0.0603\\
\textsf{PSS}$(Y\sim X)$    & 0.0553 & 0.0704 & 0.0553 & 0.0754 & 0.0452 & 0.0302 & 0.0603 & 0.0452\\
\textsf{PSS}(Omnibus)      & 0.1005 & 0.1156 & 0.1106 & 0.1558 & 0.1005 & 0.0704 & 0.1005 & 0.1005\\
\cmidrule(lr){1-1}         \cmidrule(lr){2-3}\cmidrule(lr){4-5}\cmidrule(lr){6-7}\cmidrule(lr){8-9}
\textsf{FMDD}$(X\sim Y)$   & 0.0553 & 0.0603 & 0.0653 & 0.0653 & 0.0653 & 0.0603 & 0.0553 & 0.0603\\
\textsf{FMDD}$(Y\sim X)$   & 0.0452 & 0.0452 & 0.0503 & 0.0503 & 0.0553 & 0.0603 & 0.0603 & 0.0603\\
\textsf{FMDD}(Omnibus)     & 0.0553 & 0.0653 & 0.0704 & 0.0704 & 0.0754 & 0.0804 & 0.0754 & 0.0754\\
\cmidrule(lr){1-1}         \cmidrule(lr){2-3}\cmidrule(lr){4-5}\cmidrule(lr){6-7}\cmidrule(lr){8-9}
\textsf{wavHSIC}           & 0.0402 & 0.0402 & 0.0553 & 0.0653 & 0.0352 & 0.0302 & 0.0352 & 0.0302\\
median$\{\beta_X\}$        & 1.014 & 1.020 & 0.996 & 0.997 & 1.024 & 1.030 & 1.006 & 1.008        \\
median$\{\beta_Y\}$        & 0.752 & 0.765 & 0.750 & 0.754 & 0.774 & 0.783 & 0.770 & 0.772        \\
  \bottomrule
\end{tabular}
\end{table}
\begin{table}[ht!] 
\caption{%
  Empirical powers for %
  all methods 
  under Setting 2 {with correlated noise}.
  The last two rows provide the medians of the selected $\beta_X$ and $\beta_Y$ for \textsf{wavHSIC}. 
}
\label{tab:case2c}
\centering
\scriptsize
\begin{tabular}{l|rrrrrrrr}
  \toprule
Setting 2 with             & \multicolumn{4}{c}{$n=50$} & \multicolumn{4}{c}{$n=200$}\\
                           \cmidrule(lr){2-5}           \cmidrule(lr){6-9}
correlated noise           & \multicolumn{2}{c}{$m=64$} & \multicolumn{2}{c}{$m=256$}& \multicolumn{2}{c}{$m=64$} & \multicolumn{2}{c}{$m=256$}\\
\cmidrule(lr){1-1}         \cmidrule(lr){2-3}\cmidrule(lr){4-5}\cmidrule(lr){6-7}\cmidrule(lr){8-9}
Power                      & SNR=4 & SNR=8 & SNR=4 & SNR=8 & SNR=4 & SNR=8 & SNR=4 & SNR=8\\
\cmidrule(lr){1-1}         \cmidrule(lr){2-3}\cmidrule(lr){4-5}\cmidrule(lr){6-7}\cmidrule(lr){8-9}
\textsf{Pearson}           & 0.0854 & 0.0854 & 0.0804 & 0.0804 & 0.1508 & 0.1407 & 0.1256 & 0.1307\\
\textsf{dnm}               & 0.0653 & 0.0653 & 0.0754 & 0.0704 & 0.1558 & 0.1608 & 0.1558 & 0.1457\\
\textsf{gtemp}             & 0.0653 & 0.0653 & 0.0905 & 0.0905 & 0.0854 & 0.0804 & 0.0754 & 0.0854\\
\textsf{dCov-c}            & 0.1005 & 0.0955 & 0.0854 & 0.0854 & 0.2663 & 0.2714 & 0.2714 & 0.2814\\
\textsf{FPCA}              & 0.0854 & 0.0804 & 0.0754 & 0.0804 & 0.1658 & 0.1759 & 0.1809 & 0.1809\\
\cmidrule(lr){1-1}         \cmidrule(lr){2-3}\cmidrule(lr){4-5}\cmidrule(lr){6-7}\cmidrule(lr){8-9}
\textsf{KMSZ}              & 0.5427 & 0.5628 & 0.5126 & 0.5327 & 1.0000 & 1.0000 & 1.0000 & 1.0000\\
\textsf{KMSZ-p}            & 0.8241 & 0.8141 & 0.8191 & 0.8291 & 1.0000 & 1.0000 & 1.0000 & 1.0000\\
\cmidrule(lr){1-1}         \cmidrule(lr){2-3}\cmidrule(lr){4-5}\cmidrule(lr){6-7}\cmidrule(lr){8-9}
\textsf{PSS}$(X\sim Y)$    & 0.0955 & 0.0653 & 0.0905 & 0.1005 & 0.1156 & 0.1106 & 0.1005 & 0.1156\\
\textsf{PSS}$(Y\sim X)$    & 0.0553 & 0.0653 & 0.0704 & 0.0603 & 0.0704 & 0.0653 & 0.0603 & 0.0754\\
\textsf{PSS}(Omnibus)      & 0.1407 & 0.1156 & 0.1558 & 0.1508 & 0.1809 & 0.1457 & 0.1508 & 0.1809\\
\cmidrule(lr){1-1}         \cmidrule(lr){2-3}\cmidrule(lr){4-5}\cmidrule(lr){6-7}\cmidrule(lr){8-9}
\textsf{FMDD}$(X\sim Y)$   & 0.1055 & 0.0905 & 0.0854 & 0.0854 & 0.2412 & 0.2513 & 0.2714 & 0.2714\\
\textsf{FMDD}$(Y\sim X)$   & 0.0804 & 0.0905 & 0.0704 & 0.0704 & 0.2111 & 0.2111 & 0.2412 & 0.2412\\
\textsf{FMDD}(Omnibus)     & 0.1106 & 0.1055 & 0.0905 & 0.0905 & 0.2613 & 0.2714 & 0.3166 & 0.3015\\
\cmidrule(lr){1-1}         \cmidrule(lr){2-3}\cmidrule(lr){4-5}\cmidrule(lr){6-7}\cmidrule(lr){8-9}
\textsf{wavHSIC}           & 0.9950 & 0.9950 & 0.9899 & 0.9899 & 1.0000 & 1.0000 & 1.0000 & 1.0000\\
median$\{\beta_X\}$        & 1.011 & 1.022 & 0.990 & 0.995 & 1.023 & 1.033 & 1.011 & 1.014        \\
median$\{\beta_Y\}$        & 0.749 & 0.765 & 0.757 & 0.760 & 0.769 & 0.781 & 0.764 & 0.766        \\
  \bottomrule
\end{tabular}
\end{table}
\begin{table}[h!] 
\caption{%
  Empirical powers for %
  all methods 
  under Setting 3 {with correlated noise}.
  The last two rows provide the medians of the selected $\beta_X$ and $\beta_Y$ for \textsf{wavHSIC}. 
} \label{tab:case3c}
\centering
\scriptsize
\begin{tabular}{l|rrrrrrrr}
  \toprule
Setting 3 with             & \multicolumn{4}{c}{$n=50$} & \multicolumn{4}{c}{$n=200$}\\
                           \cmidrule(lr){2-5}           \cmidrule(lr){6-9}
correlated noise           & \multicolumn{2}{c}{$m=64$} & \multicolumn{2}{c}{$m=256$}& \multicolumn{2}{c}{$m=64$} & \multicolumn{2}{c}{$m=256$}\\
\cmidrule(lr){1-1}         \cmidrule(lr){2-3}\cmidrule(lr){4-5}\cmidrule(lr){6-7}\cmidrule(lr){8-9}
Power                      & SNR=4 & SNR=8 & SNR=4 & SNR=8 & SNR=4 & SNR=8 & SNR=4 & SNR=8\\
\cmidrule(lr){1-1}         \cmidrule(lr){2-3}\cmidrule(lr){4-5}\cmidrule(lr){6-7}\cmidrule(lr){8-9}
\textsf{Pearson}           & 0.0402 & 0.0402 & 0.0603 & 0.0603 & 0.0503 & 0.0503 & 0.0452 & 0.0553\\
\textsf{dnm}               & 0.0704 & 0.0653 & 0.0704 & 0.0704 & 0.0653 & 0.0653 & 0.0754 & 0.0754\\
\textsf{gtemp}             & 0.0854 & 0.0804 & 0.0704 & 0.0603 & 0.0553 & 0.0653 & 0.0704 & 0.0804\\
\textsf{dCov-c}            & 0.1055 & 0.1106 & 0.1005 & 0.1005 & 0.0905 & 0.0804 & 0.0804 & 0.0854\\
\textsf{FPCA}              & 0.0854 & 0.0905 & 0.1005 & 0.1005 & 0.0704 & 0.0653 & 0.0754 & 0.0704\\
\cmidrule(lr){1-1}         \cmidrule(lr){2-3}\cmidrule(lr){4-5}\cmidrule(lr){6-7}\cmidrule(lr){8-9}
\textsf{KMSZ}              & 0.0151 & 0.0101 & 0.0302 & 0.0352 & 0.1256 & 0.1357 & 0.1357 & 0.1407\\
\textsf{KMSZ-p}            & 0.0905 & 0.0854 & 0.1256 & 0.1256 & 0.1709 & 0.1759 & 0.1809 & 0.1960\\
\cmidrule(lr){1-1}         \cmidrule(lr){2-3}\cmidrule(lr){4-5}\cmidrule(lr){6-7}\cmidrule(lr){8-9}
\textsf{PSS}$(X\sim Y)$    & 0.0905 & 0.0905 & 0.0754 & 0.0754 & 0.0653 & 0.0804 & 0.0503 & 0.0653\\
\textsf{PSS}$(Y\sim X)$    & 0.0704 & 0.0653 & 0.0754 & 0.0854 & 0.0503 & 0.0603 & 0.0503 & 0.0402\\
\textsf{PSS}(Omnibus)      & 0.1558 & 0.1558 & 0.1457 & 0.1457 & 0.1156 & 0.1407 & 0.1005 & 0.0955\\
\cmidrule(lr){1-1}         \cmidrule(lr){2-3}\cmidrule(lr){4-5}\cmidrule(lr){6-7}\cmidrule(lr){8-9}
\textsf{FMDD}$(X\sim Y)$   & 0.0804 & 0.0804 & 0.0905 & 0.0905 & 0.0653 & 0.0754 & 0.0704 & 0.0754\\
\textsf{FMDD}$(Y\sim X)$   & 0.1005 & 0.1005 & 0.1005 & 0.0955 & 0.0754 & 0.0754 & 0.0754 & 0.0754\\
\textsf{FMDD}(Omnibus)     & 0.1106 & 0.1055 & 0.1005 & 0.1055 & 0.0804 & 0.0854 & 0.0754 & 0.0854\\
\cmidrule(lr){1-1}         \cmidrule(lr){2-3}\cmidrule(lr){4-5}\cmidrule(lr){6-7}\cmidrule(lr){8-9}
\textsf{wavHSIC}           & 0.4221 & 0.4472 & 0.4422 & 0.4422 & 0.9849 & 0.9849 & 0.9899 & 0.9899\\
median$\{\beta_X\}$        & 1.022 & 1.030 & 0.986 & 0.987 & 1.019 & 1.029 & 0.996 & 0.998        \\
median$\{\beta_Y\}$        & 0.799 & 0.810 & 0.800 & 0.802 & 0.801 & 0.807 & 0.795 & 0.798        \\
  \bottomrule
\end{tabular}
\end{table}

\section{Real Data Application}\label{sec:data}

We applied our proposed method to study human brain functional
connectivity using the MEG dataset collected by the HCP. 
MEG measures magnetic fields generated by human neuronal activities with a high
temporal resolution. Before source reconstruction, the signals from all MEG
sensors outside head were preprocessed 
 following the HCP MEG pipeline reference
 (\url{www.humanconnectome.org/software/hcp-meg-pipelines}) and the preprocessed data are publicly accessible from the HCP website.
 To obtain the electric activity signals from cortex regions,
 we applied the source reconstruction procedure of MEG signals 
 to the cerebral cortex atlas provided by \citet{Glas2016}
 using the linearly constrained minimum variance beamforming method in the MATLAB package \texttt{FieldTrip}.

To study the functional dependency between cortex regions under some motor activities,
we focused on motor task trials where subjects moved their right hands.
There were $n=61$ subjects in the trials. For each subject,
$8,004$ signal curves were obtained by denoising and source reconstruction procedures with
around 75 repeated trials.
Within each trial, the signals were recorded about every $2$ ms from $-1.2$ to
$1.2$ seconds, where the time $0$ is the starting time of the motion.
Since the motion in each trial usually lasts no longer than about $0.75$ seconds
and typically a subject finished the previous movement and received a new cue
between times $-0.25$ and $0$ of the next trial, we considered the time domain
$[-0.2521, 0.7525]$ which covers the time period of interest, with
$m=512$ sampled time points in total.

We applied the proposed method \textsf{wavHSIC} to perform an independence test for every pair of the MEG signals. To implement \textsf{wavHSIC}, we chose the CDJV wavelet basis functions with vanishing moment $D=4$ %
which leads to $\alpha\approx 1.6179$. %
For each signal, the smoothness parameter $\beta$ was selected by the method in Section \ref{sec:select}. 
For comparison, we also provided the results for the model-based test
\textsf{KMSZ}, \textsf{KMSZ-p} and two model-free tests, \textsf{Pearson} and \textsf{FPCA}.
\textsf{KMSZ}, \textsf{KMSZ-p} and \textsf{FPCA} were based on top FPC scores which cumulatively account for 
95\% of the variation of each signal. The p-value for testing the independence between each pair of signals were obtained by 
1,999 permutations for \textsf{wavHSIC}, \textsf{FPCA} and \textsf{KMSZ-p}.
{
We did not include \textsf{PSS} and \textsf{FMDD} here due to their extended computing
times. See Table \ref{tab:time} below for an illustration.}

\begin{figure}[h!]
\vspace{-0.6cm}
  \centering
  \includegraphics[width=0.6\textwidth]{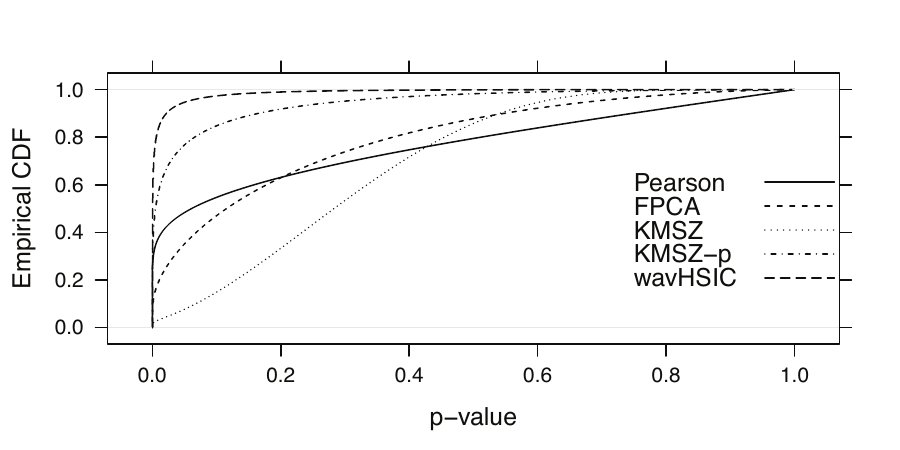}
\vspace{-0.4cm}
  \caption{Empirical cumulative distribution function for the p-values for testing the independence between every pair of the 8,004
  signals for each method.}
  \label{fig:RHecdf}
\vspace{-0.5cm}
\end{figure}

The empirical cumulative distribution functions for the p-values of the five
methods are given in Figure \ref{fig:RHecdf}, which shows that \textsf{wavHSIC}
is more sensitive to detecting connectivity than the other methods.
To evaluate and compare the five methods at the presence of multiple testing, we
set the same discovery rate at 60\% to control the number of edges, or sparsity, 
of each brain connectivity network, which is important in evaluating the reliability 
of brain network metrics \citep[e.g.][]{VanSD10,Tsai18}. 
In this analysis, we focus on sensorimotor areas 4, 3a, 3b, 1 and 2 on the left
and right hemispheres as illustrated in Figure \ref{fig:RH_wavHSIC_adjmat} (c) which are
most related to motor task trials \citep{Glas2016}.
With a controlled discovery rate, we expect an excellent connectivity detection method 
to identify plenty of edges within these areas.

\begin{figure}[hb!]
  \centering
  \begin{subfigure}[b]{0.3\textwidth}
    \centering
    \includegraphics[width=\textwidth]{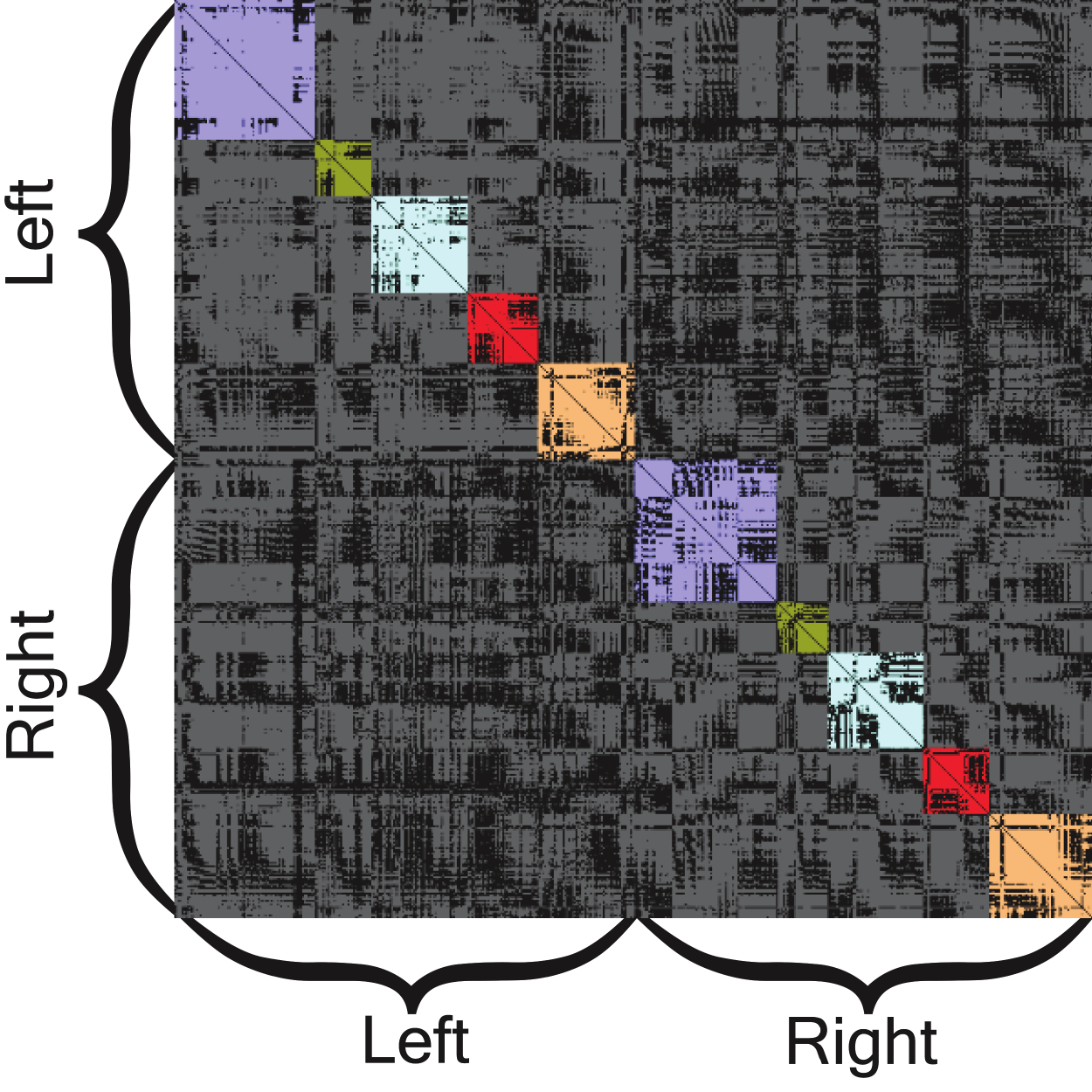}
    \caption{\textsf{Pearson}}
  \end{subfigure}
  \begin{subfigure}[b]{0.3\textwidth}
    \centering
    \includegraphics[width=\textwidth]{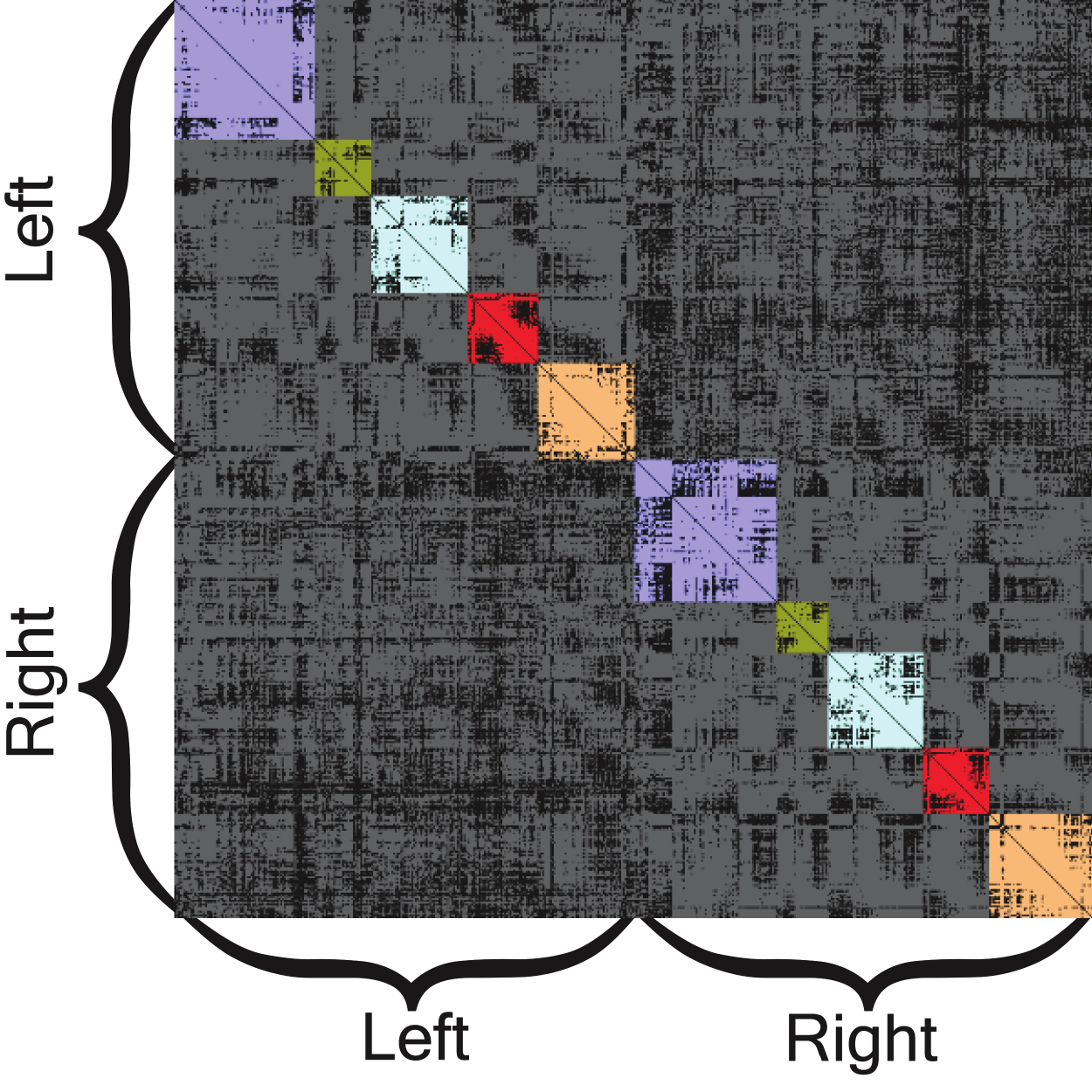}
    \caption{\textsf{FPCA}}
  \end{subfigure}
  \\
  \begin{subfigure}[b]{0.3\textwidth}
    \centering
    \includegraphics[width=\textwidth]{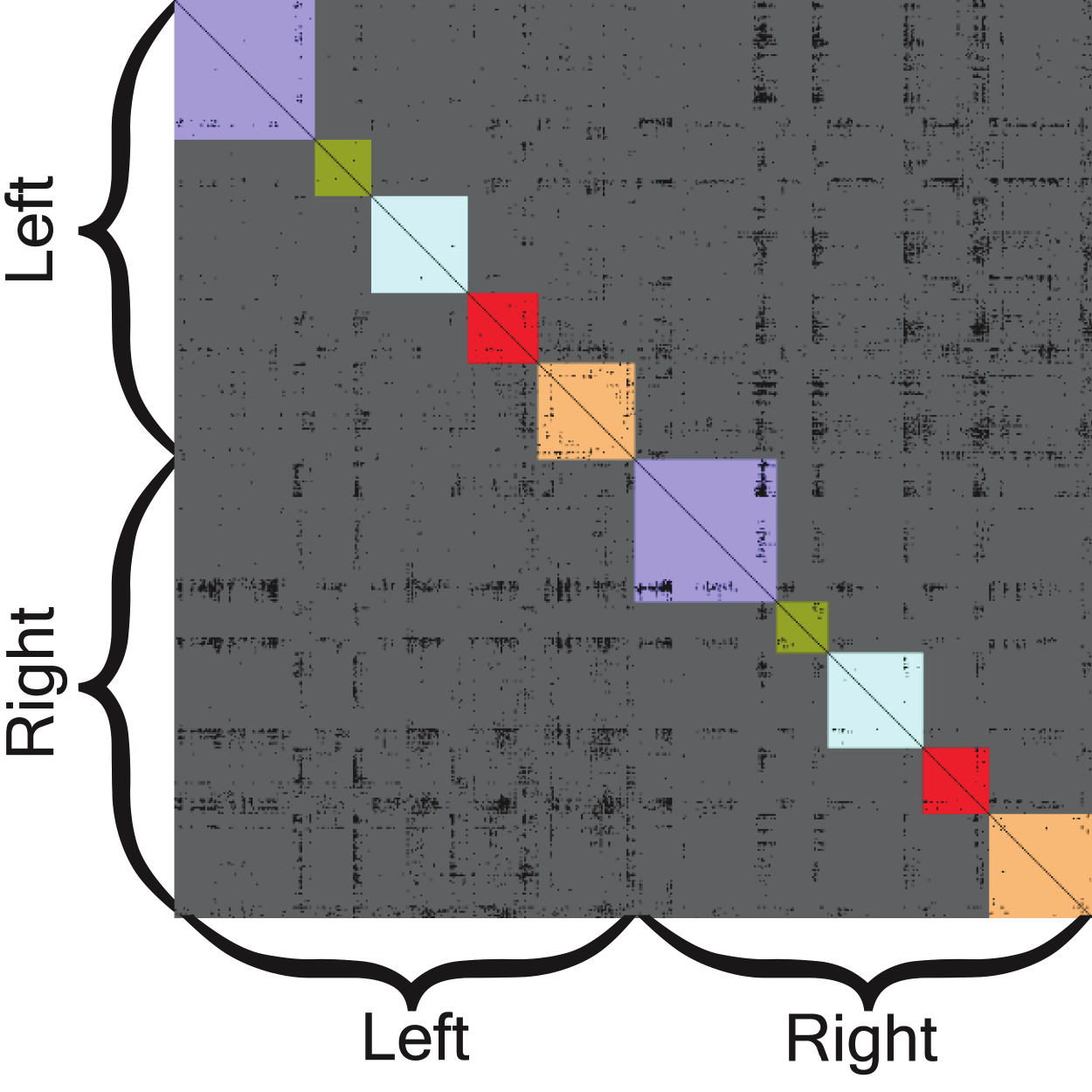}
    \caption{\textsf{KMSZ}}
  \end{subfigure}
  \begin{subfigure}[b]{0.3\textwidth}
    \centering
    \includegraphics[width=\textwidth]{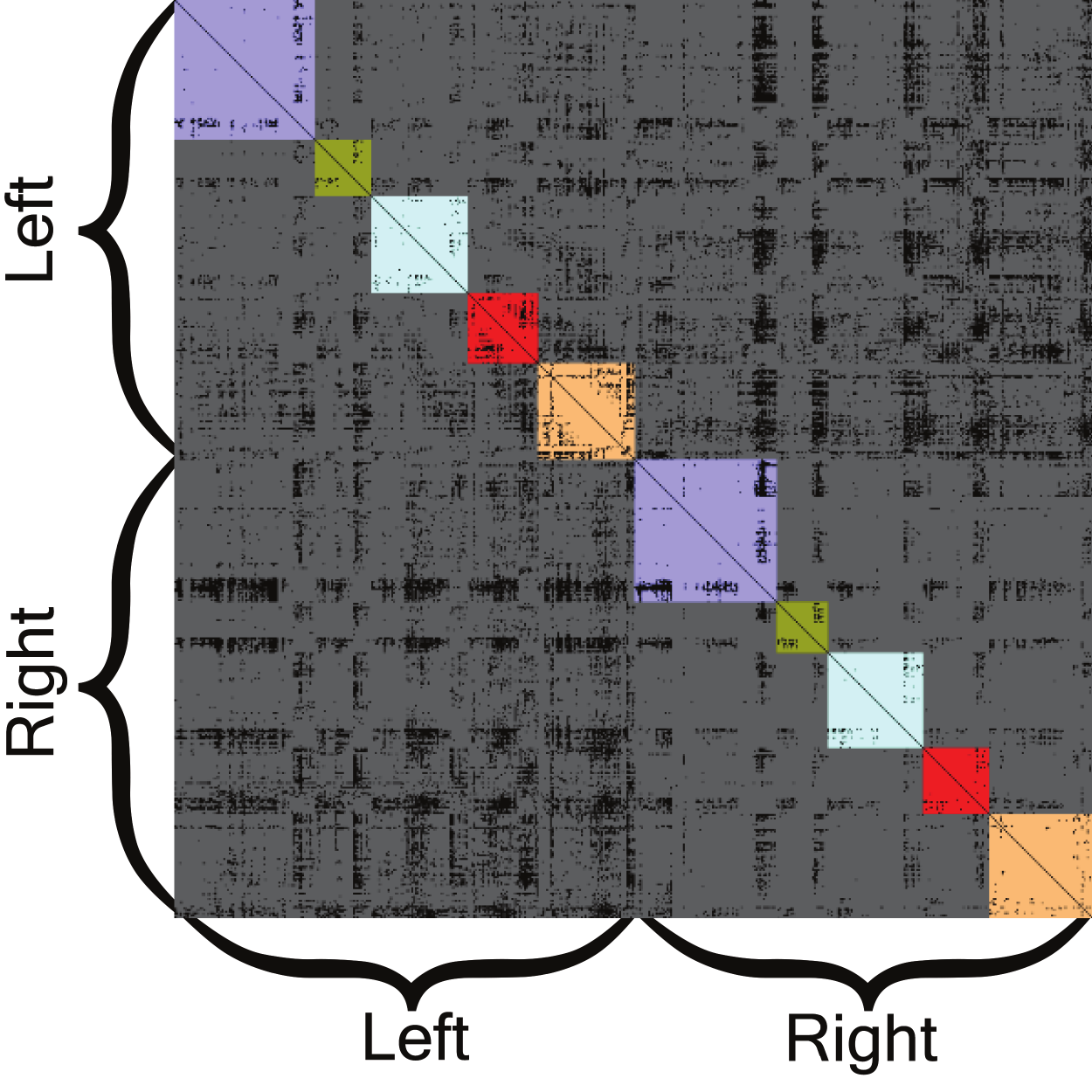}
    \caption{\textsf{KMSZ-p}}
  \end{subfigure}

\begin{subfigure}[b]{\textwidth}
\centering
\raisebox{.6\height}{\fcolorbox{black}{Area4}{\hspace{2mm}}} Area 4
\raisebox{.6\height}{\fcolorbox{black}{Area3a}{\hspace{2mm}}} Area 3a
\raisebox{.6\height}{\fcolorbox{black}{Area3b}{\hspace{2mm}}} Area 3b
\raisebox{.6\height}{\fcolorbox{black}{Area1}{\hspace{2mm}}} Area 1
\raisebox{.6\height}{\fcolorbox{black}{Area2}{\hspace{2mm}}} Area 2
\end{subfigure}
  \caption{Functional connectivity networks of the five sensorimotor areas in the left and right hemispheres. 
  In the adjacency matrices in (a), (b), (c), (d) obtained by the four methods respectively, 
  a bright entry indicates significant dependency between the corresponding signal pairs while a dark one indicates otherwise.}
    \label{fig:RHadjmat}
\vspace{-0.6cm}
\end{figure}

\begin{figure}[ht!]
  \centering
  \begin{subfigure}[b]{0.3\textwidth}
    \centering
    \includegraphics[width=\textwidth]{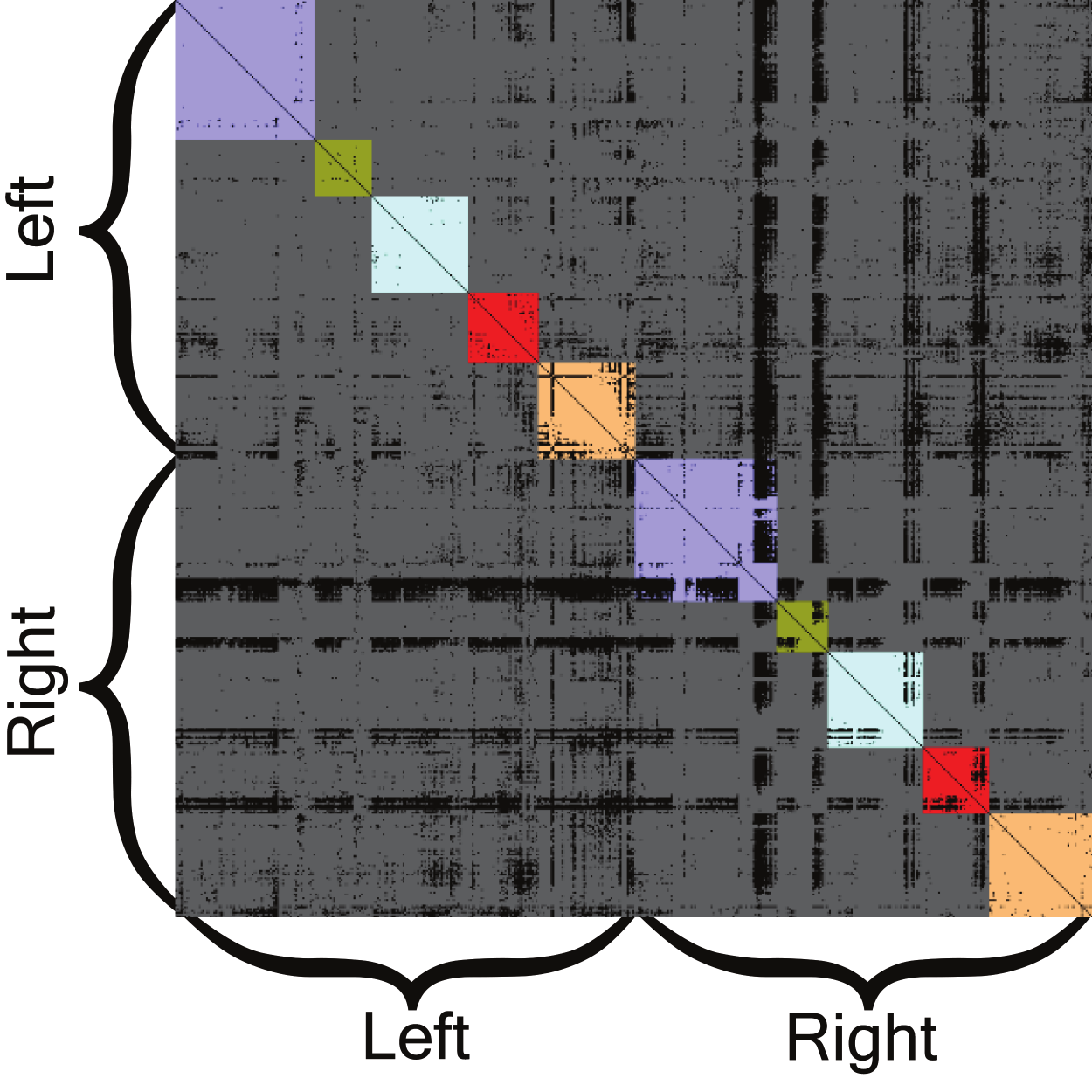}
    \caption{\textsf{wavHSIC}}
  \end{subfigure}
  \begin{subfigure}[b]{0.14\textwidth}
    \includegraphics[width=0.875\textwidth]{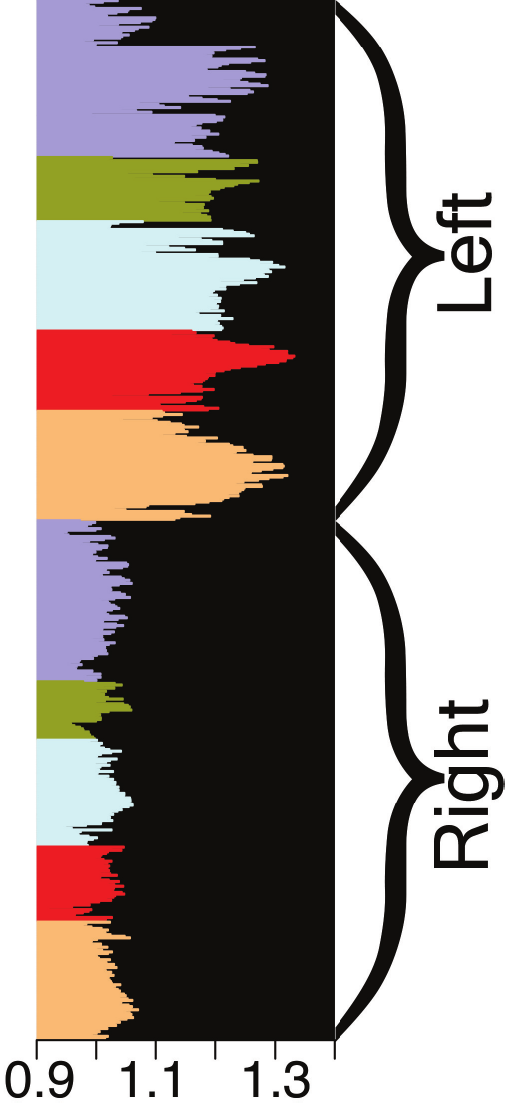}
    \vspace{15pt}
    \caption{$\beta$}
  \end{subfigure}
  \begin{subfigure}[b]{0.23\textwidth}
    \centering
    \includegraphics[width=1.1\textwidth]{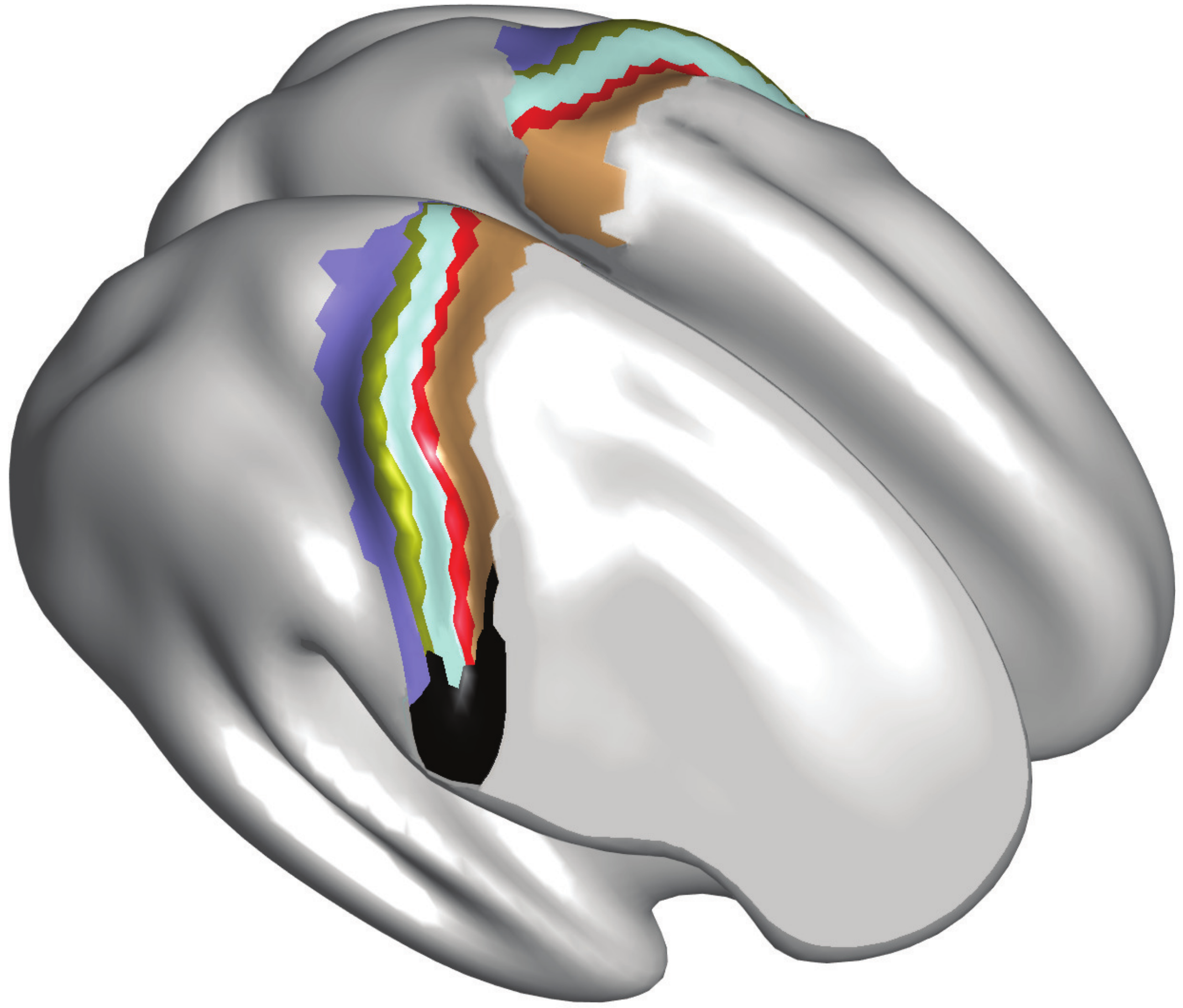}
    \vspace{5pt}
    \caption{Brain Cortex}
  \end{subfigure}
\\
  \begin{subfigure}[b]{0.3\textwidth}
    \centering
    \includegraphics[width=\textwidth]{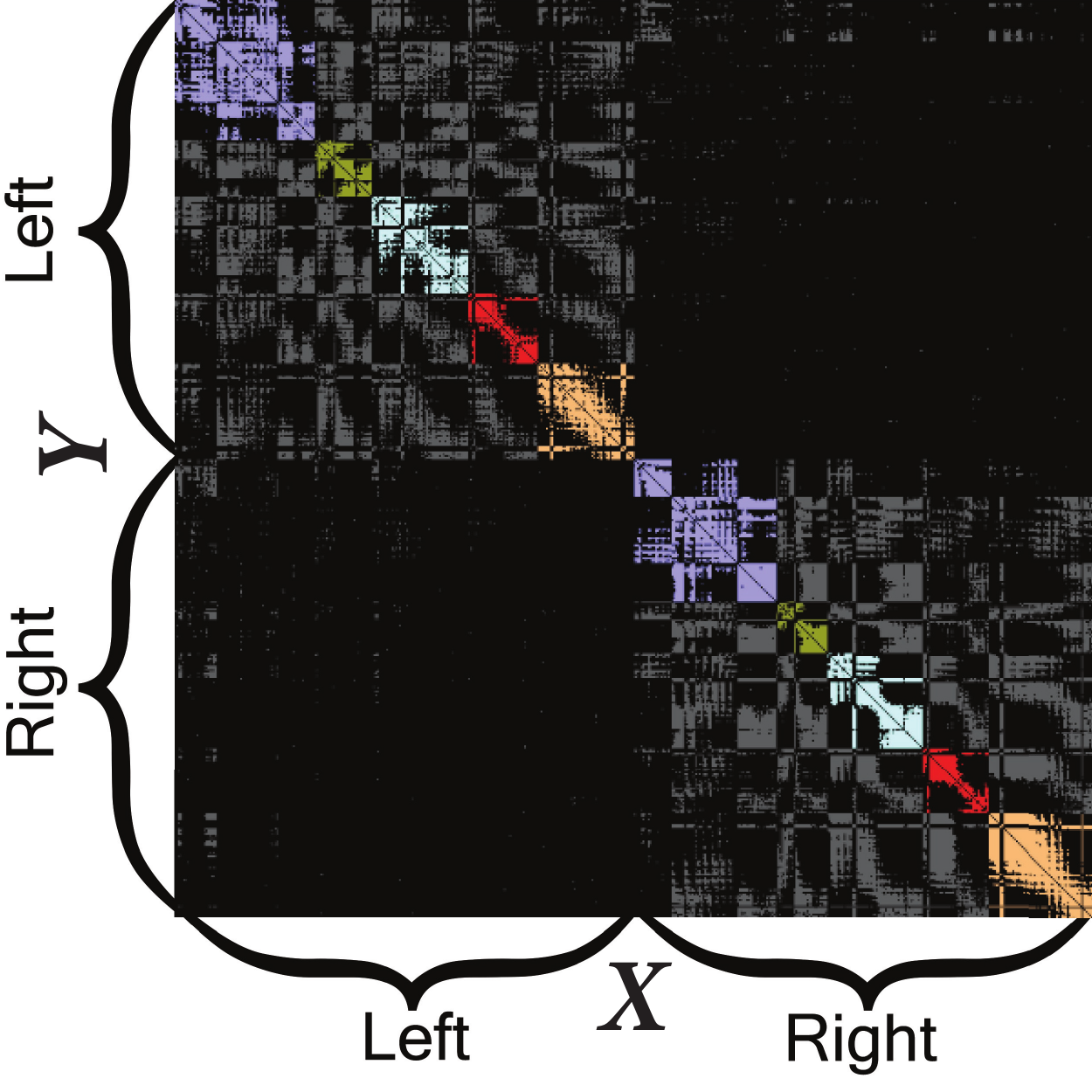}
    \caption{(LF of $X$, LF of $Y$)}
  \end{subfigure}
  \begin{subfigure}[b]{0.3\textwidth}
    \centering
    \includegraphics[width=\textwidth]{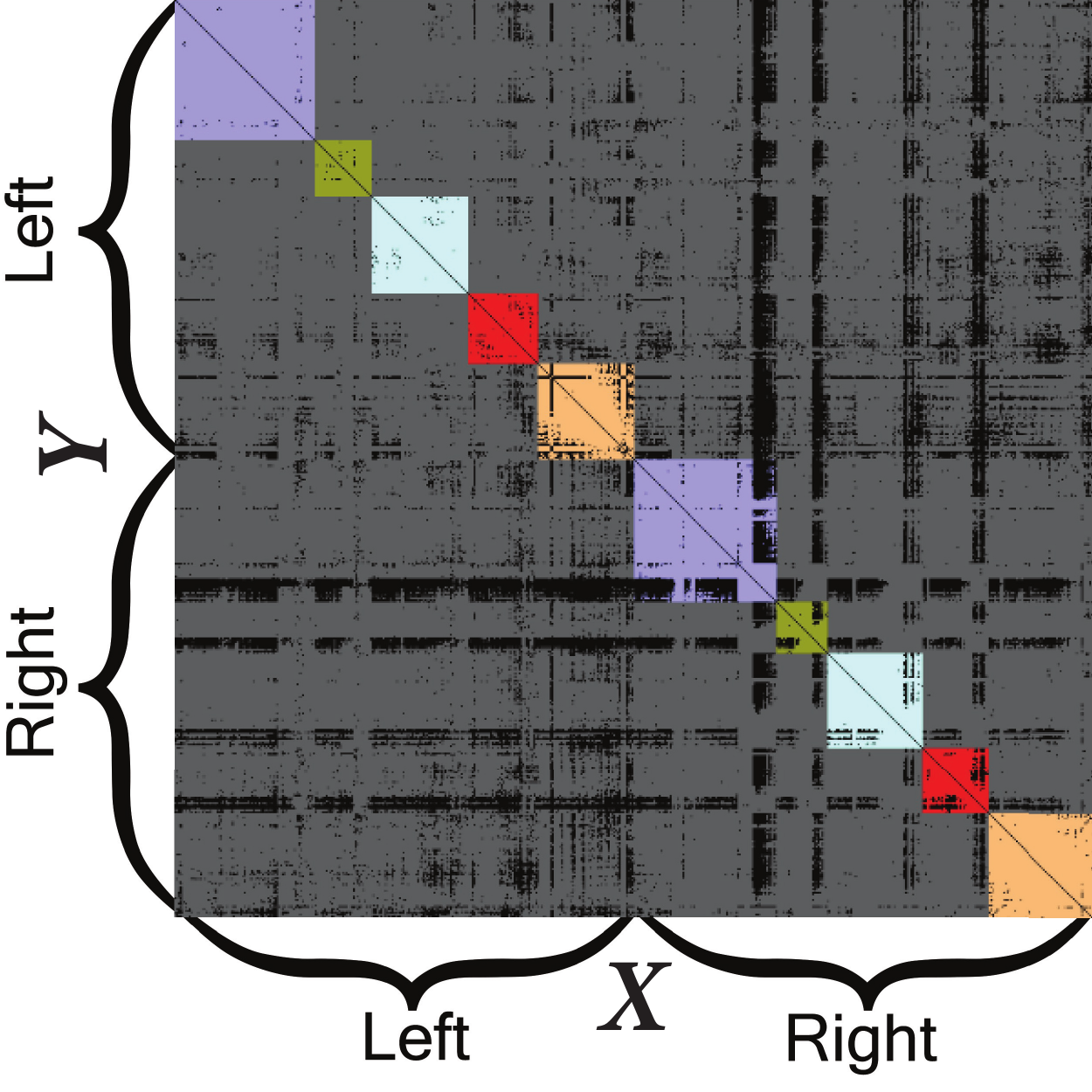}
    \caption{(HF of $X$, HF of $Y$)}
  \end{subfigure}
  \begin{subfigure}[b]{0.3\textwidth}
    \centering
    \includegraphics[width=\textwidth]{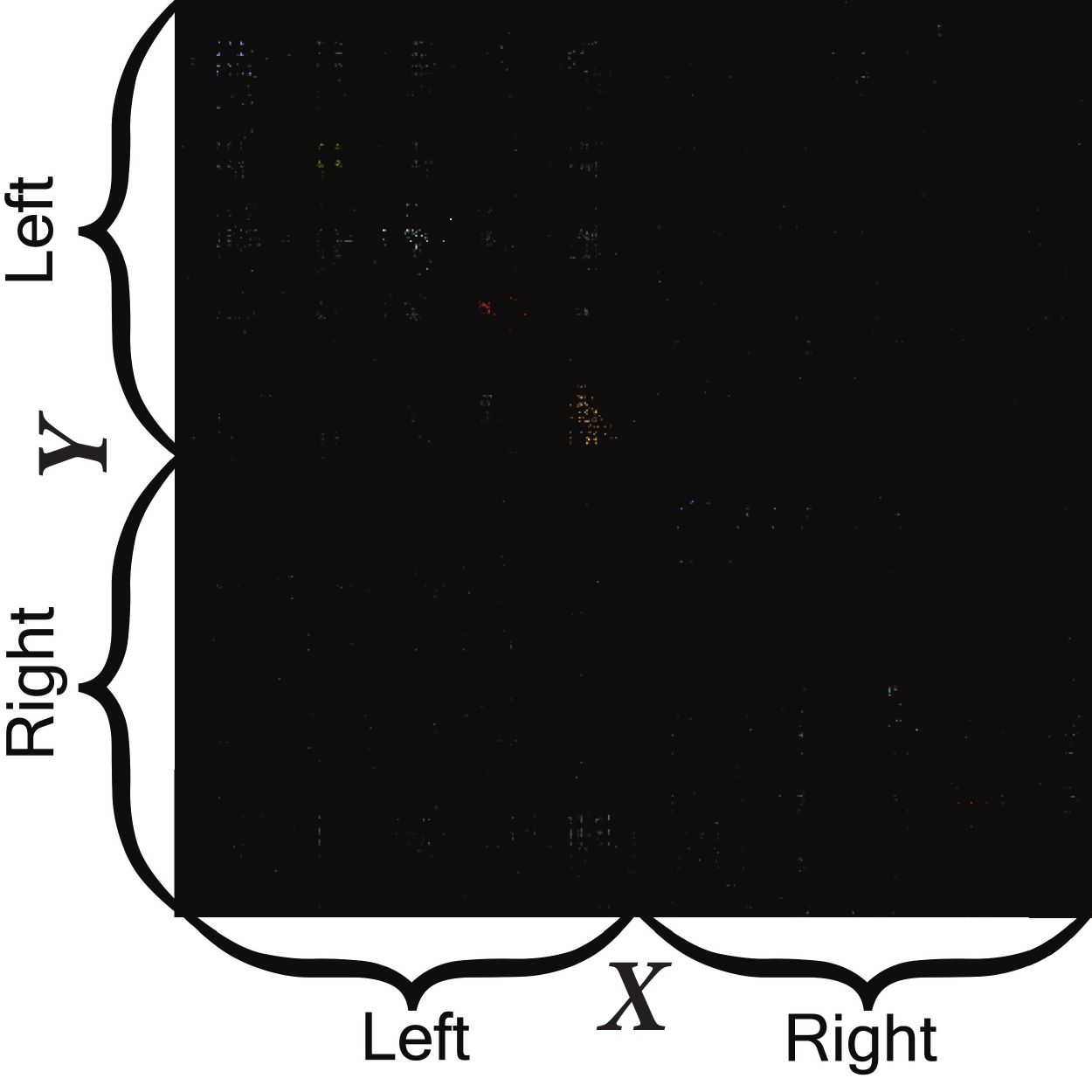}
    \caption{(LF of $X$, HF of $Y$)}
  \end{subfigure}

\begin{subfigure}[b]{\textwidth}
\centering
\raisebox{.6\height}{\fcolorbox{black}{Area4}{\hspace{2mm}}} Area 4
\raisebox{.6\height}{\fcolorbox{black}{Area3a}{\hspace{2mm}}} Area 3a
\raisebox{.6\height}{\fcolorbox{black}{Area3b}{\hspace{2mm}}} Area 3b
\raisebox{.6\height}{\fcolorbox{black}{Area1}{\hspace{2mm}}} Area 1
\raisebox{.6\height}{\fcolorbox{black}{Area2}{\hspace{2mm}}} Area 2
\end{subfigure}
\caption{
Functional connectivity networks of the five sensorimotor areas in the left and right hemispheres with the same color scheme in Figure \ref{fig:RHadjmat}.
The adjacency matrix (a) is obtained by \textsf{wavHSIC} with the smoothness parameters $\beta$, selected 
by the method in Section \ref{sec:select}, illustrated in the barplot (b).
The black subregion in (c) corresponds to face and eye portions
and the rest of the colored area corresponds to upper limbs, trunk and lower limbs portions in the right hemisphere.
The adjacency matrices in (d), (e), (f) are obtained by applying \textsf{wavHSIC} to low($\leq 4$Hz)/high($>4$Hz)-pass-filtered signals with the same %
$\beta$ values in (b) and the same p-value threshold in (a).
}
    \label{fig:RH_wavHSIC_adjmat}
\vspace{-0.4cm}
\end{figure}

{Figures \ref{fig:RHadjmat} and \ref{fig:RH_wavHSIC_adjmat} (a) provide} the functional connectivity networks within these
sensorimotor areas obtained by the five methods. The nodes in
each area were ordered from the superio-medial cortex to infero-lateral cortex following 
the atlas ``atlas\_MMP1.0\_4k.mat'' in \texttt{FieldTrip}. 
Compared with \textsf{KMSZ}, \textsf{KMSZ-p} and \textsf{wavHSIC}, \textsf{Pearson} and
\textsf{FPCA} are substantially less sensitive to detecting functional
connectivity and their corresponding networks are less structured (see Figure
\ref{fig:RHadjmat} (a) and (b)). This demonstrates the superior performances of
both \textsf{KMSZ}, \textsf{KMSZ-p} and \textsf{wavHSIC} in identifying connectivity patterns
within these areas which are anatomically connected and functionally related to
the motion task trials. Different from the overall homogeneous pattern in the
network for \textsf{KMSZ}, several structured dark strips appear in the network
obtained by \textsf{KMSZ-p} and \textsf{wavHSIC} within {sensorimotor
areas 4, 3a, 3b and 1 in the right hemisphere (see Figures \ref{fig:RHadjmat} (c--d) and {\ref{fig:RH_wavHSIC_adjmat} (a)})}.
These dark strips are much clearer in {Figure \ref{fig:RH_wavHSIC_adjmat} (a) than in Figure \ref{fig:RHadjmat} (d)}.
This indicates that \textsf{wavHSIC} can {more clearly} identify two sub-areas {in sensorimotor areas 4, 3a,
3b and 1 in the right hemisphere}, the top left (TL) and bottom right (BR)
corners respectively in these corresponding colored squares as in {Figure
\ref{fig:RH_wavHSIC_adjmat} (a)}. The signals within these four TL sub-areas or within
these four BR sub-areas are strongly connected, while the connectivities between
these TL and BR sub-areas are generally weak. According to \citet{Glas2016}, the
four BR sub-areas in the same hemisphere correspond to face and eye portions
while the four TL sub-areas correspond to upper limbs, trunk and lower limbs
portions. Since the motor task involved in this dataset is raising the right
hand, the connectivity patterns detected by \textsf{wavHSIC} are intuitively and
anatomically interpretable.

{
Next we illustrate how to identify dependency structures between and within different 
frequency bands using \textsf{wavHSIC}. Explicitly, we first split the denoised wavelet 
coefficients of each brain signal into two parts, the low-frequency part (LF, $j\leq 3$) 
and high-frequency part (HF, $j>3$), which approximately correspond to the Delta band ($\leq 4$Hz)
and the Theta to the Ultra-Gamma bands ($>4$Hz) respectively \citep[e.g.,][]{buzsaki2006rhythms}. %
Then for each pair of signals $(X,Y)$ as illustrated in Figure \ref{fig:RH_wavHSIC_adjmat} (d), (e) and (f), we applied \textsf{wavHSIC} to (LF of $X$, LF of $Y$), 
(HF of $X$, HF of $Y$), and (LF of $X$, HF of $Y$) respectively. 
Their corresponding functional connectivity patterns are shown in 
Figure \ref{fig:RH_wavHSIC_adjmat} (d), (e) and (f) respectively. Note that the results for (HF of $X$, LF of $Y$) are included in Figure \ref{fig:RH_wavHSIC_adjmat} (f) by switching the roles of $X$ and $Y$. 
Apparently, the network in Figure \ref{fig:RH_wavHSIC_adjmat} (e) is very similar to that 
in Figure \ref{fig:RH_wavHSIC_adjmat} (a), which indicates that the functional dependency 
induced by this motor task mainly lies at high frequencies. Moreover, 
Figure \ref{fig:RH_wavHSIC_adjmat} (f) shows that there is essentially no dependency between 
the low-frequency and high-frequency signals. Lastly, Figure \ref{fig:RH_wavHSIC_adjmat} (d) 
reveals that some dependency can be detected at low frequencies, but only within the same hemisphere. 
This is probably due to the fact that functional Delta oscillations appear to be implicated in the 
synchronization of brain activity with autonomic functions of vegetative nervous system, but is 
not affected by a specific task \citep{knyazev2012eeg}. 
}

To compare computing times of these methods together with \textsf{PSS} and \textsf{FMDD}, we randomly selected \textit{one} pair of signals 
and then 
repeatedly executed each of them 20 times 
{on a Windows 10 desktop with AMD Ryzen7 3800X CPU and 16GB RAM.} 
A summary of their averaged computing times (in seconds) is given in Table \ref{tab:time}.
{The long computational times of \textsf{PSS} and \textsf{FMDD} make it difficult to study dependency between \textit{every} pair and create corresponding functional connectivity networks, so we did not include them in the analysis above.}

\begin{table}[H]
  {
  \caption{Mean computing times (in seconds) based on \textit{one} randomly selected pair of signals for the seven methods in comparison. The values in parentheses are standard deviations.}
  \label{tab:time}
  \centering
  \scriptsize
  \begin{tabular}{c|ccccccc}
    \toprule
    Method & \textsf{Pearson} & \textsf{FPCA} & \textsf{KMSZ} & \textsf{KMSZ-p} & \textsf{wavHSIC}  & \textsf{PSS} & \textsf{FMDD}\\
    \midrule
    Time & $0.003 (0.002)$ & $0.103 (0.072)$ & $0.020 (0.003)$ & $0.158 (0.082)$ & $0.162 (0.017)$ & $43.076 (0.728)$ & $15.280 (0.253)$ \\
    \bottomrule
  \end{tabular}
  }
\end{table}

\section{Discussion} \label{sec:disc}

In this paper, we propose a model-free wavelet-based independence test for 
two random functions of which sample paths belong to  
possibly different Besov spaces. 
Our method is built upon HSIC endowed with characteristic kernels, 
which is zero if and only if the two random functions are independent. 
Since the Besov space with wavelet basis functions provides an effective modeling environment for
sample paths with various levels of smoothness, HSIC with characteristic kernels
induced by wavelet coefficients is capable of capturing the dependency at different
frequencies. Therefore, the proposed method is especially powerful when the two random functions are dependent only at high frequencies, as demonstrated in Section \ref{sec:simu}. If the dependency is strong at low frequencies, our simulation not presented here shows that the proposed method is not substantially advantageous over \textsf{FPCA}.

In the application to MEG functional connectivity, 
the proposed method by construction is only able to identify the \emph{unconditional} 
dependency between two signal curves. 
Although metrics that reflect unconditional functional connectivity are 
still widely used in neuroscience \citep[see, e.g.,][for a review]{Marzetti2019}, 
a conditional independence measure or test will be more convincing to identify the 
functional connectivity between two signal curves given all others in the brain.
To address this problem, there have been some advances in functional graphical models. 
Most of the existing methods 
reply on either Gaussianity 
\citep[e.g.,][]{Zhu2016Bayesian,Qiao2019Functional,qiao2020doubly,zapata2019partial,solea2020copula,zhao2021high}
or regression models \citep[e.g.][]{lundborg2021conditional}, while a few exceptions assume additive structures \citep[e.g.,][]{li2018nonparametric,Lee2021Conditional,solea2021nonparametric}. Developing a conditional independence test with these assumptions relaxed would be an interesting future research topic.

\bibliographystyle{chicago}
{\linespread{0.7}\selectfont\bibliography{references}}

\clearpage

\setcounter{section}{0}
\setcounter{subsection}{0}
\setcounter{equation}{0}
\setcounter{figure}{0}
\setcounter{table}{0}
\setcounter{footnote}{0}
\setcounter{thm}{0}
\setcounter{defi}{0}
\setcounter{lem}{0}
\setcounter{coro}{0}
\setcounter{prop}{0}
\setcounter{ex}{0}
\setcounter{rmk}{0}
\setcounter{asmp}{0}

\renewcommand{\thesection}{S\arabic{section}}
\renewcommand{\thesubsection}{S\arabic{section}.\arabic{subsection}}
\renewcommand{\theequation}{S\arabic{equation}}
\renewcommand{\thetable}{S\arabic{table}}
\renewcommand{\thethm}{S\arabic{thm}}
\renewcommand{\thedefi}{S\arabic{defi}}
\renewcommand{\thelem}{S\arabic{lem}}
\renewcommand{\thecoro}{S\arabic{coro}}
\renewcommand{\theprop}{S\arabic{prop}}
\renewcommand{\theex}{S\arabic{ex}}
\renewcommand{\thermk}{S\arabic{rmk}}
\renewcommand{\theasmp}{S\arabic{asmp}}

\providecommand{\theHsection}{}
\providecommand{\theHsubsection}{}
\providecommand{\theHequation}{}
\providecommand{\theHfigure}{}
\providecommand{\theHtable}{}
\providecommand{\theHfootnote}{}
\providecommand{\theHthm}{}
\providecommand{\theHdefi}{}
\providecommand{\theHlem}{}
\providecommand{\theHcoro}{}
\providecommand{\theHprop}{}
\providecommand{\theHex}{}
\providecommand{\theHrmk}{}
\providecommand{\theHasmp}{}
\renewcommand{\theHsection}{supp.section.\arabic{section}}
\renewcommand{\theHsubsection}{supp.subsection.\arabic{section}.\arabic{subsection}}
\renewcommand{\theHequation}{supp.equation.\arabic{equation}}
\renewcommand{\theHfigure}{supp.figure.\arabic{figure}}
\renewcommand{\theHtable}{supp.table.\arabic{table}}
\renewcommand{\theHfootnote}{supp.footnote.\arabic{footnote}}
\renewcommand{\theHthm}{supp.thm.\arabic{thm}}
\renewcommand{\theHdefi}{supp.defi.\arabic{defi}}
\renewcommand{\theHlem}{supp.lem.\arabic{lem}}
\renewcommand{\theHcoro}{supp.coro.\arabic{coro}}
\renewcommand{\theHprop}{supp.prop.\arabic{prop}}
\renewcommand{\theHex}{supp.ex.\arabic{ex}}
\renewcommand{\theHrmk}{supp.rmk.\arabic{rmk}}
\renewcommand{\theHasmp}{supp.asmp.\arabic{asmp}}

\renewcommand{\GamX}{{\bm \Gamma^{X}}}
\renewcommand{\GamhX}{{\bm \Gamma^{\hX}}}
\renewcommand{\cGamX}{\check{\bm \Gamma}^{X}}
\renewcommand{\tGamX}{\tilde{\Gamma}^{\mX}}
\renewcommand{\GamY}{{\bm \Gamma^{Y}}}
\renewcommand{\GamhY}{{\bm \Gamma^{\hY}}}
\renewcommand{\cGamY}{\check{\bm \Gamma}^{Y}}
\renewcommand{\tGamY}{\tilde{\Gamma}^{\mY}}
\renewcommand{\bkx}{{\bm \kappa}_{\mathcal{X}}}
\renewcommand{\bky}{{\bm \kappa}_{\mathcal{Y}}}
\renewcommand{\bkz}{{\bm \kappa}_{\mathcal{Z}}}
\renewcommand{\ray}[1]{{\color{magenta}\bf [Raymond: #1]}}

\begin{center}
{\large\bf SUPPLEMENTARY MATERIAL}
\end{center}
\vspace{-0.3cm}
The supplementary material includes background materials on distance-induced
characteristic kernels and Besov spaces, technical proofs of Theorems
\ref{thm:BesovNegType}--\ref{thm:perm} and additional simulations.
\vspace{-0.1cm}

\def\spacingset#1{\renewcommand{\baselinestretch}%
{#1}\small\normalsize} \spacingset{1}

\linespread{1.5}

\section{Background Materials}
\subsection{Distance-Induced Characteristic Kernels}
\label{sec:metric}

Characteristic kernels are required to construct HSIC for two random functions under the RKHS framework. 
Such a kernel can be generated by a semi-metric of strong negative type.

\begin{defi}[%
Strong Negative Type Semi-Metric]
\label{defi:negatype}
A semi-metric $\rho:\mZ\times\mZ\rightarrow
  [0,\infty)$ defined on a non-empty set $\mZ$ 
  is of negative type if  $\sum_{i=1}^n\sum_{j=1}^n \alpha_i\alpha_j\rho(z_i,z_j)\leq 0$ 
  for all $z_1, \dots, z_n \in \mZ$ and $\alpha_1,\dots,\alpha_n\in \R$ such that 
  $\sum_{i=1}^n\alpha_i =0$, $n\geq 2$. 
Furthermore, it is of strong negative type if for any two probability
measures $P$ and $P'$ on $\mZ$ such that 
$$\int_{\mZ} \rho(z,z_0) \dd P(z), \int_{\mZ} \rho(z,z_0) \dd P'(z) <\infty$$ 
for some $z_0\in \mZ$, we have
$\int_{\mZ}\int_{\mZ} \rho(z_1,z_2) \,\dd (P-P')(z_1)\dd (P-P')(z_2)=0$ if and only if $P=P'$.
\end{defi}
Proposition \ref{prop:dist-induce} shows that a kernel induced by a strong negative type semi-metric is characteristic.
\begin{prop}%
\label{prop:dist-induce}
  Let $\rho$ be a semi-metric defined on $\mZ$ and $z_0\in \mZ$. 
  The induced kernel
   $\kappa_{\rho}\left(z,z^{\prime}\right) = \rho(z,z_0) +
  \rho \left( z^{\prime},z_0 \right) - \rho \left( z,z^{\prime} \right)$,
   $z, z^{\prime} \in \mZ$,
   is symmetric and positive definite.
   Moreover, {for all probability measures $P$ such that
   $\int_{\mZ}\rho(z,z_0) d P(z)<\infty$ for some $z_0\in\mZ$,}
   $\kappa_{\rho}$ is characteristic if and only if $\rho$ is of strong negative type.
\end{prop}
Obviously distance-induced kernels are symmetric. For the proof of Proposition
\ref{prop:dist-induce}, see Lemma 2.1 of \citet{BergCR84} for positive definiteness
and \citet{Lyon13} and \citet{SejdSG13} for the characteristic property.
Since the set $\mZ$ of interest often contains zero, in this paper we always set $z_0=0$ for any 
distance-induced kernel $\kappa_{\rho}$ for simplicity and convenience.

\subsection{Besov Spaces and Norms}\label{sec:besov}

The Besov space is a generalization of the Sobolev space, which is widely used in nonparametric regression under the RKHS framework. %
A Besov space $\Bpqa[0,1], p,q,\alpha>0,$ contains all functions of which Besov norm $\nmB{\cdot}$ is finite. Explicitly, %
with any integer $r \ge 1$, %
define the $r$th order difference of a function $f$ by
$$
\Delta_h^r(f,x) = \sum_{k=0}^r \binom{r}{k} (-1)^{r-k} f(x+kh), %
$$
and its $r$th order modulus of continuity by
$$
\omega_r(f,t)_p = \sup_{0\leq h\leq t}\norm{\Delta_h^r(f,\cdot)\mid_{[0,1-rh]}}_{L^p},
$$
where $\Delta_h^r(f,\cdot)\mid_{[0,1-rh]}$ represents $\Delta_h^r(f,\cdot)$ restricted on $[0,1-rh]$ and $\norm{\cdot}_{L^p}$ is the $L^p$ norm. Then the Besov norm of $f$ is defined by
$$
  \nmB{f} = \norm{f}_{L^p} + \left| f \right|_{B_{p,q}^{\alpha}}, \quad
  \text{where} \quad \left| f \right|_{B_{p,q}^{\alpha}} = \left[
    \int_{0}^{\infty}\left\{ \frac{\omega_r(f,t)_p}{t^{\alpha}} \right\}^q
    \frac{\dd t}{t} \right]^{\frac{1}{q}}.
$$

For the same $\alpha$, the Besov norms generated by different values of $r > \alpha$ are equivalent 
when $p > 1$ \citep{DevoL93}. In this paper we always assume $p > 1$ and $r = \lfloor\alpha\rfloor + 1$ where $\lfloor
\alpha \rfloor$ is the greatest integer less than or equal to $\alpha$.

The Besov norm (semi-norm) 
generalizes some traditional smoothness measures, such as 
the Sobolev semi-norm $|\cdot|_{W_p^k}$ 
\begin{equation*}
|f|_{W_p^k} = \left( \int_0^1 \left| D^k f \right|^p\dd x \right)^{1/p}, ~~ 1\leq p \leq \infty,
\end{equation*}
where $D^k$ is $k$th order weak-derivative operator.

\section{Technical Proofs} 
\subsection{Proof of Theorem \ref{thm:BesovNegType}}
We first list two lemmas %
on some properties of negative type semi-metrics, which will be needed in the proof of Theorem \ref{thm:BesovNegType}.

\begin{defi}[Radial Positive Definite Function]
  A real function $F$ defined on $[0,\infty)$ is called radial positive definite on the
  semi-metric space $(\mZ,\rho)$ if $F$ is continuous and
\begin{equation*}
\sum_{j=1}^n\sum_{k=1}^n F(\rho(z_j,z_k))c_j c_k\geq 0, 
\end{equation*}
for all choices of $n \geq 1$ points $z_1,\dots,z_n \in \mZ$.
We denote the set of all radial positive definite functions by $\RPD(\mZ)$.
\end{defi}  

\begin{lem}%
\label{lem:Wells4.4}
  The following hold in any semi-metric space $\mZ$.
\begin{enumerate}[(a)]
\item $\RPD(\mZ)$ is never empty.
\item If $F_1,F_2\in\RPD(\mZ)$, then $F_1\cdot F_2\in \RPD(\mZ)$.
\item If $F_j\in \RPD(\mZ)$ and $0\leq c_j < \infty$, $j=1,\dots,n$, then
  $\sum_{j=1}^nc_jF_j\in\RPD(\mZ)$.
  \item If $F_j\in \RPD(\mZ),j=1,2,\dots$ and the $F_j$ converge point-wise to a
    continuous limit $F$, then $F\in \RPD(\mZ)$.
  \item For space $(L^p,\norm{\cdot}_p)$, $(\ell^p,\norm{\cdot}_p)$ with $0<p\leq
    2$, then $\exp(-t^{\alpha})$ is RPD for $0<\alpha\leq p$. %
\end{enumerate}
\end{lem} 
Lemma \ref{lem:Wells4.4} is a combination of Theorems 4.4 and 4.10 of \citet{WellsW2012}. %

\begin{lem}[Theorem 4.5, \citet{WellsW2012}] \label{lem:Wells4.5}
  In a semi-metric space $(\mZ,\rho)$, the following are equivalent:
\begin{enumerate}[(a)]
\item $\rho$ is of negative type;
\item the function $\exp (-\lambda t)$ belongs to $\RPD(\mZ,\rho)$ for
  $\lambda>0$;
\item $(\mZ,\rho^{1/2})$ is {isometrically} embeddable in a Hilbert space.
\end{enumerate}
\end{lem}

\begin{lem}[Theorem 4.7, \citet{WellsW2012}] \label{lem:Wells4.7}
If semi-metric $\rho$ is of negative type on $\mZ$, then $\rho^r$ is of negative type for any $0<r<1$.
\end{lem}

\begin{proof}[Proof of Theorem \ref{thm:BesovNegType}]
  By Proposition \ref{prop:dist-induce}, it suffices to prove that $\rho_{b^{\alpha}_{p,q}}$ is of 
  strong negative type.
Lemmas \ref{lem:Wells4.4} (e) and \ref{lem:Wells4.5} (a) ensure that 
$\barrho_j(f,g) :=\norm{\thetam_j^f-\thetam_j^g}_p^q,j=-1,0,1,\dots$ are of
negative type for $q\leq p \leq 2$. 
By Lemma \ref{lem:Wells4.5}, the function $F_j(t) = \exp(-2^{sjq}t)$
belongs to $\RPD(B_{p,q}^{\alpha'}[0,1])$, where $s=\alpha+1/2-1/p$.
For any finite product, by Lemma \ref{lem:Wells4.4} (b)
\begin{equation}
\label{eq:RPDfinite}
\prod_{j=-1}^n F_j(\barrho_j) = \exp \left\{ -\sum_{j=-1}^n 2^{sjq}\barrho_j \right\}
\end{equation}
belongs to $\RPD(B_{p,q}^{\alpha'}[0,1])$. Lemma \ref{lem:Wells4.4} (d) ensures the continuous
sequence limit of \eqref{eq:RPDfinite}, i.e., $\exp(-\rho_{b^{\alpha}_{p,q}})\in \RPD (B_{p,q}^{\alpha'}[0,1])$ as $n\rightarrow\infty$.
Therefore $\sum_{j\geq -1}2^{sjq}\barrho_j$ is of negative type on $B_{p,q}^{\alpha'}[0,1]$.
By Lemma \ref{lem:Wells4.5} (c), the $\left(B_{p,q}^{\alpha'}[0,1],
  \left(\sum_{j\geq -1}2^{sjq}\barrho_j\right)^{1/2}\right)$ is a metric space
isometrically embeddable in a Hilbert space. By the same procedure of Remark 3.19 in \citet{Lyon13},
the map
$$P \mapsto \left( f \mapsto \int_{B_{p,q}^{\alpha'}[0,1]}\left(\sum_{j\geq
    -1}2^{sjq}\barrho_j\right)^{r/2}(f,g) \dd P(g) \right)$$
is injective for any $r \in (0,\infty)\backslash 2\N$, where $\N$ is the set of natural numbers
\citep{Lind86}. The result folllows from the fact that $(\sum_{j\geq -1}2^{sjq}\barrho_j)^{r/2}$ is of negative
type for any $r\in(0,2)$ by Lemma \ref{lem:Wells4.7}.
\end{proof}

\subsection{Proof of Theorem %
\ref{thm:denoise}}

\begin{proof}[Proof of Theorem %
\ref{thm:denoise}]
Here we prove a more general result where $\tau_j$ and $j_{\#}^Z$ involved in the penalty $\pen_j(k) = k\zeta_Z \{1+\sqrt{2(1+2\varsigma_Z)\log (\tau_j m_j/k)}\}^2$ in Step 1 are replaced by 
$\tau_j = \tau_Z 2^{2\beta_Z'(j-j_{\#}^Z)_+}$ and 
  $j_{\#}^Z=(1+(\varsigma_Z+1/2)/\beta_Z')/(\alpha_Z+\varsigma_Z+1/2) \cdot
  \log_2 \delta_Z^{-1}$ for any $\beta_Z<\beta_Z'\leq\alpha_Z$ respectively. Apparently Theorem \ref{thm:denoise} is a special case where $\beta_Z'=\alpha_Z$.

For notational simplicity, %
we omit the subscript $Z$ and subject index $i$ in all terms; namely we replace 
$\delta_{Z,j}$ by $\delta_j$, $\delta_Z$ by $\delta$, $\varsigma_Z$ by
$\varsigma$, $C_Z$ by $C$, $\alpha_Z$ by $\alpha$, $\beta_Z$ by $\beta$ and $\beta_Z'$ by $\beta'$ respectively. We further replace $\thetam^{e_i^Z}$ by $\thetam$ and $\theta_{j,k}^{e_i^Z}$ by $\theta_{j,k}$ respectively.

We first decompose the loss function by
\begin{equation*}
\E \left(\nmbb{\hthetam-\thetam}^2\mid \thetam \right) = \sum_{j\geq -1} 2^{2\beta j} \E\left( \norm{\hthetam_j - \thetam_j}^2\mid \thetam_j\right).
\end{equation*}
By Theorem 11.11 in \citet{John19}, %
there exist constants $a(\zeta)$ and $b(\zeta)$ that depend on $\zeta$, 
and $M_j = M_j(\varsigma,\tau_j)$ that depends on $\varsigma$ and $\tau_j$ such that
\begin{equation*}
\E \left(\norm{\hthetam_j - \thetam_j}_2^2\mid \thetam_j\right) \leq b(\zeta)\xi_1 M_j \delta_j^2 + a(\zeta) \mR_j(\thetam_j, \delta_j),
\end{equation*}
where $\mR_j(\thetam_j,\delta_j)=\min_{K\subseteq \{0,\dots,2^{j-1}\}}\left[ \sum_{k\notin
K} \theta_{j,k}^2 + \delta_j^2 \pen_j(\norm{\thetam_j}_0) \right]$.
Therefore,
\begin{equation*}
\E \left(\nmbb{\hthetam-\thetam}^2\mid\thetam\right) = b(\zeta)\xi_1 \sum_j 2^{2\beta j}M_j \delta_j^2+ a(\zeta) \sum_j2^{2\beta j}\mR_j(\thetam_j,\delta_j), 
\end{equation*}
and it suffices to study the upper bounds of $\text{(I)}=\sum_j 2^{2\beta j}M_j \delta_j^2$ and $\text{(II)}=\sum_j2^{2\beta j}\mR_j(\thetam_j,\delta_j)$ respectively. 

\paragraph{Bound of (I)}
By (11.67) in \citet{John19}, %
$
M_j\leq \tau^{-1} c_{\zeta,\tau} 2^{-2\varsigma j} 2^{-2\beta'(j-j_{\#})_+}
$, where $c_{\zeta,\tau}$ is a constant that depends on $\zeta$ and $\tau$.
Thus 
\begin{align*}
\text{(I)}= \sum_j 2^{2\beta j} M_j \delta_j^2 &\leq \tau^{-1} c_{\varsigma,\tau} \delta^2 \left( \sum_{j=-1}^{j_{\#}}2^{2\beta j} + 2^{2\beta' j_{\#}}\sum_{j>j_{\#}} 2^{-2(\beta'-\beta)j}\right)\\
&= \tau^{-1} c_{\varsigma,\tau} \delta^2 \left( \sum_{j=-1}^{j_{\#}}2^{2\beta j} + 2^{2\beta j_{\#}}\sum_{j>j_{\#}} 2^{-2(\beta'-\beta)(j-j_{\#})}\right)\\
&= \tau^{-1} c_{\varsigma,\tau} \delta^2 \left( 2^{-2\beta}+\frac{2^{2\beta(j_{\#}+1)}-1}{2^{2\beta}-1} + 2^{2\beta j_{\#}}\sum_{j=1}^{\infty} 2^{-2(\beta'-\beta)(j-j_{\#})}\right)\\
&\leq \tau^{-1} c_{\varsigma,\tau} \delta^2 \left( 2^{-2\beta}+\frac{2^{2\beta}(\delta^{-w})^{2\beta}}{2^{2\beta}-1} + (\delta^{-w})^{2\beta}/(1-2^{-2(\beta'-\beta)})\right)\\
&\leq c_{\varsigma,\tau,\beta,\beta'} \delta^{2(1-w\beta)},
\end{align*}
where $w=(\alpha+\varsigma+1/2)^{-1} [1+(\varsigma+1/2)/\beta]$ and $c_{\zeta,\tau}$ is a constant that depends on $\varsigma,\tau,\beta$ and $\beta'$. %

\paragraph{Bound of (II)}
According to (11.40) in \citet{John19}, %

\begin{equation*}
\sup_{\thetam_j:\nmba{\thetam}\leq C} \mR_j(\thetam_j,\delta_j) \leq c_{\zeta,\xi_1,\varsigma}\log \tau_j r_j(C_j,\delta_j),
\end{equation*}
where $C_j=2^{-\alpha j}, m_j=2^j$, $c_{\zeta,\xi_1,\varsigma}$ is a constant that 
depends on $\zeta, \xi_1$ and $\varsigma$, and
\begin{equation*}
  r_j(C_j,\delta_j)=
  \begin{cases}
    C_j^2, &\text{if } C_j\leq \delta_jm_j^{1/2},\\
    m_j\delta_j^2, &\text{if } C_j\geq \delta_jm_j^{1/2}.
  \end{cases} 
\end{equation*}
Notice that $\log \tau_j = \log\tau + 2\beta'(\log 2)(j-j_{\#})_+$, so we have
\begin{eqnarray} \label{eqn:II}
\text{(II)}&\leq c_{\zeta,\xi_1,\varsigma} \left\{ (\log\tau) \sum_{j\geq -1} 2^{2\beta j} r_j(C_j,\delta_j) + 2\beta'(\log 2)\sum_{j > j_{\#}}(j-j_{\#})2^{2\beta j}r_j(C_j,\delta_j) \right\} \nonumber\\
  &=c_{\zeta,\xi_1,\varsigma} \Bigg\{ (\log\tau) \sum_{j\geq -1} Q_j+ 2\beta'(\log 2)\sum_{j > j_{\#}}(j-j_{\#})Q_j \Bigg\}, 
\end{eqnarray}
where $Q_j=2^{2\beta j}r_j(C_j,\delta_j)$. Next we handle $\text{(III)}=\sum_{j\geq -1} Q_j$ and $\text{(IV)}=\sum_{j > j_{\#}}(j-j_{\#})Q_j$ individually.

\begin{itemize}
\item $\text{(III)}=\sum_{j\geq -1} Q_j$. We calculate $Q_j$ respectively for
  $j\geq -1$.
Define $j_{*}=(\alpha+\varsigma+1/2)^{-1}\log_2(C/\delta)$. %
\begin{enumerate}[1$^{\circ}$]
  \item When $j\leq j_{*}$, $C_j\geq \delta_j m_j^{1/2}$, so that
\begin{equation*}
Q_j = 2^{2\beta j}m_j \delta_j^2=2^{(2\beta+2\varsigma+1)j}\delta^2.
\end{equation*}
\item When $j\geq j_{*}$, $C_j\leq \delta_j m_j^{1/2}$, so that
\begin{equation*}
Q_j = 2^{2\beta j}C_j^2=2^{-2(\alpha-\beta)j}C^2.
\end{equation*}
\end{enumerate}
Combining 1$^{\circ}$ and 2$^{\circ}$, we have
\begin{equation*}
  Q_j =
  \begin{cases}
    Q^{*} 2^{(2\beta+2\varsigma+1)(j-j_{*})}, & j\leq j_{*},\\
    Q^{*} 2^{-(\alpha-\beta)(j-j_{*})}, & j\geq j_{*},
  \end{cases}
\end{equation*}
where $Q^{*} = C^{2(1-r)}\delta^{2r}$ with $r=(\alpha-\beta)/(\alpha+\varsigma+1/2)$.
Therefore, (III)$\leq c_1 Q^{*}$.

\item $\text{(IV)}=\sum_{j > j_{\#}}(j-j_{\#})Q_j$. 
When $m$
is sufficiently large, $\delta \asymp m^{-1/2}\rightarrow 0$, 
and $j_{\#} > j_{*}$ since $1+(\varsigma+1/2)/\beta'>1$. 
Thus for $m$ large enough,
\begin{equation*}
\text{(IV)}\leq\sum_{j\geq j_{*}}(j-j_{*})Q_j = Q^{*}\sum_{j\geq j^{*}} 2^{-(\alpha-\beta)(j-j_{*})} \leq c_2 Q^{*}.
\end{equation*}

\end{itemize}

Hence by (\ref{eqn:II}), (II)$\leq c_3 C^{2(1-r)}\delta^{2r}$ where the constant 
$c_3$ depends on $c_1,c_2, \zeta, \xi_1,\varsigma,\tau$ and $\beta'$. %

Combining the upper bounds for (I) and (II) respectively, we have
\begin{equation*}
\sup_{\thetam:\nmbb{\thetam}\leq C}\E \left(\nmbb{\hthetam-\thetam}^2 \mid \thetam \right)\leq b(\zeta)\xi_1 c_{\varsigma,\tau,\beta,\beta'}\delta^{2(1-w\beta)}+ c_3C^{2(1-r)}\delta^{2r}=O(\delta^{2r}),
\end{equation*}
since 
\begin{equation*}
  1-w\beta = 1- \frac{(1+(\varsigma+1/2)/\beta')\beta}{\alpha+\varsigma+1/2} \geq 1- \frac{(1+(\varsigma+1/2)/\beta)\beta}{\alpha+\varsigma+1/2} = \frac{\alpha-\beta}{\alpha+\varsigma+1/2}=r.
\end{equation*}
\end{proof}

\subsection{Proof of Theorem \ref{thm:wkconvD}}

We first present a lemma that will be used to prove Theorem \ref{thm:wkconvD}.

\begin{lem}%
\label{lem:wkconv}
Let $\left\{ (X_i(\cdot),Y_i(\cdot)
\right\}_{i=1}^n$ be i.i.d. fully observed random samples from probability measure $P_{XY} = P_XP_Y$
defined on $\mX\otimes\mY$. 
Then as $n \rightarrow \infty$, 
\begin{equation}
\label{eq:wkconv}
n\gamma(P_{n,XY},\kx,\ky)\wkconv \sum_{r=1}^{\infty}\sum_{s=1}^{\infty}\mu_r\nu_s N_{rs}^2,
\end{equation}
where $N_{rs}\sim N(0,1),r,s\in\N$ are i.i.d. and $\left\{ \mu_r
\right\}_{r=1}^{\infty}$ and $\left\{\nu_s\right\}_{s=1}^{\infty}$ are eigenvalues of the
integral kernel operators $S_{\ckx}$ and $S_{\cky}$, respectively.
If $P_{XY}\neq P_X P_Y$, then $n\gamma(P_{n,XY},\kx,\ky)\rightarrow\infty$ in probability as $ n\rightarrow \infty$. 
\end{lem} 
Lemma \ref{lem:wkconv} is exactly Theorem 33 of \citet{SejdSG13}, which provides the weak convergence result of HSIC for fully observed random functions. %

\begin{proof}[Proof of Theorem \ref{thm:wkconvD}]
  According to Lemma \ref{lem:wkconv}, it suffices to prove that the difference
  between HSIC based on original curves $\{X_i(\cdot),Y_i(\cdot)\}_{i=1}^n$ and
  HSIC based on denoised curves $\{\hX_i,\hY_i\}_{i=1}^n$ is $o_p(1/n)$, where
  $\{\hX_i,\hY_i\}_{i=1}^n$ are obtained by Step 1 in Section \ref{sec:method}.
By Definition \ref{def:HSIC},
\begin{align*}
& n\left|\gamma(P_{n,XY},\kx,\ky) - \gamma(P_{n,\hX\hY},\kx,\ky)\right| = n^{-1}\left|\norm{\bkx^{\top}\bH\bky}_{\Hkxy}^2 - \norm{\hbkx^{\top}\bH\hbky}_{\Hkxy}^2\right|\\
=& n^{-1}\left|\norm{\bkx^{\top}\bH\bky}_{\Hkxy} - \norm{\hbkx^{\top}\bH\hbky}_{\Hkxy}\right|  \left(\norm{\bkx^{\top}\bH\bky}_{\Hkxy} + \norm{\hbkx^{\top}\bH\hbky}_{\Hkxy}\right)\\
\leq & n^{-1}\norm{\bkx^{\top}\bH\bky - \hbkx^{\top}\bH\hbky}_{\Hkxy} \left(\norm{\bkx^{\top}\bH\bky}_{\Hkxy} + \norm{\hbkx^{\top}\bH\hbky}_{\Hkxy}\right)\\
\leq & 2 n^{-1/2} \norm{\bkx^{\top}\bH\bky - \hbkx^{\top}\bH\hbky}_{\Hkxy}\times n^{-1/2}\norm{\bkx^{\top}\bH\bky}_{\Hkxy} + n^{-1}\norm{\bkx^{\top}\bH\bky - \hbkx^{\top}\bH\hbky}_{\Hkxy}^2%
\end{align*}
where $\bkx^{\top} = \left[ \kx(\cdot,X_1),\dots,\kx(\cdot,X_n) \right]$,
$\bky^{\top} = \left[ \ky(\cdot,Y_1),\dots,\ky(\cdot,Y_n) \right]$,
$\hbkx^{\top} = \left[ \kx(\cdot,\hX_1),\dots,\kx(\cdot,\hX_n) \right]$,
$\hbky^{\top} = \left[ \ky(\cdot,\hY_1),\dots,\ky(\cdot,\hY_n) \right]$.

By \eqref{eq:wkconv},
\begin{equation}
\label{eq:esckxHky}
n^{-1/2} \norm{\bkx^{\top}\bH\bky}_{\Hkxy} \wkconv \sqrt{\sum_{r=1}^{\infty}\sum_{s=1}^{\infty}\mu_r\nu_s N_{rs}^2} = O_p(1), 
\end{equation}
so it suffices to prove that 
$
\norm{\bkx^{\top}\bH\bky - \hbkx^{\top}\bH\hbky}_{\Hkxy} = o_p\left(n^{1/2}\right).
$

Notice that $\norm{\bkx^{\top}\bH\bky - \hbkx^{\top}\bH\hbky}_{\Hkxy}$ can be bounded by
the following inequality:
\begin{align*}
&\quad\norm{\bkx^{\top}\bH\bky - \hbkx^{\top}\bH\hbky}_{\Hkxy} = \norm{\bkx^{\top}\bH \left( \bky - \hbky \right) +  \left( \bkx - \hbkx \right)\bH \hbky^{\top}}_{\Hkxy}\\
  &\leq \norm{\bkx^{\top}\bH \left( \bky - \hbky \right)}_{\Hkxy} + \norm{\left( \bkx - \hbkx \right)\bH\hbky^{\top}}_{\Hkxy} \\
  &\leq \norm{\bkx^{\top}\bH \left( \bky - \hbky \right)}_{\Hkxy} + \norm{ \left( \bkx - \hbkx \right)\bH\bky^{\top}}_{\Hkxy} + \norm{ \left( \bkx - \hbkx \right)\bH\left( \bky - \hbky \right)^{\top}}_{\Hkxy} \\
  &= \tr^{\frac{1}{2}} \left( \GamX \bH \ip{\bky - \hbky,\bky^{\top} - \hbky^{\top}}_{\Hky} \bH \right) + \tr^{\frac{1}{2}} \left( \GamY \bH \ip{\bkx - \hbkx,\bkx^{\top} - \hbkx^{\top}}_{\Hkx} \bH \right)\\ 
  &\quad + \tr^{\frac{1}{2}} \left( \ip{\bkx - \hbkx,\bkx^{\top} - \hbkx^{\top}}_{\Hkx} \bH \ip{\bky - \hbky,\bky^{\top} - \hbky^{\top}}_{\Hky} \bH \right)\\
  &= \tr^{\frac{1}{2}} \left( \cGamX \ip{\bky - \hbky,\bky^{\top} - \hbky^{\top}}_{\Hky} \right) + \tr^{\frac{1}{2}} \left( \cGamY \ip{\bkx - \hbkx,\bkx^{\top} - \hbkx^{\top}}_{\Hkx} \right)\\
  &\quad + \tr^{\frac{1}{2}} \left( \ip{\bkx - \hbkx,\bkx^{\top} - \hbkx^{\top}}_{\Hkx} \bH \ip{\bky - \hbky,\bky^{\top} - \hbky^{\top}}_{\Hky} \bH \right)\\
  &\leq \tr^{\frac{1}{2}} (\cGamX) \tr^{\frac{1}{2}}(\ip{\bky - \hbky,\bky^{\top} - \hbky^{\top}}_{\Hky})  + \tr^{\frac{1}{2}} (\cGamY) \tr^{\frac{1}{2}} (\ip{\bkx - \hbkx,\bkx^{\top} - \hbkx^{\top}}_{\Hkx}) \tag{*} \label{eq:star} \\ 
  &\quad +\tr^{\frac{1}{2}} (\ip{\bkx - \hbkx,\bkx^{\top} - \hbkx^{\top}}_{\Hkx})\tr^{\frac{1}{2}}(\ip{\bky - \hbky,\bky^{\top} - \hbky^{\top}}_{\Hky}) = o_p(n^{1/2}),
\end{align*}
where $\cGamX = \bH\GamX \bH$ and $\cGamY = \bH\GamY \bH$ are centered Gram matrices.

In \eqref{eq:star} we used the fact that for symmetric positive definite matrices $\bA$ and $\bB$,
\begin{equation*}
\tr \bA\bB = \ve (\bA)^{\top} \ve(\bB) \leq \norm{\bA}_F \norm{\bB}_F = \sqrt{\tr \bA^2 \tr \bB^2} \leq \tr \bA \tr \bB.
\end{equation*}
The last equation holds due to the facts below with $(\mZ,Z,z)=(\mX,X,x)$ or $(\mY,Y,y)$:
\begin{itemize}
\item 
$\tr(\cGamZ) = O_p(n)$ because $\int_{\mZ}\ckz(z,z)\dd P_Z(z)<\infty$ which is ensured
by the assumptions in Theorem \ref{thm:denoise}. %
\item
$\tr\ip{\bkz - \hbkz,\bkz^{\top} - \hbkz^{\top}}_{\Hkz} =
\sum_{i=1}^n\norm{\kz(\cdot,Z_i) - \kz(\cdot,\hZ_i)}_{\Hkz}^2 = o_p(1)$, because
\begin{align*}
&\norm{\kz(\cdot,Z_i) - \kz(\cdot,\hZ_i)}_{\Hkz}^2 = \kz(Z_i,Z_i) + \kz(\hZ_i,\hZ_i) - 2\kz(Z_i,\hZ_i)\\
=& 2 \nmbbZ{Z_i} + 2 \nmbbZ{\hZ_i} - 2 \left( \nmbbZ{Z_i} + \nmbbZ{\hZ_i} - \nmbbZ{Z_i - \hZ_i} \right)= 2 \nmbbZ{Z_i - \hZ_i},
\end{align*}
and $\nmbbZ{Z_i - \hZ_i} = o_p(n^{-1}), i=1,\dots,n$ ensured by 
Theorem \ref{thm:denoise} and (\ref{eq:smoothrate}) in Theorem \ref{thm:wkconvD}. %
\end{itemize}

\end{proof}

\subsection{Proof of Theorem \ref{thm:perm}}
We first introduce a few notations. To perform a permutation test, let $\mS(n) = \{\sigma_1,\dots,\sigma_{n!}\}$
be the cyclic group of $\{1,\dots,n\}$. For a permutation $\sigma$ randomly
selected from $\mS(n)$, let $\gamma(P_{n,\hX\hY}^{\sigma},\kx,\ky) =
n^{-2}\tr(\GamhX \bH\GamhY(\sigma) \bH)$, where $\GamhY(\sigma)$ is generated
by $\GamhY$ with rows and columns permuted according to $\sigma$. Let $R$ be the rank of $\gamma(P_{n,\hX\hY},\kx,\ky)$ in all 
possible permuted HSICs.
Then we reject $H_0:
P_{XY}=P_XP_Y$ if $p_{\hX\hY} = R/n!\leq \alpha$, where $p_{\hX\hY}$ denotes the 
p-value of the permutation test enumerating all permutations and $\alpha$ is the
level of significance.

In practice, it is impractical to %
consider all permutations from
$\mS(n)$. Hence we use a Monte-Carlo approximation by randomly choosing $B$ permutations
$\sigma_1,\dots,\sigma_B\in \mS(n) \backslash \{\text{\rm{id}}\}$ where
$\text{\rm{id}}$ refers to no permutation 
and calculating
$\gamma(P_{n,\hX\hY},\kx,\ky),\gamma(P_{n,\hX\hY}^{\sigma_1},\kx,\ky),\dots,\gamma(P_{n,\hX\hY}^{\sigma_B},\kx,\ky)$.
With a notational abuse, let $R$ be the rank of $\gamma(P_{n,\hX\hY},\kx,\ky)$ and we reject $H_0$ if
$\hat p_{\hX\hY} = R/(B+1)\leq \alpha$, where $\hat p_{\hX\hY}$ is the p-value
of the permutation test enumerating a finite sample of size $B$ from $\mS(n)$. 

If the value of $\gamma(P_{n,\hX\hY},\kx,\ky)$ repeats
in
$\{\gamma(P_{n,\hX\hY}^{\sigma_1},\kx,\ky),\dots,\gamma(P_{n,\hX\hY}^{\sigma_B},\kx,\ky)\}$
for several times with
$B\leq n!$, the rank $R$ of $\gamma(P_{n,\hX\hY},\kx,\ky)$ is determined by the
following two ways proposed by \citet{RindSS20}. 
\begin{itemize}
\item Breaking ties at random: $R$ is distributed uniformly on ranks of
  $\gamma(P_{n,\hX\hY}^{\sigma},\kx,\ky)$ that have
  the same value of $\gamma(P_{n,\hX\hY},\kx,\ky)$;
\item Breaking ties conservatively: $R$ is the largest among ranks of
  $\gamma(P_{n,\hX\hY}^{\sigma},\kx,\ky)$ that have the same value of $\gamma(P_{n,\hX\hY},\kx,\ky)$.
\end{itemize}

Next we list two lemmas which will be useful to prove Theorem \ref{thm:perm}.

\begin{lem}%
  \label{lem:perm1}
For $\sigma$ randomly selected from $\mS(n)$,
  $\gamma(P_{n,\hX\hY},\kx,\ky)\rightarrow 0$ in probability as $n \rightarrow \infty$. 
\end{lem}
Lemma \ref{lem:perm1} is a direct application of Theorem 3 of \citet{RindSS20} for $d=2$. 

\begin{lem}%
  \label{lem:perm2}
  Suppose that the alternative
  hypothesis $H_1: P_{XY}\neq P_XP_Y$ is true and noises are i.i.d. Let $\{t_n^1(\hmD)\geq \dots \geq
  t_n^{n!}(\hmD)\}$ be ordered values of HSIC computed on all permutations of
  denoised curves
  $\{\gamma(P_{n,\hX\hY}^{\sigma_1},\kx,\ky),\dots,\gamma(P_{n,\hX\hY}^{\sigma_{n!}},\kx,\ky)\}$.
  Let $a = \lfloor n!\alpha \rfloor$ for any level of significance $\alpha\in(0,1)$. Then
  $t_n^a (\hmD)\rightarrow 0$ in probability as $n \rightarrow \infty$.
\end{lem}
Lemma \ref{lem:perm2} is a direct application of Theorem 4 of \citet{RindSS20} for $d=2$.

\begin{proof}[Proof of Theorem \ref{thm:perm}]
Denote the fully observed dataset by 
$\mD = \{(X_i,Y_i): i=1,\dots,n\}$ and the %
denoised %
dataset by $\hmD = \{(\hX_i,\hY_i): i=1,\dots,n\}$. For a permutation
$\sigma\in \mS(n)$, denote the permuted datasets by $\sigma(\mD)$ and $\sigma(\hmD)$, 
resulting in permuted HSIC $\gamma(P_{n,XY}^{\sigma},\kx,\ky)$ and
$\gamma(P_{n,\hX\hY}^{\sigma},\kx,\ky)$ respectively.

\paragraph{\textbf{If $H_0: P_{XY}=P_XP_Y$ is true,}} then for any $\sigma\in\mS(n)$, $\mD$ and $\sigma(\mD)$ have the
same distribution and $\hmD$ and $\sigma(\hmD)$ have the same distribution due to the facts that 
the noise across subjects are i.i.d 
and that the denoising procedure in Section \ref{sec:method} is separately for each subject. For
$B$ permutations $\sigma_1,\dots,\sigma_B$ randomly selected from $\mS(n)
\backslash \{\text{\rm{id}}\}$,
$(\mD,\sigma_1(\mD),\dots,\sigma_B(\mD))$
is an exchangeable vector, and thus $\left(\gamma(P_{n,\hX\hY},\kx,\ky),\gamma(P_{n,\hX\hY}^{\sigma_1},\kx,\ky),\dots,\gamma(P_{n,\hX\hY}^{\sigma_B},\kx,\ky)\right)$ is exchangeable.

By breaking ties at random, each entry is equally likely to have any given rank,
so the rank of $\gamma(P_{n,\hX\hY},\kx,\ky)$ is uniformly distributed in
$\{1,\dots,B\}$. Therefore the type I error rate can be controlled for any level of significance $\alpha\in (0,1)$. Breaking ties conservatively can result in
an even smaller Type I error rate.

\vspace{5pt}

\paragraph{
\textbf{If $H_1: P_{XY}\neq P_XP_Y$ is true,}} then by the definition of $t_n^a(\hmD)$ in Lemma
\ref{lem:perm2}, we reject
$H_0: P_{XY}=P_XP_Y$ if $\gamma(P_{n,\hX\hY},\kx,\ky) > t_n^a(\hmD)$. For any $\alpha\in (0,1)$,
\begin{equation*}
\lim_{n\rightarrow\infty}P(p_{\hX\hY}\leq\alpha)\geq\lim_{n\rightarrow\infty}P(\gamma(P_{n,\hX\hY},\kx,\ky) > t_n^a(\hmD))=1,
\end{equation*}
since
$\gamma(P_{n,\hX\hY},\kx,\ky)\rightarrow\gamma(P_{XY},\kx,\ky)>0$ in probability as $n \rightarrow \infty$ 
by the proof of Theorem \ref{thm:wkconvD}. 

For a finite number $B$ of permutations, the p-value $\hat p_{\hX\hY} = (1+U)/(B+1)$ where
$U\sim\text{Binomial}(B, p_{\hX\hY})$. If $U=0$, then $\hat p_{\hX\hY} =
1/(B+1)\leq \alpha$ and we reject the null hypothesis. Since
$P(p_{\hX\hY}\leq\eps_1)\geq 1-\eps_2$ for some $\eps_1,\eps_2>0$. For $n$ large
enough, we have
\begin{align*}
  P(\hat p_{\hX\hY}) &\geq P(\hat p_{\hX\hY} = 1/(B+1)\mid p_{\hX\hY}\leq \eps_1) P(p_{\hX\hY}\leq \eps_1)\\
  &\geq (1-\eps_1)^B(1-\eps_2).
\end{align*}
Then the consistency of the permutation test is proved by letting $\eps_1,\eps_2\rightarrow 0$.
\end{proof}

\section{Additional Simulation}
\subsection{Performance of \textsf{wavHSIC} for Irregular Design}
In this section, we present the results of a simulation study where 
subjects are not measured at the same regular grid with $m=2^{J+1}$ for some integer $J$. 

Similar to Section \ref{sec:simu}, we had 199 simulation runs and in each simulation run 
$\{(X_i(t),Y_i(t)): t\in [0, 1], i=1, \ldots, n\}$ where $n=50$ or $200$ were generated under Settings 1--3. %
For each subject $i$, $i=1,\dots,n$, the numbers of measurements per subject, $m_i^X$ and $m_i^Y$, were both sampled from either $\text{DiscreteUnif}\{50, \dots, 70\}$ or $\text{DiscreteUnif}\{220,\dots, 280\}$. Given $m_i^X$ and $m_i^Y$, the measurement times $\{T_{il}^X:
l=1\dots,m_i^X\}$ and $\{T_{il}^Y: l=1\dots,m_i^Y\}$ were sampled independently on $\text{ContinuousUnif}[0,1]$. Since the number of measurements per subject and measurement times may be different across subjects, their notations here have an additional subscript ``i'' compared to those in Section \ref{sec:method}. 
We added white Gaussian noise to all measurements
with signal-to-noise ratio SNR=4 or 8. Therefore, the observed data were
$\{\tX_i(T_{il}^X)=X_i(T_{il}^X)+e_{il}^X: l=1,\dots,m_i^X\}$ and
$\{\tY_i(T_{il}^Y)=Y_i(T_{il}^Y)+e_{il}^Y: l=1,\dots,m_i^Y\}$.

Before the two steps in Section \ref{sec:twostep}, we performed the linear interpolation method by
\citet{KovaS00} %
to interpolate data onto a common and regular grid of $[0,1]$ with $m=2^{J+1}$ for some integer $J$. 
When $m_i^X,m_i^Y\sim \text{DiscreteUnif}\{50, \dots, 70\}$, we chose $m=64$; when $m_i^X,m_i^Y\sim
\text{DiscreteUnif}\{220, \dots, 280\}$, we chose $m=256$.
The results are given in Table \ref{tab:additional}. Compared with Tables \ref{tab:case1} - \ref{tab:case3}, \textsf{wavHSIC} now performs slightly worse in controlling the Type I error rate and achieving a high power, but it is overall satisfactory. 

\begin{table}[htb!] 
\caption{
Rejection rates of \textsf{wavHSIC} when subjects are not measured at the same dyadic grid with $m=2^{J+1}$. 
Medians of selected $\beta_X$ and $\beta_Y$ are provided for each setting. 
} 
\label{tab:additional}
\centering
\scriptsize

\begin{tabular}{ll|rrrrrrrr}
\toprule
          & & \multicolumn{4}{c}{$n=50$} & \multicolumn{4}{c}{$n=200$}\\
              \cmidrule(lr){3-6}           \cmidrule(lr){7-10}
          & & \multicolumn{2}{c}{$m\sim 50-70$} & \multicolumn{2}{c}{$m\sim 220-280$}& \multicolumn{2}{c}{$m\sim 50-70$} & \multicolumn{2}{c}{$m\sim 220-280$}\\
              \cmidrule(lr){3-4}\cmidrule(lr){5-6}\cmidrule(lr){7-8}\cmidrule(lr){9-10}
          &                     & SNR=4 & SNR=8 & SNR=4 & SNR=8 & SNR=4 & SNR=8 & SNR=4 & SNR=8                                         \\
          \midrule
          & Type I error rate   & 0.0553 & 0.0503 & 0.0955 & 0.0905 & 0.0452 & 0.0452 & 0.0402 & 0.0603 \\
Setting 1 & median$\{\beta_X\}$ & 1.066 & 1.126 & 0.972 & 0.988 & 1.072 & 1.131 & 0.990 & 1.005         \\
          & median$\{\beta_Y\}$ & 0.791 & 0.866 & 0.723 & 0.745 & 0.807 & 0.871 & 0.742 & 0.764         \\
          \midrule
          & Power               & 0.8291 & 0.9296 & 0.9648 & 0.9598 & 1.0000 & 1.0000 & 1.0000 & 1.0000 \\
Setting 2 & median$\{\beta_X\}$ & 1.059 & 1.124 & 0.979 & 0.994 & 1.073 & 1.133 & 0.989 & 1.009         \\
          & median$\{\beta_Y\}$ & 0.788 & 0.860 & 0.723 & 0.745 & 0.801 & 0.870 & 0.738 & 0.758         \\
          \midrule
          & Power               & 0.1859 & 0.2563 & 0.2814 & 0.2714 & 0.5578 & 0.7739 & 0.8291 & 0.8040 \\
Setting 3 & median$\{\beta_X\}$ & 1.074 & 1.133 & 0.961 & 0.981 & 1.066 & 1.124 & 0.974 & 0.995         \\
          & median$\{\beta_Y\}$ & 0.830 & 0.906 & 0.777 & 0.794 & 0.830 & 0.896 & 0.769 & 0.789         \\
\bottomrule
\end{tabular}
\end{table}

\subsection{Simulation Settings in \citet[][Supplementary Material]{LeeZS20}}

{
In this section, we run an additional simulation study under the same settings 
in \citet[][Supplementary
Material, Section 1.2]{LeeZS20} to compare our method \textsf{wavHSIC} with \textsf{PSS} and \textsf{FMDD}. We also include 
\textsf{KMSZ}, \textsf{KMSZ-p}, \textsf{dCov-c} and \textsf{FPCA} here due to their competitive performances shown in Section \ref{sec:simu}. 
Here we use the same strategies for tuning parameters as in Section \ref{sec:simu}. For \textsf{wavHSIC} in following examples, we perform linear interpolation method by 
\citet{KovaS00} to interpolate data onto a regular grid of $[0,1]$ with $m=2^{J+1}=64$.

\begin{ex} \citep[][Supplementary Material, Example 1]{LeeZS20} %
  We generated functional response $Y$ by a quadratic form of covariate $X$, 
\begin{equation*}
Y_i(t) = c \cdot \left\{ X_i(t)^2 - 1 \right\} + \eps_i(t),
\end{equation*}
where $X_i$ and $\eps_i$, $i=1,\dots,n$ are independent Brownian motion and Brownian bridge, respectively.
$X$ is independent of $Y$ when $c=0$, while the alternative is satisfied when $c=0.5$. Sampling points
are $t=1/200, 3/200, \dots, 199/200$, with sample size $n=40$ or $100$. The results are given in Table \ref{tab:ExS1}. 
\end{ex}

Table \ref{tab:ExS1} shows that \textsf{KMSZ}, \textsf{PSS}$(Y\sim X)$, \textsf{FMDD}$(Y\sim X)$ 
perform essentially the same as that in \citet[][Supplementary Material, Table 1]{LeeZS20}.
Even the tests \textsf{PSS}$(X\sim Y)$ and \textsf{FMDD}$(X\sim Y)$ with 
the response and covariate switched 
can control type I error rates when $c=0$, but when $c=0.5$ their powers are much lower than that 
of \textsf{PSS}$(Y\sim X)$ and of \textsf{FMDD}$(Y\sim X)$ respectively. Two omnibus tests \textsf{PSS}(Omnibus) 
and \textsf{FMDD}(Omnibus) cannot control type I error probabilities when $c=0$. 
For two distance covariance methods, \textsf{dCov-c} cannot control type I error rate well when $\alpha=0.05,0.01$,
while \textsf{FPCA} has an accurate size for any combination of $(\alpha,n)$ when $c=0$.
When $c=0.5$, the powers of these two methods are uniformly better than or comparable with \textsf{PSS} and \textsf{FMDD}.
Our \textsf{wavHSIC} can almost always control the type I error rates when $c=0$ and is uniformly more powerful than all 
the other methods for all $(\alpha,n)$ when $c=0.5$.

\begin{ex}\citep[][Supplementary
Material, Example 2]{LeeZS20} We generate
\begin{align*}
  X_i(t)  &= \frac{4}{\pi}\sum_{k=1,3,\dots,21} Z_{i,k} \sin (2\pi kt),\\
  Y_i(t)  &= \frac{4}{\pi}\sum_{k=3,5,7,9} Z_{i,k}^2 \sin (2\pi kt) + 4\eps_i(t),
\end{align*}
where $Z_{i,k}, k=1,\dots, 21$, $i=1,\dots,n$ are i.i.d. $N(0,1)$ random variables and
$\eps_i(t)$, $i=1,\dots,n$ are standard Brownian bridges on $[0,1]$. Sampling points are $t=1/200, 3/200, \dots, 199/200$, with 
sample size $n=40$ or $100$. The results are given in Table \ref{tab:ExS2}. 
\end{ex}

Table \ref{tab:ExS2} shows that \textsf{KMSZ}, \textsf{PSS}$(Y\sim X)$, \textsf{FMDD}$(Y\sim X)$ 
perform almost the same as those in \citet[][Supplementary Material, Table 2]{LeeZS20}. Permutation
based \textsf{KMSZ-p} performs better than \textsf{KMSZ} when the sample size $n$ is small, but for small
nominal levels $\alpha=0.05$ or $0.01$, the powers of \textsf{KMSZ-p} are not as good as those of \text{KMSZ} for 
$n=100$. Similar to Table \ref{tab:ExS1}, the tests \textsf{PSS}$(X\sim Y)$ and \textsf{FMDD}$(X\sim Y)$ 
have much lower powers than \textsf{PSS}$(Y\sim X)$ and \textsf{FMDD}$(Y\sim X)$ respectively. 
Between the two distance covariance based methods, \textsf{FPCA} performs better than \textsf{dCov-c}.
\textsf{FPCA} performs better than other model-based methods for 
$n=40$, and its powers
lie between \textsf{PSS} and \textsf{FMDD} when $n=100$.
Our proposed method \textsf{wavHSIC} has uniformly
higher powers than the other methods. Interestingly, the median of $\beta_X$ are always 0 by our tuning
parameter selection strategy, which indicates %
that the distance variances across low to high frequencies for $X(t)$
are successfully detected as equally distributed.
}

\begin{table}[htb!] 
\caption{
{Rejection rates of six test methods for Example S1. 
For \textsf{wavHSIC}, medians of selected $\beta_X$ and $\beta_Y$ are provided for each setting. 
}
} 
  \label{tab:ExS1}
  \centering
  \scriptsize
{
\begin{tabular}{l|rrrrrr}
\toprule
$c=0$ & \multicolumn{2}{c}{$\alpha=0.1$} & \multicolumn{2}{c}{$\alpha=0.05$} &\multicolumn{2}{c}{$\alpha=0.01$}\\
Type I error rate& $n=40$ & $n=100$ & $n=40$ & $n=100$ &$n=40$ & $n=100$ \\
\cmidrule(lr){1-1} \cmidrule(lr){2-3} \cmidrule(lr){4-5} \cmidrule(lr){6-7}
\textsf{dCov-c} &                    0.1106 & 0.0955 & 0.0804 & 0.0653 & 0.0251 & 0.0553 \\
\textsf{FPCA} &                      0.1055 & 0.0955 & 0.0452 & 0.0603 & 0.0101 & 0.0101 \\
\cmidrule(lr){1-1} \cmidrule(lr){2-3} \cmidrule(lr){4-5} \cmidrule(lr){6-7}
\textsf{KMSZ} &                      0.0352 & 0.0754 & 0.0101 & 0.0302 & 0.0000 & 0.0050 \\
\textsf{KMSZ-p} &                    0.1156 & 0.0854 & 0.0553 & 0.0452 & 0.0000 & 0.0151 \\
\cmidrule(lr){1-1} \cmidrule(lr){2-3} \cmidrule(lr){4-5} \cmidrule(lr){6-7}
\textsf{PSS}$(Y\sim X)$ &            0.1407 & 0.1005 & 0.0704 & 0.0402 & 0.0402 & 0.0050 \\
\textsf{PSS}$(X\sim Y)$ &            0.0854 & 0.1106 & 0.0302 & 0.0503 & 0.0050 & 0.0000 \\
\textsf{PSS}(Omnibus) &              0.2060 & 0.2010 & 0.1005 & 0.0905 & 0.0452 & 0.0050 \\
\cmidrule(lr){1-1} \cmidrule(lr){2-3} \cmidrule(lr){4-5} \cmidrule(lr){6-7}
\textsf{FMDD}$(Y\sim X)$ &           0.1005 & 0.0905 & 0.0653 & 0.0653 & 0.0151 & 0.0151 \\
\textsf{FMDD}$(X\sim Y)$ &           0.1206 & 0.0704 & 0.0603 & 0.0452 & 0.0101 & 0.0151 \\
\textsf{FMDD}(Omnibus) &             0.1256 & 0.1005 & 0.0804 & 0.0653 & 0.0151 & 0.0201 \\
\cmidrule(lr){1-1} \cmidrule(lr){2-3} \cmidrule(lr){4-5} \cmidrule(lr){6-7}
\textsf{wavHSIC}  &                  0.1055 & 0.0955 & 0.0352 & 0.0553 & 0.0050 & 0.0101 \\
median$\{\beta_X\}$ &                0.957  & 0.958  & 0.957  & 0.958  & 0.957  & 0.958  \\
median$\{\beta_Y\}$ &                1.624  & 1.629  & 1.624  & 1.629  & 1.624  & 1.629  \\
\midrule
$c=0.5$ & \multicolumn{2}{c}{$\alpha=0.1$} & \multicolumn{2}{c}{$\alpha=0.05$} &\multicolumn{2}{c}{$\alpha=0.01$}\\
Power & $n=40$ & $n=100$ & $n=40$ & $n=100$ &$n=40$ & $n=100$ \\
\cmidrule(lr){1-1} \cmidrule(lr){2-3} \cmidrule(lr){4-5} \cmidrule(lr){6-7}
\textsf{dCov-c} &                    0.8141 & 1.0000 & 0.7286 & 0.9950 & 0.5477 & 0.9950 \\
\textsf{FPCA} &                      0.9598 & 1.0000 & 0.8995 & 1.0000 & 0.5678 & 0.9899 \\
\cmidrule(lr){1-1} \cmidrule(lr){2-3} \cmidrule(lr){4-5} \cmidrule(lr){6-7}
\textsf{KMSZ} &                      0.2362 & 0.3015 & 0.1256 & 0.2261 & 0.0352 & 0.0754 \\
\textsf{KMSZ-p} &                    0.3719 & 0.3367 & 0.2412 & 0.2714 & 0.1055 & 0.1055 \\
\cmidrule(lr){1-1} \cmidrule(lr){2-3} \cmidrule(lr){4-5} \cmidrule(lr){6-7}
\textsf{PSS}$(Y\sim X)$ &            0.4925 & 1.0000 & 0.3417 & 1.0000 & 0.1608 & 0.9347 \\
\textsf{PSS}$(X\sim Y)$ &            0.1005 & 0.0955 & 0.0553 & 0.0452 & 0.0101 & 0.0201 \\
\textsf{PSS}(Omnibus) &              0.5327 & 1.0000 & 0.3719 & 1.0000 & 0.1709 & 0.9347 \\
\cmidrule(lr){1-1} \cmidrule(lr){2-3} \cmidrule(lr){4-5} \cmidrule(lr){6-7}
\textsf{FMDD}$(Y\sim X)$ &           0.6734 & 1.0000 & 0.3970 & 0.9799 & 0.0704 & 0.6332 \\
\textsf{FMDD}$(X\sim Y)$ &           0.1709 & 0.0955 & 0.0854 & 0.0553 & 0.0201 & 0.0101 \\
\textsf{FMDD}(Omnibus) &             0.6734 & 1.0000 & 0.3970 & 0.9799 & 0.0704 & 0.6332 \\
\cmidrule(lr){1-1} \cmidrule(lr){2-3} \cmidrule(lr){4-5} \cmidrule(lr){6-7}
\textsf{wavHSIC}  &                  1.0000 & 1.0000 & 1.0000 & 1.0000 & 0.8492 & 1.0000 \\
median$\{\beta_X\}$ &                0.957  & 0.958  & 0.957  & 0.958  & 0.957  & 0.958  \\
median$\{\beta_Y\}$ &                1.278  & 1.266  & 1.278  & 1.266  & 1.278  & 1.266  \\
\bottomrule
\end{tabular}
}
\end{table}

\begin{table}[htb!] 
\caption{
{Rejection rates of six test methods for Example S1. 
For \textsf{wavHSIC}, medians of selected $\beta_X$ and $\beta_Y$ are provided for each setting. 
}
} 
  \label{tab:ExS2}
  \centering
  \scriptsize
{
\begin{tabular}{l|rrrrrr}
\toprule
& \multicolumn{2}{c}{$\alpha=0.1$} & \multicolumn{2}{c}{$\alpha=0.05$} &\multicolumn{2}{c}{$\alpha=0.01$}\\
Power & $n=40$ & $n=100$ & $n=40$ & $n=100$ &$n=40$ & $n=100$ \\
\cmidrule(lr){1-1} \cmidrule(lr){2-3} \cmidrule(lr){4-5} \cmidrule(lr){6-7}
 \textsf{dCov-c} &                    0.3266 & 0.7035 & 0.2362 & 0.5980 & 0.1407 & 0.3719 \\
 \textsf{FPCA} &                      0.5779 & 0.9045 & 0.4573 & 0.8392 & 0.2864 & 0.6281 \\
\cmidrule(lr){1-1} \cmidrule(lr){2-3} \cmidrule(lr){4-5} \cmidrule(lr){6-7}
 \textsf{KMSZ} &                      0.1910 & 0.0804 & 0.1910 & 0.3317 & 0.0955 & 0.2211 \\
 \textsf{KMSZ-p} &                    0.3568 & 0.3668 & 0.2563 & 0.2764 & 0.1055 & 0.1307 \\
\cmidrule(lr){1-1} \cmidrule(lr){2-3} \cmidrule(lr){4-5} \cmidrule(lr){6-7}
 \textsf{PSS}$(Y\sim X)$ &            0.0352 & 0.5477 & 0.1256 & 0.6734 & 0.1910 & 0.7487 \\
 \textsf{PSS}$(X\sim Y)$ &            0.0151 & 0.0050 & 0.0603 & 0.0503 & 0.1156 & 0.1005 \\
 \textsf{PSS}(Omnibus) &              0.0503 & 0.5528 & 0.1759 & 0.6834 & 0.2714 & 0.7638 \\
\cmidrule(lr){1-1} \cmidrule(lr){2-3} \cmidrule(lr){4-5} \cmidrule(lr){6-7}
 \textsf{FMDD}$(Y\sim X)$ &           0.5528 & 0.9950 & 0.3568 & 0.9598 & 0.1055 & 0.5980 \\
 \textsf{FMDD}$(X\sim Y)$ &           0.1709 & 0.1357 & 0.1005 & 0.0905 & 0.0151 & 0.0201 \\
 \textsf{FMDD}(Omnibus) &             0.5528 & 0.9950 & 0.3568 & 0.9598 & 0.1055 & 0.5980 \\
\cmidrule(lr){1-1} \cmidrule(lr){2-3} \cmidrule(lr){4-5} \cmidrule(lr){6-7}
 \textsf{wavHSIC}  &                  0.9598 & 1.0000 & 0.8844 & 1.0000 & 0.6131 & 0.9950 \\
 median$\{\beta_X\}$ &                0.000  & 0.000  & 0.000  & 0.000  & 0.000  & 0.000  \\
 median$\{\beta_Y\}$ &                0.618  & 0.579  & 0.618  & 0.579  & 0.618  & 0.579  \\
\bottomrule
\end{tabular}
}
\end{table}

\end{document}